\documentclass[aps,pre,twocolumn,amsmath,amssymb,nofootinbib,superscriptaddress,floatfix,showpacs]{revtex4-1}  
\usepackage{amsfonts}
\usepackage{amsmath}
\usepackage{graphicx}
\usepackage{dcolumn}
\usepackage{bm}

\usepackage[pdftex,colorlinks,citecolor=blue,linkcolor=blue,bookmarks=false,urlcolor=blue]{hyperref} %
\usepackage{comment}
\usepackage{datetime}
\usepackage{color}

\newcommand{\bea}{\begin{eqnarray}}
\newcommand{\eea}{\end{eqnarray}}

\begin{document}

\preprint{Work completed in 2006}
\title{Gene-Mating Dynamic Evolution Theory I: \\
Fundamental assumptions, exactly solvable models and analytic solutions}
\author{Juven C. Wang} \email{juven@ias.edu} 
\affiliation{Department of Physics, Massachusetts Institute of Technology, Cambridge, MA 02139, USA}
\affiliation{Perimeter Institute for Theoretical Physics, Waterloo, ON, N2L 2Y5, Canada}
\affiliation{School of Natural Sciences, Institute for Advanced Study, Einstein Drive, Princeton, NJ 08540, USA }
\affiliation{Center of Mathematical Sciences and Applications, Harvard University, MA 02138, USA}
\affiliation{Department of Physics, Center for Theoretical Physics,
and Leung Center for Cosmology and Particle Astrophysics,
National Taiwan University,
Taipei 10617, Taiwan}

\author{Jiunn-Wei Chen} \email{jwc@phys.ntu.edu.tw}
\affiliation{Department of Physics, Center for Theoretical Physics,
and Leung Center for Cosmology and Particle Astrophysics,
National Taiwan University,
Taipei 10617, Taiwan}
\date{Work completed in 2006. Work reported in 2014.}


\begin{abstract} 
Fundamental properties of macroscopic gene-mating dynamic evolutionary systems are investigated. 
A model is studied  
to describe a large class of systems within population genetics.
We focus on a single locus, any number of alleles in a two-gender dioecious population.
Our governing equations are time-dependent continuous differential equations
labeled by a set of parameters, where each
parameter stands for a population percentage carrying certain common 
genotypes. 
The full parameter space consists of all allowed parameters of these genotype frequencies. 
Our equations are uniquely derived 
from four fundamental assumptions within any population:
(1) a closed system; (2) average-and-random mating process (mean-field behavior); 
(3) Mendelian inheritance; 
(4) exponential growth and exponential death.
Even though our equations are nonlinear with time-evolutionary dynamics, 
we have obtained 
an exact analytic time-dependent solution 
and an exactly solvable model. 
Our findings are summarized from phenomenological and mathematical viewpoints.
From the phenomenological
viewpoint, 
any initial parameter of genotype frequencies
of a closed
system will eventually approach a stable fixed point. 
Under time
evolution,  
we show (1) the monotonic behavior of genotype frequencies, 
(2) any genotype or allele that appears in the population will never become extinct, 
(3) the Hardy-Weinberg law,
and (4) the global stability without chaos 
in the parameter space. 
To demonstrate the experimental evidence for our theory, as an example, we show a 
mapping from the data of blood type genotype frequencies 
of world ethnic groups 
to our stable fixed-point solutions.
From the mathematical
viewpoint, our highly symmetric governing equations 
result in continuous global stable equilibrium
solutions: 
These solutions altogether consist of a continuous curved
manifold as a subspace of the whole parameter space of 
genotype frequencies. 
This fixed-point 
manifold is a global stable attractor known as the Hardy-Weinberg manifold, 
attracting any initial point in any Euclidean
fiber bounded within the genotype frequency space 
to the fixed point where this fiber is attached. 
The stable
base manifold and its attached fibers form a fiber bundle, which fills in the whole
genotype frequency
space completely. 
We can define the genetic distance of two populations as their geodesic distance on the equilibrium manifold.
In addition, the modification of our theory under the process of natural selection and mutation is 
addressed.
\end{abstract}

\maketitle






\tableofcontents

\section{\label{sec:level1}Introduction}

Time evolution of population percentages labeled by biological traits 
or genetic characteristics 
in a population
system is a main subject of studies for population genetics and evolutionary
biology (\cite{{Wright},{Fisher},{Crowbook},book1, book2,book3} and references therein). 
Their governing laws may involve classical genetics, which can be traced back to Mendel's seminal work \cite{Mendel}.
Both the birth rate and death rate of a system will change the
population number. Also the mating mechanism and inheritance laws will determine
the percentage weight of newborn biological traits. 
%
There are abundant past researches and advance studying 
population genetic questions and equations ( 
\cite{{Wright},{Fisher},{Crowbook},book1, book2,book3}). 
For example, Wright and other pioneer researchers \cite{{Wright}} favor use of discrete difference equations to label each generation and the reproduction.
However, the difference equation does not manifest the dynamical properties as transparently as the continuous differential equations,
pioneered by Fisher, Crow and Kimura, {\it et al.} \cite{{Fisher},{Crowbook}}.
In this work, we will formulate a set of exactly solvable nonlinear differential equations with a first-order time derivative to address
population genetic questions. \\

Out of curiosity, the authors had posed the following questions to ourselves many years ago:\footnote{This model presented here is originated from an independent thought of the first author during his undergrad freshman year.
The governing equations and model are derived in 2003.
Substantial 
work is completed and exact solutions are found in 2006.
The manuscript presented here is a late update of our 2006's work 
aiming to contribute to the academic literature.}\\ 

\noindent
$\bullet$ How should we characterize an ecological or macroscopic biological system consisting 
of a large population of many living beings within a mathematical framework?\\ 
\noindent
$\bullet$ How can we characterize the time-dependent evolution of genetic diversity and genotype percentage of a population driven by the mating process and population growth?\\
\noindent
$\bullet$ 
Is there a mathematical definition of genetic distance between two population groups even within the same species under the biological taxonomy?
For example, can we quantitatively define the mathematical genetic distance of different ethnic groups of Homo sapiens, human beings? 
What would be the genetic distance of ethnic groups: Taiwanese, Cantonese, Japanese, Jewish, Irish people, etc.? 
Similarly, can we quantitatively define the relative genetic distances of other animal or plant species (within the same species), such as Darwin's finches, studied in Charles Darwin's ``\emph{On the Origin of Species}'' \cite{origin}?\\

To our amusement, however,
even without any necessity of deep biological knowledge, 
we can independently derive, from scratch, a set of exactly solvable governing equations describing a universal large class of systems addressing the issues of population genetics and evolutionary biology:\\

\noindent
$\bullet$ Our model contains time-dependent governing equations together with our four fundamental assumptions. \\
$\bullet$ Our theory incorporates exactly solvable models and their mathematical properties describing population genetic systems. \\
$\bullet$ 
Our theory provides an answer to each of the questions we posed above. 
\\

To the best of our knowledge, the analysis closest to ours in the literature is Ref.\cite{Nagylaki}. 
In Ref.\cite{Nagylaki}, some exact solvable models within a gene pool are analyzed, where the monotonic evolutionary behavior is found.
Our model is similar to Ref.\cite{Nagylaki}'s model: Ref.\cite{Nagylaki}'s studies a single locus for an arbitrary number of alleles with or without distinguishing the sex
(e.g. monoecious or dioecious organisms\footnote{See the comparison of 
models of one-gender monoecious population, two-gender dioecious population, and multi-gender population in \cite{gene2}, and also in Nagylaki and Crow (1974) \cite{Nagylaki}.};
we study a single locus, arbitrary number alleles in a two-gender dioecious population.
Though Ref.\cite{Nagylaki} 
reaches results similar to ours, we find several new ingredients: \\ 
{\bf (1)} The stable equilibrium solutions as a manifold.\\ 
{\bf (2)} The parameterization of the equilibrium manifold.\\ 
{\bf (3)} The experimental evidence of the model, such as the blood type of human ethnic groups.
We present the {experimental data} of phenotype (O$_{exp}$,A$_{exp}$,B$_{exp}$,AB$_{exp}$)
in Table.\ref{Table1}, \ref{Table3}, \ref{Table4}. 
We present our corresponding {theoretical prediction} of genotype (O$_{t}$,AA$_{t}$, Ai$_{t}$,BB$_{t}$,Bi$_{t}$,AB$_{i}$) in
Table.\ref{Table2}, \ref{Table5}, \ref{Table6}. See Sec.\ref{sec:fig} for figures.\\
{\bf (4)} The proposal to define the genetic distance of two populations as their geodesic distance on the equilibrium manifold in the genotype frequency space.\\
{\bf (5)} The exact analytic solution in terms of a Euclidean fiber bundle.\\
{\bf (6)} Our work may be viewed as 
a unified framework
combining the exact analytic solutions of Ref.\cite{Nagylaki} with 
the stability analysis of the Hardy-Weinberg law \cite{{Hardy},{Weinberg}}, together with a proper parameterization of stable equilibrium manifold.\\
We believe that our work adds value to the literature. \\

We will work out our model step by step. 
To characterize a population genetic system, the first step is to find a good biological \emph{parameter} to label the system.
The biological trait can be a genotype or phenotype.
Here we will use the genotype, the genetic makeup of an individual.
The genotype contains a set of choices of possible alleles. 
We will focus on a genotype determined by a single locus and an arbitrary number of alleles.
We will use the \emph{genotype frequency}, the number of individuals with a given genotype normalized by the total number of individuals in the population. 
%
A set of genotype frequencies provides the \emph{normalized parameter} to label the given population.
Throughout the text, we may also term this \emph{genotype frequency} concept as ``\emph{percentage parameter}'' or simply ``\emph{parameter}'' of the given population.
The full parameter space consists of all allowed genotype frequencies of the population.
Under governing principles, the {genotype frequency} can evolve under the time evolution.
\\



\begin{figure}[!t] 
\begin{flushleft}
\includegraphics[scale=.32]{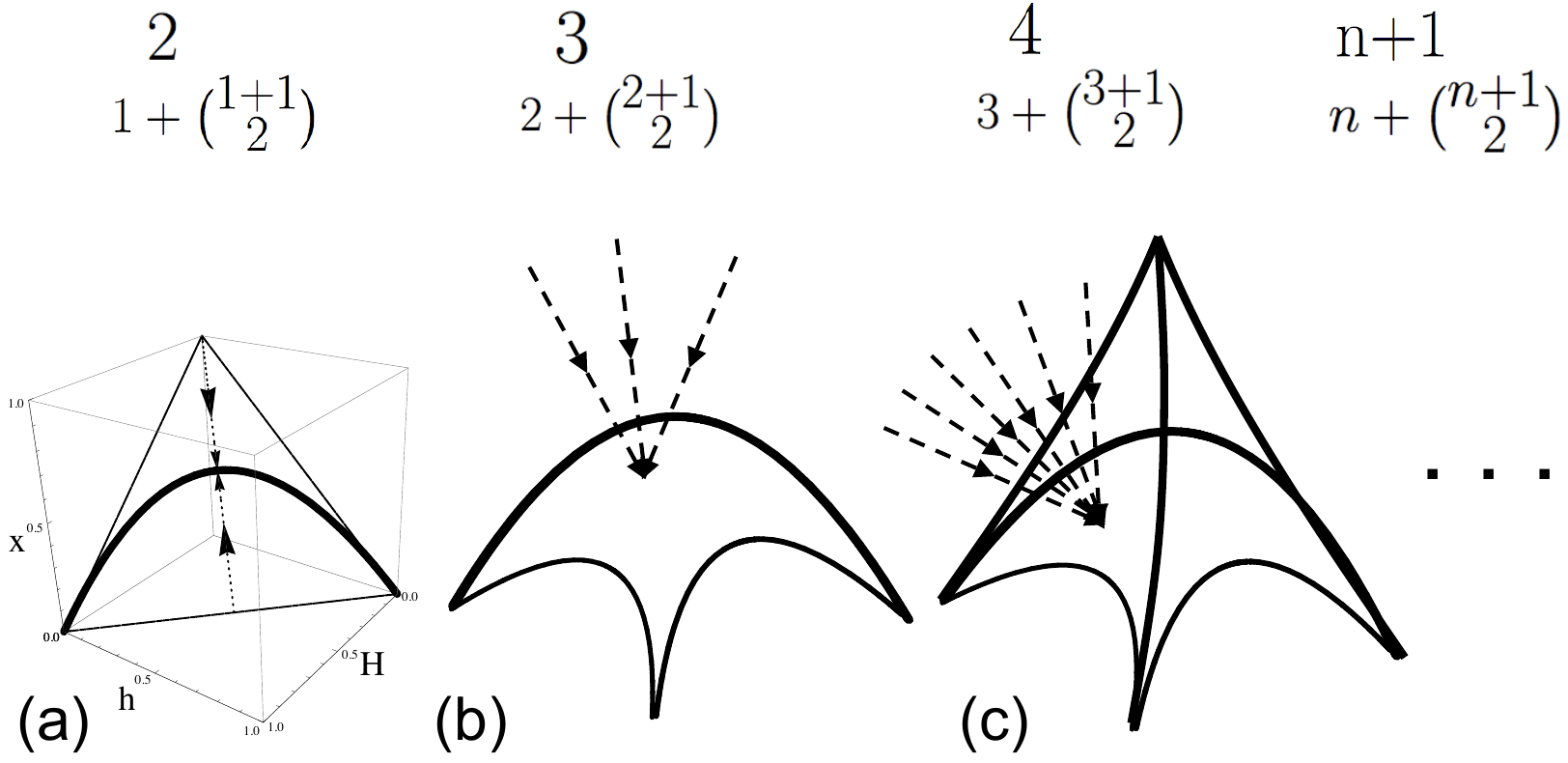} 
\end{flushleft}
\caption{The geometrical illustration of exact analytic solutions of our model, see Sec.\ref{sec:level3}, Sec.\ref{sec:level4} for details.
For a total number $(n+1)$ of alleles, denoted as $n+1$ in the top row, such as n dominant genes and 1 recessive gene. 
The full dimension of the whole \emph{percentage parameter} space of genotype frequency 
is
$n+{n+1 \choose 2}$, shown in the second row.
The stable solutions altogether consist of a $n$-dimensional continuous curved
manifold as a subspace of the whole 
parameter space. 
This $n$-dimensional fix-point manifold is a global stable attractor under time evolution, 
attracting any initial point in a ${n+1 \choose 2}$-dimensional Euclidean
fiber
to the fixed point where this fiber is attached. 
The Euclidean fiber here is spanned by the dashed arrows, with its dimensionality of ${n+1 \choose 2}$: 1 in (a), 3 in (b), 6 in (c).
The stable
manifold and its attached fibers form the $n+{n+1 \choose 2}$-dimensional fiber bundles, filling in the whole
parameter space completely. Figure (a) illustrates the example presented in Sec.\ref{sec:1-1hair}.
Figure (b) illustrates the example presented in Sec.\ref{sec:2-1G}; Figure (c) illustrates a more generic case, presented in Sec.\ref{sec:level4}.} 
\label{fig:01} 
\end{figure}

In this work, we start from four fundamental assumptions in Sec.\ref{sec:level2}. 
In Sec.\ref{sec:level3}, \ref{sec:level4}: 
we derive the dynamical governing equations under time evolution.  
We solve the time-dependent 
exact analytic solution (see Fig.\ref{fig:01} for an illustration) 
from a set of coupled nonlinear differential equations with first-order time derivatives. 
We are able to show global
stability, monotonic evolution, 
and no chaos for any population system. 
We prove that any genotype or allele that ever presents in the population will never become extinct
and also derive the Hardy-Weinberg law \cite{{Hardy},{Weinberg}} in Sec.\ref{sec:level4}.\\ 

\begin{widetext}
\begin{figure}[!h] 
\includegraphics[scale=.4]{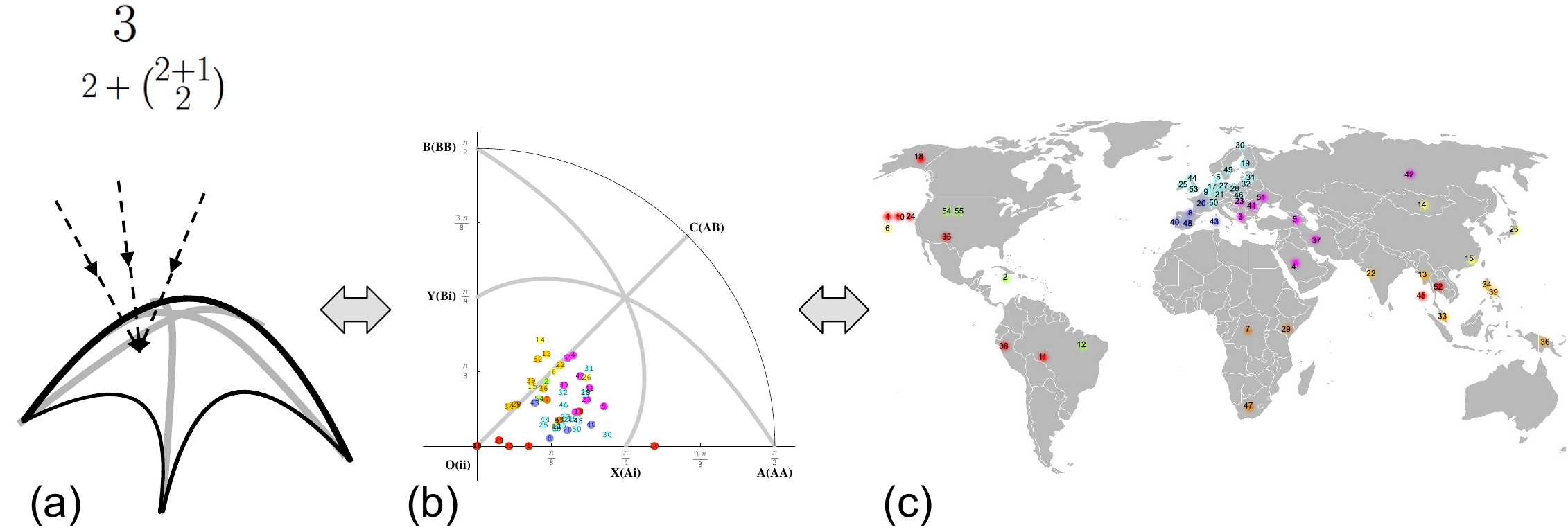} 
\caption{Here we take the blood type evolution model (3 alleles) presented in Sec.\ref{sec:2-1G} as an example.
In this work, we establish 
the mapping from (a) the stable equilibrium curved manifold (see Fig.\ref{fig:01}) 
to (b) the stable fixed-point quadrant parametrized by $(\theta_1,\theta_2)$ defined in Sec.\ref{sec:2-1G}, see FIG.\ref{fig:10},\ref{fig:11}. 
Our mapping parameterization may find its use for a correspondence to (c) the geographical map of ethnic groups in the world, see FIG.\ref{fig:9}. 
We can define the genetic distance of ethnic groups using the method presented in Sec.\ref{sec:2-1G}.} 
\label{fig:02} 
\end{figure}
\end{widetext}

In addition to the analytical work, 
we also found experimental data strongly consistent with our study to confirm the validity of our theory.
Specifically, we examine the experimental data 
of the
genotype frequency of blood type \cite{data} 
for ({\bf 1}) different ethnic groups and ({\bf 2}) different countries, in Sec.\ref{sec:2-1G}. 
With strong evidence and some expected error bars,
the data can be rigorously fitted into our stable equilibrium solutions of governing equations. 
The stable equilibrium manifold can be mapped to 
a continuous two-dimensional quadrant map.  
This gives a panorama of how each ethnic group relates to another one. See FIG.\ref{fig:02} for an illustration. 
Our result demonstrates that the 
laws of inheritance tend to 
approach stable equilibrium 
rather than with the usual 
chaos or complexity occurring in other nonlinear systems. 
In addition to this standard case, 
we
further briefly analyze the case incorporating natural selection and mutation. \\

In Sec.~\ref{subsec:distance}, we propose that 
a geodesic distance on the stable fixed-point manifold as a measure of the genetic distance.
We comment that the genetic distances
have also been  
studied and proposed in the past in the influential works of 
\cite{cavalli1967phylogenetic}
\cite{nei1972genetic}
\cite{antonelli1977geometry1},
\cite{antonelli1977geometry2},
\cite{antonelli1977geometry3},
\cite{antonelli1978geometry4},
\cite{reynolds1983estimation},
and see also a recent review \cite{dougan2016genetic}.
However, it is worthwhile to mention that since we have provided an exact analytic 
stable fixed-point manifold parametrized by 
Eq.(\ref{eq:Equiln}) for the general case (Also the Eq.(\ref{eq:Equil}) and Eq.(\ref{eq:Equil2}) for more specific cases),
we can precisely solve the geodesic as a continuous path connecting any two points on the stable fixed-point manifold.
This has a better advantage to define a continuous measure of genetic distances, while some of the previous work
use the discretized measures can only achieve a discretized distance measurement for genetic distances of different populations of the same given species, 
and cannot obtain a continuous varying distance measurement.

\onecolumngrid
\twocolumngrid

We anticipate that our analytical solutions may be a good starting basis for understanding 
more sophisticated models. 
With the experimental evidence between the   
the world-ethnic-group blood type genotype frequency data and our stable solutions, 
we believe that 
our stable solutions and exact dynamical analytic solutions 
 can be applied to other macroscopic population genetic systems (including human ethnic groups as well as other animals or plants), 
as long as the system is approximately obeying the assumptions we made. 
We hope that our parameterization of analytic stable fixed-point solutions can be an efficient map to characterize the stabilized genotype frequencies. 
Our mapping may find its home 
in the textbooks of population genetics and evolutionary
biology in the 
future.\\


\section{\label{sec:level2}Four assumptions from fundamental views}

Here we start from scratch, by introducing the four fundamental assumptions used to construct our model.
Our purpose lies in knowing how the genotype frequencies
evolve under the mating process of a certain population system.

First, 
we focused on a
system where the mating process is approximately {\bf closed} inside itself.
Hence, the newborn generation is totally a production from the old generation of the
system. We should be aware that it does not matter whether the population
locates together geographically. 
In this way, we could associate our model to different ethnic groups of
humans, because the mating process of each ethnic group is approximately
closed in each system. Hybrids between different ethnic groups are usually minorities
compared to the majority ethnic groups. 

Second, to determine the mating
procedure, we assume the mating is overall 
{\bf average and random}, which is intuitively reasonable under mean-field approximation.
This assumption would not be far away from reality, considering the balance within an overall
large population system. 
This assumption states no preference for the mating in a certain
system. For example, we consider a system of $n$ genotypes. 
Suppose $P_{i}$ is the population amount of the genotype $i$, and assume a
half-male-half-female population. The random mating mechanism of the male
amount $P_{k}/2$ mates with all types of the female amount $P_{i}/2$ is $\frac{P_{k}}{2%
}(\sum \frac{P_{i}/2}{P/2})=\frac{P_{k}}{2}(\sum \frac{P_{i}}{P})$, where $P$
is the total population.\ We interpret this expression as the male amount
multiplied by the probability to meet certain types of female. On the other
hand, we have the female amount $P_{k}/2$ to meet all types of the male amount $%
P_{i} $ . Notice that the interchangeable $P_{k}P_{k}$ term counts only once (male $k$
with female $k$); however, $P_{k}P_{j}$ $(j\neq k)$ counts twice (male $k$
with female $j$, and male $j$ with female $k$). The only mating dependence
on a certain genotype 
is the population number of that
genotype 
- if the population number for a certain genotype 
is
large, then, it will have a greater chance to mate and also to be mated with by
other genotype; 
and vice versa. Overall, this mechanism gives quadratic forms to the governing equations.
The mating within the same ethnic groups or in a local geographical region plausibly obeys this
second assumption. 

Third, we consider the simplest inheritance
law, the classical genetics: Mendelian inheritance, to relate the
transmission of hereditary traits (alleles) 
from the parents to the newborn
generation. Although there is much new progress in genetics study,
Mendelian inheritance is still a primary principle 
for capturing the main properties of heredity. 

Fourth, in order to make the system dynamic under the
time evolution, there must be a guideline for population growth. Here we
consider the exponential growth --- Malthusian growth
model. Again, this law 
adumbrates the main feature of population growth
with 
some acceptable deviations from the experiment. We will find 
later that these assumptions are beneficial for having analytically exactly solvable solutions.

One can apply the model to other animal or plant mating systems under
a similar inheritance law, if it basically satisfies the above assumptions.

We summarize our {\bf four assumptions} as follows: \\ 

\noindent (a) The mating of gene-holders is (approximately) {\bf closed} in a
population system.\\

\noindent (b) The mating probability for a certain genotype 
to another genotype,
due to {\bf average-and-random} mating mechanism, is proportional to
the product of their population (quadratic form). It evolves under the {\bf mean-field} assumption.\\

\noindent (c) The probability of a genotype 
for the newborn generation obeys the
{\bf Mendelian inheritance}.\\

\noindent (d) The accumulation of human population obeys the {\bf exponential growth law}, for both the birth and the death processes.\\

We should be aware of the fourth assumption that the exponential growth
includes the balance between the newborn and the dead. We denote the birth rate $%
k_{b} $ and the death rate $k_{d}$. The overall net growth rate is $%
k=k_{b}-k_{d}$. 

To derive the governing differential equations, implicitly we have another hidden assumption: the continuous limit.
Throughout our work, we will take 
the continuous time $t$.
The discrete positive integer population number can be treated approximately as a continuous real number 
in a large population. 

\section{\label{sec:level3}Model}

\subsection{Model of 2 Alleles: 1 Dominant Gene and 1 Recessive Gene \label{sec:1-1hair}} 

\subsubsection{Governing equations for population}

We first consider the simplest model of our theory, which begins with
two alleles and a single locus, such as one dominant gene $A$ and one recessive gene $a$. We take the hairstyle being curly or straight as an 
example, which roughly obeys this kind of inheritance law. Dominant gene $A$
shows the curly property; only inheritance of no dominant gene shows the
straight property. Curl hair owners indicate their genotype 
must be either $%
AA$ or $Aa$. Straight hair owners' genotype 
must be $aa$. 
Below we denote the \emph{population parameters}: $H$ for $AA$, $x$ for $Aa$, and $h$ for $aa$.
$H$ stands for the number of people in the population carrying $AA$, 
$x$ stands for the number of people in the population carrying $Aa$,
and $h$ stands for the number of people in the population carrying $aa$.

We now can 
write down the governing differential equations 
under a time $t$ evolution of population number for hairstyle based on 4 assumptions in Sec.\ref{sec:level2}:
\begin{equation} \label{eq:pop1-1}
\left\{ 
\begin{array}{l}
\frac{dH}{dt}=k_{b}\frac{1}{P}(H^{2}+Hx+\frac{x^{2}}{4})-k_{d}H, \\ 
\frac{dh}{dt}=k_{b}\frac{1}{P}(h^{2}+hx+\frac{x^{2}}{4})-k_{d}h, \\ 
\frac{dx}{dt}=k_{b}\frac{1}{P}(2Hh+Hx+hx+\frac{x^{2}}{2})-k_{d}x.%
\end{array}%
\right.
\end{equation}
We denote the total population to be $H+h+x=P$.

These equations have a first-order time derivative $\frac{d}{dt}$, associating the left-hand-side (LHS) population changes to the right-hand-side (RHS) birth 
and death effects. Linear terms on the RHS represent the death amount per unit
time. Quadratic forms on the RHS represent the newborn population per unit
time following Mendelian inheritance. Taking $\frac{H^{2}}{P}$ of $\frac{dH}{dt}$, for example, it represents
our 
second assumption that the probability for male $H$ to meet female $H$ is $\frac{H}{P}
$. The offspring of those parents is definitely $H$. So there must be a
term $\frac{H^{2}}{P}$ in the RHS of the equation $\frac{dH}{dt}$, up to a constant factor.
Now considering the $Hx$ term, it can be either
male $H$ to meet female $x$, or male $x$ to meet female $H$. So the overall
effect must be $2Hx$ for the RHS newborn generation. 
The $2Hx$ is separated into two parts: $Hx$
for $\frac{dH}{dt}$ and $Hx$ for $\frac{dx}{dt}$, because Mendelian
inheritance shows the offspring of $H$ and $x$ parents has a half
probability to be $H$ and another half to be $x$. Similarly, we can
determine all the other coefficients of quadratic terms by the same
arguments. The sum on the newborn part of the RHS must be all the
possibilities for any pair of genes (alleles), 
$(H+h+x)\frac{(H+h+x)}{P}$, and
this is equal to $P$. We notice that the sum of RHS is $\frac{dP}{dt}$, LHS is 
$(k_{b}-k_{d})P$. This perfectly obeys assumption (d): $\frac{dP}{dt}%
=(k_{b}-k_{d})P=kP$. We next check that the model is self-consistent, and uniquely determined by our four assumptions listed in Sec.\ref{sec:level2}.

\subsubsection{Governing equations for percentages}

Now we revise our equations via normalizing each gene carrier population by the total population,
using the 
\emph{percentage parameters}, so called \emph{genotype frequencies}.
Consider, $\frac{d}{dt}(\frac{G}{P})=$ $\frac{1}{P}\frac{dG}{dt}-G\frac{dP/dt%
}{P^{2}}$, where $G$ is any one of $H,h,x$. This turns out to be $\frac{d}{dt%
}(\frac{G}{P})=\frac{1}{P}[k_{b}\frac{1}{P}(\text{quadratic form})-k_{d}G]-G\frac{%
(k_{b}-k_{d})P}{P^{2}}=k_{b}(\frac{\text{quadratic form}}{P^{2}}-\frac{G}{P})$. Now
by redefining $\frac{G}{P}\rightarrow G$, we have the governing equations for
\emph{genotype frequencies}: 
\begin{equation} \label{eq:Gene1-1}
\left\{ 
\begin{array}{l}
\frac{dH}{dt}=k_{b}(H^{2}+Hx+\frac{x^{2}}{4}-H), \\ 
\frac{dh}{dt}=k_{b}(h^{2}+hx+\frac{x^{2}}{4}-h), \\ 
\frac{dx}{dt}=k_{b}(2Hh+Hx+hx+\frac{x^{2}}{2}-x).%
\end{array}%
\right. %
\end{equation}
We notice that the death rate $k_{d}$ takes no effect on the percentage
governing equations. That is because the death effect does not rearrange the
{percentage parameters},
every percentage parameter just universally dies away. Although
the \emph{population parameter} changes under death effect, the \emph{genotype frequency}
does not.

For both Eq.(\ref{eq:pop1-1}) and Eq.(\ref{eq:Gene1-1}), there is a permutation symmetric group S${}_2$ symmetry by exchanging $H\leftrightarrow h$.

\subsubsection{Equilibrium solutions}

We should be aware that there is a constraint $H+h+x=1$, and the limited
range for each parameter. The total degree of freedom is $3-1=2$. 
Namely the whole parameter space of \emph{population parameters} are 3 dimensional,
but the whole parameter space of \emph{percentage parameters}, the  \emph{genotype frequencies}, are 2 dimensional.  

Our aim 
is to realize the time evolution properties of this dynamical
model. We first narrow down to see the equilibrium solution, which is the
solution fixed in the parameter space without evolving under time evolution. The
solution are solved by setting the RHS algebraic equations Eq.(\ref{eq:Gene1-1}) to be zero. Among all expressions
of the algebraic solution, we find the following one is the most appropriate representation: 
%
\begin{equation} \label{eq:Equil}
\left\{ 
\begin{array}{l}
H_{eq}(\theta )=\frac{\cos ^{2}(\theta _{})}{(\cos (\theta _{})+\sin
(\theta _{}))^{2}}, \\ 
h_{eq}(\theta )=\frac{\sin ^{2}(\theta _{})}{(\cos (\theta _{})+\sin
(\theta _{}))^{2}}, \\ 
x_{eq}(\theta )=\frac{2\sin (\theta _{})\cos (\theta _{})}{(\cos (\theta
_{})+\sin (\theta _{}))^{2}}.%
\end{array}%
\right.
\end{equation}
It gives an one-to-one mapping between the equilibrium solution $%
(H_{eq},h_{eq},x_{eq})$ and $\theta$, and shows the solution is a
1-dimensional continuous manifold.
This representation also shows the symmetry of $H$ and $h$ in governing equations relates
to the symmetry of $H$ and $h$ in equilibrium solution. The symmetry of $%
\theta _{1}\leftrightarrow \frac{\pi }{2}-\theta _{1}$ relates to 
the S${}_2$ symmetry of $H\leftrightarrow h$.
The S${}_2$ symmetry also results in a number of time-independent \emph{conserved quantities} under 
Eq.(\ref{eq:Gene1-1}). 
We find, for example, $\frac{d(H-h)}{dt}=\frac{d(2H+x)}{dt}=\frac{d(2h+x)}{dt}=0$.
By the constraint $H+h+x=1$, the three quantities $H-h$, $2H+x$, and $2h+x$ are indeed the same equivalent conserved quantity.
We can say the S${}_2$ symmetry results in 1 \emph{conserved quantity}, 
thus the dimensionality of fixed-point solutions is 1, here parametrized by $\theta$.

\subsubsection{From the linear stability analysis to the exact analytic time-dependent dynamical solution}

By doing the linear stability analysis on the dynamical 
Eq.(\ref{eq:Gene1-1}) with a
small perturbation around this equilibrium solution Eq.(\ref{eq:Equil}), 
we find that it is stable with eigenvalues $0$, $-1$  \cite{RG} for the entire two-dimensional parameter space. 
The
first eigenvector for the eigenvalue 0 corresponds to the \emph{marginal} tangent direction \cite{RG} of $1$-dimensional
equilibrium solution. Surprisingly, the second eigenvector for the eigenvalue 1 is \emph{irrelevant and stable perturbation} \cite{RG}
along the fixed
direction $\widehat{H}+\widehat{h}-2\widehat{x}$. We define a new basis $%
\widehat{s}=$ $\widehat{H}+\widehat{h}-2\widehat{x}$, and $s$ is the
coordinate along $\widehat{s}.$\bigskip

It is
well-known that nonlinear equations may have complicated global properties, such as chaotic behaviors. 
Usually the numerical simulation is required, and there are
seldom cases which analytic solution for time evolution is exactly
solvable. 
However, below we will show how to obtain the exact analytic time-dependent solution of Eq.(\ref{eq:Gene1-1}).
Our method is to change the description of parameters in
Cartesian coordinates $(H,h,x)$ to parameters in curvilinear coordinates $(\theta, s)$. We transform 
$H,h,x$ to $\theta ,s$ by the following:
\begin{equation} \label{eq:G1-1para}
\left\{ 
\begin{array}{l}
H(\theta, s)=H_{eq}(\theta )+s, \\ 
h(\theta, s)=h_{eq}(\theta )+s, \\ 
x(\theta, s)=x_{eq}(\theta )-2s.%
\end{array}%
\right.
\end{equation}
The transformation is well-defined for a 2-dimensional one-to-one mapping with an
inverse function: 
\begin{equation} \label{eq:G1-1inverse}
\left\{ 
\begin{array}{l}
\theta =\tan ^{-1}(\frac{2h+x}{2H+x}), \\ 
s=Hh-\frac{x^{2}}{4}.%
\end{array}%
\right.
\end{equation}

We could substitute this inverse transformation into the equilibrium
solution Eq.(\ref{eq:Equil}) 
to obtain the equilibrium solution $H_{eq},h_{eq},x_{eq}$ as the
parameterizations of $H,h,x$:
\begin{equation}
\left\{ 
\begin{array}{l}
H_{eq}=\frac{1}{4}(2H+x)^{2}, \\ 
h_{eq}=\frac{1}{4}(2h+x)^{2},\\ 
x_{eq}=\frac{1}{2}(2H+x)(2h+x).%
\end{array}%
\right.
\end{equation}
This indicates two lessons. First, once we know the original set of genotype frequencies 
$H,h,x$ as the initial condition, remarkably we can deduce the final equilibrium $H_{eq},h_{eq},x_{eq}$.
Second, $(2H+x)$ and $(2h+x)$, these two numbers determine the final equilibrium.
This implies, for the same number set of $(2H+x)$ and $(2h+x)$, even for
different $H,h,x$, their final equilibriums are the same.

Substituting the reparameterization 
Eq.(\ref{eq:G1-1para}) 
to the governing equations Eq.(\ref{eq:Gene1-1}), this not
only decouples the parameters, but also decodes the equations to one
time-dependent equation and the other time-independent equation:  
\begin{equation}  \label{eq:G1-1linear}
\left\{ 
\begin{array}{l}
\frac{d}{dt}s=-k_{b}s, \\ 
\frac{d }{dt}\theta=0.%
\end{array}%
\right.
\end{equation}
Both are
exactly solvable. 
Remarkably, {\bf we have transformed the nonlinear coupled differential equations Eq.(\ref{eq:Gene1-1}) to the linear decoupled differential equations Eq.(\ref{eq:G1-1linear})}.

We foresee in advance the equilibrium solution is a global attractor,
attracting all the points on the line direction of a certain given initial $\theta$ to the
equilibrium point of the same $\theta$ in the exponential decay way along the $s$ direction. Because each
line direction for a $\theta$ is independent to each other with no
intersection, this makes our solution well-defined everywhere in the parameter space. 
The global
picture of time evolution is: 
giving any initial value in the parameter
space, there is only one corresponding $\theta$ with a line direction of $s$
connecting that equilibrium point to the initial value; the time evolution
of parameters will go along the line direction to the equilibrium point in
the exponential decay to reduce the distance away from the equilibrium point.

The method for deriving the analytic solution is the following: for any
given initial value $\widetilde{H},\widetilde{h}$ and $\widetilde{x},$ find
the corresponding set of $\widetilde{\theta },\widetilde{s}.$ By Eq.$(7)$,We
then have $H(t)\widehat{H}+h(t)\widehat{h}+x(t)\widehat{x}=H_{eq}(\widetilde{%
\theta })\widehat{H}+h_{eq}(\widetilde{\theta })\widehat{h}+x_{eq}(%
\widetilde{\theta })\widehat{x}+\widetilde{s}e^{-k_{b}t}(\widehat{H}+%
\widehat{h}-2\widehat{x})$. Because the transformation between old and new
coordinate is well-defined, we can further solve the new equations and
replace the parameters from $\theta ,s$ to $H,h,x$ by Eq.(\ref{eq:G1-1inverse}). 

\begin{figure}[!h] 
\begin{center} 
\includegraphics[scale=.6]{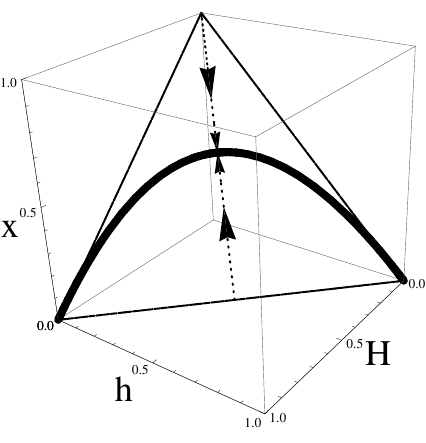} 
\end{center}
\caption{The illustration of time-dependent exact analytic solutions  
Eq.(\ref{Eq.:exact1-1}) in the genotype frequency parameter space for the model of 2 alleles. 
The thick black curve stands for the 1-dimensional equilibrium solution Eq.(\ref{eq:Equil}). 
The dashed arrow direction indicates the 1-dimensional fiber direction $\widehat{s}=$ $\widehat{H}+\widehat{h}-2\widehat{x}$,
where every point along the fiber line will be attracted to the thick black curve under time evolution.}
\label{fig:03} 
\end{figure}

Hence, we achieve our exact analytic solution: 
\begin{equation} \label{Eq.:exact1-1}
\left\{ 
\begin{array}{l}
H(t)=H_{eq}(\tan ^{-1}(\frac{2\widetilde{h}+\widetilde{x}}{2\widetilde{H}+%
\widetilde{x}}))+(\widetilde{H}\widetilde{h}-\frac{\widetilde{x}^{2}}{4}%
)e^{-k_{b}t}, \\ 
h(t)=h_{eq}(\tan ^{-1}(\frac{2\widetilde{h}+\widetilde{x}}{2\widetilde{H}+%
\widetilde{x}}))+(\widetilde{H}\widetilde{h}-\frac{\widetilde{x}^{2}}{4}%
)e^{-k_{b}t}, \\ 
x(t)=x_{eq}(\tan ^{-1}(\frac{2\widetilde{h}+\widetilde{x}}{2\widetilde{H}+%
\widetilde{x}}))-2(\widetilde{H}\widetilde{h}-\frac{\widetilde{x}^{2}}{4}%
)e^{-k_{b}t}.%
\end{array}%
\right.
\end{equation}
The illustration of Eq.(\ref{Eq.:exact1-1}) is shown in Fig.\ref{fig:03}. The equilibrium solutions coincide with the De Finetti diagram \cite{book2}.

Intuitively we would like to compare this model and its analytic solution to experimental numerical data.
It will be more appropriate to fit the experimental data into the equilibrium solution, because we can imagine that the genotype frequencies 
have been more-or-less stabilized within a closed population
under the sufficient amount of time-evolution.

For this 1-Dominant-Gene-and-1-Recessive-Gene (hairstyle) model, the observables of available
experimental data are the phenotypes, the representative biological characters, 
of curl ($AA$ and $Aa$) 
and straight ($aa$) hairs. 
The two phenotype frequencies with one constraint, has 1
degree of freedom; this is the same as the 1 degree of freedom of the
continuous equilibrium solution. The fitting can be perfectly with no error
bar; however, this is due to the same degree of freedom of correspondence,
rather than the evidence of successful description of this model. 
Hence, the hairstyle
experimental data cannot illustrate the validity of the theory.

In the next, we shall turn to the next-simplest model: 3 alleles, such as 2 dominant genes and 1 recessive gene case, for
example, the blood-type model, to check the experimental evidence of our theory.

\subsection{Model of 3 Alleles: 2 Dominant Genes and 1 Recessive Gene: Blood Type
Evolution Model  \label{sec:2-1G} } 

\subsubsection{Governing equations and exact analytic solutions}

For the model of 3 alleles, such as two dominant genes $A$, $B$ and one recessive gene $i$ of blood types, the
representative phenotypes 
are type A: $AA$ and $Ai$, type B: $%
BB$ and $Bi$, type AB: $AB$, and type O: $ii$.\\

$\left\{ 
\begin{array}{l}
\text{Type O: }o\text{ represents }ii, \\ 
\text{Type A: }%
\begin{array}{c}
a\text{ represents }AA, \\ 
x\text{ represents }Ai,%
\end{array}
\\ 
\text{Type B: }%
\begin{array}{c}
b\text{ represents }BB, \\ 
y\text{ represents }Bi,%
\end{array}
\\ 
\text{Type AB: }c\text{ represents }AB.%
\end{array}%
\right. $\\
There are four kinds of representative phenotypes 
(A,\thinspace B,\thinspace AB,\thinspace O), and six parameters of genotype population  $%
(o,a,b,x,y,c)$,  for a total population $o+a+b+x+y+c=P$. Here we follow the similar derivation 
as the previous model in Sec \ref{sec:1-1hair}. to
write down the governing equations for the \emph{population parameters}: %
%
%
\begin{equation}
\label{eq:pop2-1}
\left\{ 
\begin{array}{l}
\frac{do}{dt} = k_{b}\frac{1}{P}(o^{2}+ox+oy+\frac{x^{2}}{4}+\frac{y^{2}}{4}+%
\frac{xy}{2})-k_{d}o, \\ 
\frac{da}{dt} = k_{b}\frac{1}{P}(a^{2}+ac+ax+\frac{c^{2}}{4}+\frac{x^{2}}{4}+%
\frac{cx}{2})-k_{d}a, \\ 
\frac{db}{dt} = k_{b}\frac{1}{P}(b^{2}+by+bc+\frac{y^{2}}{4}+\frac{c^{2}}{4}+%
\frac{yc}{2})-k_{d}b, \\ 
\frac{dx}{dt}= k_{b}\frac{1}{P}(2oa+ox+ax+oc+ay+\frac{xy}{2}+\frac{xc}{2}+%
\frac{yc}{2}+\frac{x^{2}}{2})
- k_{d}x, \\ 
\frac{dy}{dt} = k_{b}\frac{1}{P}(2ob+by+oy+bx+oc+\frac{yc}{2}+\frac{yx}{2}+%
\frac{cx}{2}+\frac{y^{2}}{2})-k_{d}y, \\ 
\frac{dc}{dt} = k_{b}\frac{1}{P}(2ab+ac+bc+ay+bx+\frac{cx}{2}+\frac{cy}{2}+%
\frac{xy}{2}+\frac{c^{2}}{2})-k_{d}c.%
\end{array}%
\right.
\end{equation}
%
%

By redefining $\frac{G}{P}\rightarrow G$ as the \emph{percentage parameters}, where $G$ is any of the six parameters,
we can derive the governing \emph{percentage parameters} equations
for \emph{genotype frequencies}: 
\begin{equation}
\label{eq:Gene2-1}
\left\{ 
\begin{array}{l}
\frac{do}{dt}=k_{b}(o^{2}+ox+oy+\frac{x^{2}}{4}+\frac{y^{2}}{4}+\frac{xy}{2}%
-o), \\ 
\frac{da}{dt}=k_{b}(a^{2}+ac+ax+\frac{c^{2}}{4}+\frac{x^{2}}{4}+\frac{cx}{2}%
-a), \\ 
\frac{db}{dt}=k_{b}(b^{2}+by+bc+\frac{y^{2}}{4}+\frac{c^{2}}{4}+\frac{yc}{2}%
-b), \\ 
\frac{dx}{dt}=k_{b}(2oa+ox+ax+oc+ay+\frac{xy}{2}+\frac{xc}{2}+\frac{yc}{2}+%
\frac{x^{2}}{2}-x), \\ 
\frac{dy}{dt}=k_{b}(2ob+by+oy+bx+oc+\frac{yc}{2}+\frac{yx}{2}+\frac{cx}{2}+%
\frac{y^{2}}{2}-y), \\ 
\frac{dc}{dt}=k_{b}(2ab+ac+bc+ay+bx+\frac{cx}{2}+\frac{cy}{2}+\frac{xy}{2}+%
\frac{c^{2}}{2}-c).%
\end{array}%
\right.
\end{equation}%
Now, we generalize the equilibrium solution (\ref{eq:Equil}) of Eq.(\ref{eq:Gene1-1}) 
to the
equilibrium solution of Eq.(\ref{eq:Gene2-1}):  
\begin{equation} \label{eq:Equil2}
\left\{ 
\begin{array}{l}
o_{eq}(\theta _{1},\theta _{2})=\frac{\cos ^{2}(\theta _{1})}{(\cos (\theta
_{1})+\sin (\theta _{1}))^{2}}, \\ 
a_{eq}(\theta _{1},\theta _{2})=\frac{\sin ^{2}(\theta _{1})}{(\cos (\theta
_{1})+\sin (\theta _{1}))^{2}}\frac{\cos ^{2}(\theta _{2})}{(\cos (\theta
_{2})+\sin (\theta _{2}))^{2}}, \\ 
b_{eq}(\theta _{1},\theta _{2})=\frac{\sin ^{2}(\theta _{1})}{(\cos (\theta
_{1})+\sin (\theta _{1}))^{2}}\frac{\sin ^{2}(\theta _{2})}{(\cos (\theta
_{2})+\sin (\theta _{2}))^{2}}, \\ 
x_{eq}(\theta _{1},\theta _{2})=\frac{2\sin (\theta _{1})\cos (\theta _{1})}{%
(\cos (\theta _{1})+\sin (\theta _{1}))^{2}}\frac{\cos (\theta _{2})}{\cos
(\theta _{2})+\sin (\theta _{2})}, \\ 
y_{eq}(\theta _{1},\theta _{2})=\frac{2\sin (\theta _{1})\cos (\theta _{1})}{%
(\cos (\theta _{1})+\sin (\theta _{1}))^{2}}\frac{\sin (\theta _{2})}{\cos
(\theta _{2})+\sin (\theta _{2})}, \\ 
c_{eq}(\theta _{1},\theta _{2})=\frac{\sin ^{2}(\theta _{1})}{(\cos (\theta
_{1})+\sin (\theta _{1}))^{2}}\frac{2\sin (\theta _{2})\cos (\theta _{1})}{%
(\cos (\theta _{2})+\sin (\theta _{2}))^{2}}.%
\end{array}%
\right.
\end{equation}
We can further make the above 2-dimensional equilibrium
parameterization an one-to-one mapping to $(\theta _{1},\theta _{2})$, if we define that $\theta _{1}=0,\forall
\theta _{2}$ shrinks into a point. Namely, $(\theta _{1},\theta _{2})$ can be a well-defined one-to-one reparameterization
of the equilibrium solution Eq.(\ref{eq:Equil2}).

The linear stability analysis shows this system is again locally stable with
five eigenvalues $0,0,-1,-1,-1$ \cite{RG}. Here two $0$ eigenvalues correspond to
two eigenvectors along the \emph{marginal} tangent plane \cite{RG} of 2-dimensional equilibrium solutions, three %
$-1 $ eigenvalues correspond to the \emph{stable perturbation} \cite{RG} along the three eigenvectors 
$\widehat{o}+\widehat{a}-2\widehat{x}$, 
$\widehat{o}+\widehat{b}-2\widehat{y}$, 
$\widehat{a}+\widehat{b}-2\widehat{c}$. Defining those eigenvectors as $\widehat{s}_{01},\widehat{s}%
_{02},\widehat{s}_{12}$ with coordinates $s_{01},s_{02},s_{12}$, 
we could transform $(o,a,b,x,y,c)$\ of the whole parameter space
(6-dimensions with 1 constraint $o+a+b+x+y+c=1$) to a set of 5-dimensional new coordinates $%
(\theta _{1},\theta _{2},s_{01},s_{02},s_{12})$:
\begin{equation} \label{eq:G2-1para}
\left\{ 
\begin{array}{l}
o=o_{eq}(\theta _{1},\theta _{2})+s_{01}+s_{02}, \\ 
a=a_{eq}(\theta _{1},\theta _{2})+s_{01}+s_{12}, \\ 
b=b_{eq}(\theta _{1},\theta _{2})+s_{02}+s_{12}, \\ 
x=x_{eq}(\theta _{1},\theta _{2})-2s_{01}, \\ 
y=y_{eq}(\theta _{1},\theta _{2})-2s_{02}, \\ 
c=c_{eq}(\theta _{1},\theta _{2})-2s_{12}.%
\end{array}%
\right.
\end{equation}

We have the following inverse function:
\begin{equation} \label{eq:G2-1inverse}
\;\;\;\;\;\;\;\;\;\;\;\;\left\{ 
\begin{array}{l}
\theta _{1}=\tan ^{-1}(\frac{2(a+b+c)+x+y}{2o+x+y}), \\ 
\theta _{2}=\tan ^{-1}(\frac{2b+y+c}{2a+x+c}), \\ 
s_{01}=ao+\frac{ay}{2}+\frac{oc}{2}-\frac{xb}{2}-\frac{xc}{4}-\frac{xy}{4}-%
\frac{x^{2}}{4}+\frac{cy}{4}, \\ 
s_{02}=ba+\frac{bx}{2}+\frac{ay}{2}-\frac{co}{2}-\frac{cy}{4}-\frac{cx}{4}-%
\frac{c^{2}}{4}+\frac{yx}{4}, \\ 
s_{12}=ob+\frac{oc}{2}+\frac{bx}{2}-\frac{ya}{2}-\frac{yx}{4}-\frac{yc}{4}-%
\frac{y^{2}}{4}+\frac{xc}{4}.%
\end{array}%
\right.
\end{equation}

Substitute the reparameterization Eq.(\ref{eq:G2-1para}) to Eq.(\ref{eq:Gene2-1}), we confirm 
again this system will be global
stable like the previous model in Sec.\ref{sec:1-1hair}.

\begin{equation}
\left\{ 
\begin{array}{l}
\frac{d}{dt}{s_{01}}=-k_{b}s_{01}, \\ 
\frac{d}{dt}{s_{02}}=-k_{b}s_{02},\\ 
\frac{d}{dt}{s_{12}}=-k_{b}s_{12}, \\ 
\frac{d}{dt}\theta _{1}=0, \\ 
\frac{d}{dt}\theta _{2}=0.%
\end{array}%
\right.
\end{equation}

\begin{figure}[!h] 
\begin{flushleft}
\includegraphics[scale=.28]{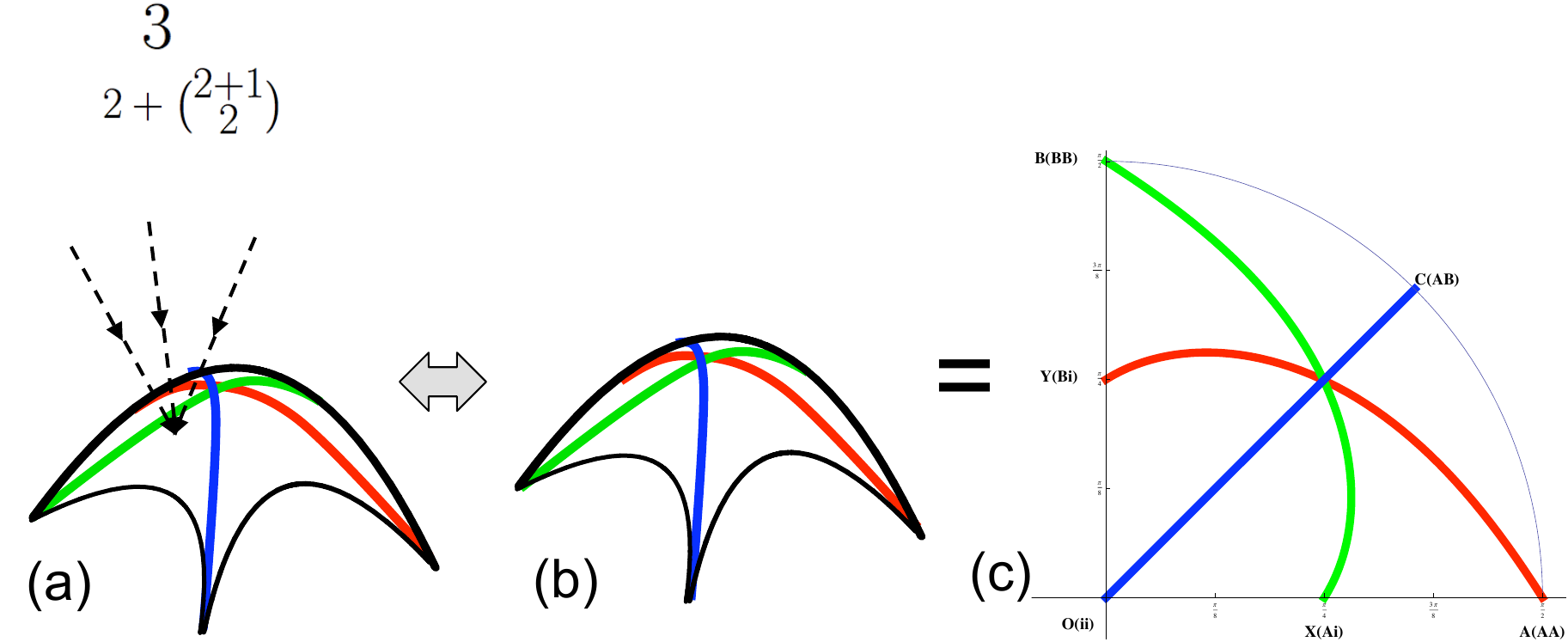} 
\end{flushleft}
\caption{(a) The illustration of time-dependent exact analytic solutions  
Eq.(\ref{Eq.:exact2-1}), solved from Eq.(\ref{eq:Gene2-1}), in the genotype frequency parameter space for the model of 3 alleles. 
The surface stands for the 2-dimensional equilibrium solution Eq.(\ref{eq:Equil2}). 
The dashed arrows indicate the 3-dimensional fiber spanned by 3 vectors: 
$\widehat{s}_{01}=\widehat{o}+\widehat{a}-2\widehat{x}$, 
$\widehat{s}_{02}=\widehat{o}+\widehat{b}-2\widehat{y}$ and 
$\widehat{s}_{12}=\widehat{a}+\widehat{b}-2\widehat{c}$,
where every point along the 3-dimensional fibers will be attracted to the 2-dimensinal stable equilibrium manifold Eq.(\ref{eq:Equil2}) under time evolution.
(b) The illustration of 2-dimensinal stable equilibrium manifold Eq.(\ref{eq:Equil2}), which is mapped to (c)
the 2-dimensional quadrant $(\theta_1 \cos(\theta_2) ,\theta_1 \sin(\theta_2))$ parametrized by $(\theta_1,\theta_2)$, detailed shown in FIG.\ref{fig:1}. } 
\label{fig:04} 
\end{figure}

The analytic solution is determined by a given set of initial values $\widetilde{o},%
\widetilde{a},\widetilde{b},\widetilde{x},\widetilde{y},\widetilde{c},$
which correspond to a set of $\widetilde{\theta }_{1},\widetilde{\theta }%
_{2},\widetilde{s}_{01},\widetilde{s}_{02},\widetilde{s}_{12}$ via Eq.(\ref{eq:G2-1inverse}).
We obtain the exact analytic time-dependent dynamical solution:%
\begin{equation} \label{Eq.:exact2-1}
\left\{ 
\begin{array}{l}
o(t)=o_{eq}(\widetilde{\theta }_{1},\widetilde{\theta }_{2})+\widetilde{s}%
_{01}e^{-k_{b}t}+\widetilde{s}_{02}e^{-k_{b}t}, \\ 
a(t)=a_{eq}(\widetilde{\theta }_{1},\widetilde{\theta }_{2})+\widetilde{s}%
_{01}e^{-k_{b}t}+\widetilde{s}_{12}e^{-k_{b}t}, \\ 
b(t)=b_{eq}(\widetilde{\theta }_{1},\widetilde{\theta }_{2})+\widetilde{s}%
_{02}e^{-k_{b}t}+\widetilde{s}_{12}e^{-k_{b}t}, \\ 
x(t)=x_{eq}(\widetilde{\theta }_{1},\widetilde{\theta }_{2})-2\widetilde{s}%
_{01}e^{-k_{b}t}, \\ 
y(t)=y_{eq}(\widetilde{\theta }_{1},\widetilde{\theta }_{2})-2\widetilde{s}%
_{02}e^{-k_{b}t}, \\ 
c(t)=c_{eq}(\widetilde{\theta }_{1},\widetilde{\theta }_{2})-2\widetilde{s}%
_{12}e^{-k_{b}t}.%
\end{array}%
\right.
\end{equation}
The illustration of Eq.(\ref{Eq.:exact2-1}) is shown in Fig.\ref{fig:04}. 

\noindent
{\bf The dimensionality}:
We shall explain the physical meaning on the dimensionality of the fibers and the stable base manifold.
For both Eq.(\ref{eq:pop2-1}) and Eq.(\ref{eq:Gene2-1}), there is a permutation symmetric group S${}_3$ symmetry by exchanging $A$, $B$, $i$ and their corresponding
genotypes.
The S${}_3$ symmetry also results in time-independent \emph{conserved quantities}, 
spanned by ${o}+{a}-2{x}$, ${o}+{b}-2{y}$ and ${a}+{b}-2{c}$,
with a constraint $o+a+b+x+y+c=1$. There are
totally 2 independent degrees of freedom.
We can say the S${}_3$ symmetry results in  
the dimensionality of fixed-point solutions is 2, here parametrized by $\theta_1,\theta_2$.


\subsubsection{Experimental evidence - data fitting} \label{sec:chart}  

As we mentioned, we would like to fit the available blood type data - blood
type population percentages as \emph{genotype frequencies}
of any ethnic group, into our equilibrium solutions on the stable 2-dimensional manifold of Eq.(\ref{eq:Equil2}).
For a set of {genotype frequencies} 
$(ii,AA,BB,Ai,Bi,AB)$, 
it totally forms a set of 5-dimensional degrees of freedom (due to 6 numbers with 1 constraint). 
We could
correlate this with $(o(\theta _{1},\theta _{2}),a(\theta _{1},\theta
_{2}),b(\theta _{1},\theta _{2}),x(\theta _{1},\theta _{2}),y(\theta
_{1},\theta _{2}),c(\theta _{1},\theta _{2}))$ for a 
2-dimensional parameterization $(\theta_{1},\theta _{2})$. 
In this way, there would be a correspondence between
5-dimensions and 2-dimensions, 
the validity of experimental fitness to our theoretical $(\theta _{1},\theta _{2})$ prediction
would be a strong 
evidence for the validity 
of our model. 
However, current
available experimental data we found is just (O$_{exp}$,A$_{exp}$,B$_{exp}$,AB$_{exp}$),
which is 3-dimensional (4 numbers with 1 constraint) $(ii,AA+Ai,BB+Bi,AB)$; we would
then correlate this to $(o(\theta _{1},\theta _{2}),a(\theta _{1},\theta
_{2})+x(\theta _{1},\theta _{2}),b(\theta _{1},\theta _{2})+y(\theta
_{1},\theta _{2}),c(\theta _{1},\theta _{2}))$ for certain 2-dimensional parameters $(\theta
_{1},\theta _{2})$. The test for the validity of this theory now becomes a
correspondence between 3-dimensions to 2-dimensions. We 
fit the 3-dimensional 
experimental data into our 2-dimensional equilibrium solutions. Although this is not as
stringent as the 5-dimensional to 2-dimensional correspondence, there remains one degree of
freedom for experimental data to deviate from our equilibrium solution. If the
error bar for this 3-dimensional to 2-dimensional correspondence is tiny, 
then our theory shows consistency 
to the level of predicting the stability of 
genotype frequencies.

Our fitting procedure is to solve $(\theta _{1},\theta _{2})$ from two
equalities:
\begin{equation} \label{eq:fitAandB}
\left\{ 
\begin{array}{c}
\text{A}_{exp}=a(\theta _{1},\theta _{2})+x(\theta _{1},\theta _{2}), \\ 
\text{B}_{exp}=b(\theta _{1},\theta _{2})+y(\theta _{1},\theta _{2}).%
\end{array}%
\right. \bigskip
\end{equation}
Then, with applicable $(\widetilde{\theta }_{1},\widetilde{\theta }_{2})$ as
solutions of equalities, we can further know $o(\widetilde{\theta }_{1},%
\widetilde{\theta }_{2})\equiv $ O$_{t}$ and $c(\widetilde{\theta }_{1},%
\widetilde{\theta }_{2})\equiv $AB$_{t}$. Since A$_{exp}=$A$_{t}$ and B$%
_{exp}=$B$_{t}$ by Eq.(\ref{eq:fitAandB}), 
we could test the differences and error bars of O
and AB\ as  follows: 
\begin{equation}
\left\{ 
\begin{array}{c}
\text{O}_{t}-\text{O}_{exp}\equiv \text{Diff}\%\text{(O}_{t}-\text{O}_{exp}%
\text{)}, \\ 
\text{AB}_{t}-\text{AB}_{exp}\equiv \text{Diff}\%\text{(AB}_{t}-\text{AB}%
_{exp}\text{)}, \\ 
\frac{\text{O}_{t}-\text{O}_{exp}}{\text{O}_{exp}}\equiv \text{Error}\%\text{%
(}\frac{\text{O}_{t}-\text{O}_{exp}}{\text{O}_{exp}}\text{)}, \\ 
\frac{\text{AB}_{t}-\text{AB}_{exp}}{\text{AB}_{exp}}\equiv \text{Error}\%%
\text{(}\frac{\text{AB}_{t}-\text{AB}_{exp}}{\text{AB}_{exp}}\text{)}.%
\end{array}%
\right. \bigskip
\end{equation}
%
%
As we mention the 2-dimensional equilibrium parameterization is an one-to-one mapping
if we define that $\theta _{1}=0,\forall \theta _{2}$ shrinks into a point.
Hence, we can introduce the $(\theta _{1}\cos (\theta _{2}),\theta _{1}\sin
(\theta _{2}))$ polar coordinates mapped from the Eq.(\ref{eq:Equil2}) of the parameter $(\theta _{1},\theta
_{2})$. In other words, we define that $\theta _{1}$ represents the radial
direction$(0\leq \theta _{1}\leq \frac{\pi }{2})$, $\theta _{2}$ represents
the angle direction$(0\leq \theta _{2}\leq \frac{\pi }{2})$ for our stable
equilibrium solution diagram, show in FIG.\ref{fig:1} 
as our blood type mapping \emph{chart}. 
This chart 
is a quadrant: one fourth of a circle. Inside the quadrant
consists of all stable fixed points mapping from the 2-dimensional manifold Eq.(\ref{eq:Equil2}).  
At three
corners of the quadrant, there are o$\equiv $O(ii), a$\equiv $A(AA), and b$\equiv $B(BB).
For specifying their Cartesian coordinates $(o,a,b,x,y,c)$, each of these is 
$100\%$ of that specified type - e.g. O(ii) $=(1,0,0,0,0,0)$, A(AA) $%
=(0,1,0,0,0,0)$, B(BB) $=(0,0,1,0,0,0)$. 
Similarly, at the midpoints of the three edge sides, we have
X(Ai) $=(\frac{1}{2},\frac{1}{4},0,\frac{1}{2},0,0)$, Y(Bi) $=(\frac{1}{2},0,%
\frac{1}{4},0,\frac{1}{2},0)$, and C(AB) $=(0,0,0,\frac{1}{4},\frac{1}{4},%
\frac{1}{2})$. The blue line represents the symmetric line 
invariant under switching $a,x$ to $b,y$. %

From Eq.(\ref{eq:Equil2}), 
we could easily see the blue line is $\theta _{2}=\frac{%
\pi }{4}$. The green curve represents the symmetry 
invariant under switching $a,c$ to $o,y$.
This gives the green curve parameterization:
\begin{equation}
\theta _{1}(\theta _{2})=\frac{(\sqrt{3+\cos (2\theta _{2})+2\sin (2\theta
_{2})}(1+\tan (\theta _{2})))}{\sqrt{2+2\sin (2\theta _{2})}}.
\end{equation}
The red curve represents the symmetry 
invariant under switching $b,c$ to $o,x$.
This gives the red curve parameterization:
\begin{equation}
\theta _{1}(\theta _{2})=\frac{(\sqrt{3-\cos (2\theta _{2})+2\sin (2\theta
_{2})}(1+\cot (\theta _{2})))}{\sqrt{2+2\sin (2\theta _{2})}}.
\end{equation}
The intersection of three color lines, is $(\frac{2}{9},\frac{2}{9},\frac{2}{%
9},\frac{1}{9},\frac{1}{9},\frac{1}{9})$, which is the most symmetric mid point.

\subsubsection{Explanation and application of the chart - relations of
different ethnic groups by geodesic distances \label{sec:G2-1geo}}

The experimental data that we implemented is mainly from two data resources \cite{data}. 
%
%
We fit the experimental data \cite{data} for both ({\bf 1}) different ethnic groups and ({\bf 2}) different countries. 
We present the {\bf experimental data} (O$_{exp}$,A$_{exp}$,B$_{exp}$,AB$_{exp}$)
in Table.\ref{Table1}, \ref{Table3}, \ref{Table4}. 
We present our corresponding {\bf theoretical prediction}, the data (O$_{t}$,AA$_{t}$, Ai$_{t}$,BB$_{t}$,Bi$_{t}$,AB$_{i}$) in
Table.\ref{Table2}, \ref{Table5}, \ref{Table6}.  \\

Data
fitting for ethnic groups is much promising, because random mating
assumption is approximately true for a certain given ethnic group. 
On the other hand, a
country may possess multi-ethnicities 
where the random mating assumption 
generally fails for the country with multi-ethnic groups.\bigskip

We also investigate the data distribution of different ethnic groups and countries
under our $(\theta _{1},\theta _{2})$ domain space. We found some facts:

\noindent
(1) American aboriginals possess large portion of Type O. \\
%
%
\noindent
(2) Europeans or Caucasians possess much more Type A than others. \\
%
\noindent
(3) Mongolians, like Buryats, possess comparably more Type B.\\
\noindent
(4) Diff\% and Error\% for islanders, like Irish and Japanese,
are smaller than races living in a larger continent, like Hindus.
Thus, we can say that the closed system assumption plays an important rule 
such that the data of islanders are more close to the stable fixed-point equilibrium solutions.


In our plots (see {Figures} in Sec.\ref{sec:fig}), $\theta _{1}$ represents the radial direction %
$(0\leq \theta _{1}\leq \frac{\pi }{2})$, $\theta _{2}$ represents the angle
direction $(0\leq \theta _{2}\leq \frac{\pi }{2})$. 
We invent the \emph{chart} to organize the data - for the blood type model, we have a \emph{quadrant} spanned by $\theta _{1}$ and $\theta _{2}$
with the horizontal and vertical coordinates as $(\theta _{1} \cos(\theta _{2}) , \theta _{1} \sin(\theta _{2}))$.
We propose the
distribution of different ethnic groups (marked with numbers) in our plots would
reveal their relative distance correlations. 
More precisely, {\bf to quantitatively define the distance between two ethnic groups, 
we propose a mathematical method to achieve this by solving the
geodesic equation} numerically from the given metric of stable-solution
manifold. 

For the model of 2 alleles studied here in Sec.\ref{sec:1-1hair},
we can write down the intrinsic metric of 1-dimensional equilibrium manifold Eq.(\ref{eq:Equil}).
This can be done by considering the infinitesimal distance $ds$ in the genotype frequency space by 
transforming the Euclidean $(H,h,x)$ space to the curved space $(\theta_1)$:
\begin{multline}
ds^2=dH^2 +dh^2+dx^2 =\big((\frac{\partial H}{\partial \theta_1})^2+(\frac{\partial h}{\partial \theta_1})^2+(\frac{\partial x}{\partial \theta_1})^2 \big)d\theta_1{}^2\;\;\;\;\;\;\; \\ 
\equiv g_{\theta_1,\theta_1} d\theta_1{}^2
\end{multline}

For example, for the model of 3 alleles studied in Sec.\ref{sec:2-1G},
we can write down the intrinsic metric of 2-dimensional equilibrium manifold Eq.(\ref{eq:Equil2}).
This can be done by considering the infinitesimal distance $ds$ in the genotype frequency space by 
transforming the Euclidean $(o,a,b,x,y,c)$ space to the curved space $(\theta_1,\theta_2)$:
\bea
&&ds^2=do^2 +da^2+db^2+dx^2+dy^2+dc^2 \nonumber \\
&&\equiv g_{\theta_1,\theta_1} d\theta_1{}^2 +
2g_{\theta_1,\theta_2} d\theta_1 d\theta_2 +g_{\theta_2,\theta_2} d\theta_2{}^2 =g_{\mu \nu} d\theta_\mu d\theta_\nu ,\;\;\;\;\;\;\;
\eea
where in the second line all the parameters are re-written in terms of $(\theta_1,\theta_2)$, e.g. 
$do=(\frac{\partial o}{\partial \theta_1}) d\theta_1 + (\frac{\partial o}{\partial \theta_2}) d\theta_2$, etc.
We propose that:\\ 

{\bf The 
genetic distance between two ethnic groups with stabilized genotype frequencies
can be 
defined as the \emph{geodesic length} by solving the geodesic equation
on the stable equilibrium manifold.}  \\ 

On one hand, we could further define
this least distance as the ``\emph{genetic distance}'' 
between any pair of two
ethnic groups.  
On the other hand, by comparing the geographical distribution of ethnic groups
and the distribution of ethnic groups on our equilibrium-solution chart, we find their 
strong correlation and certain meaningful pattern. 
The details of the data of {geodesic distance} shall be left for the future.



\begin{table*}[tbp]  
\begin{ruledtabular}
\begin{tabular}{cccccccccc}  
 Index&Ethnic (People Group)&O$_{exp}$(\%)&A$_{exp}$(\%)
&B$_{exp}$(\%)&AB$_{exp}$(\%)& $\theta_{1} $ & $\theta_{2} $ &Diff\%(O$_{t}$-O$_{exp}$)&Diff\%(AB$_{t}$-AB$_{exp}$) \\ \hline
1	&	Aborigines	&	61	&	39	&	0	&	0	&	0.2734 	&	0.0000 	&	0.0000 	&	0.0000 	\\
2	&	Abyssinians	&	43	&	27	&	25	&	5	&	0.5011 	&	0.7511 	&	-1.2547 	&	1.2547 	\\
3	&	Albanians	&	38	&	43	&	13	&	6	&	0.5487 	&	0.3314 	&	0.5169 	&	-0.5169 	\\
4	&	Arabs	&	34	&	31	&	29	&	6	&	0.6971 	&	0.7570 	&	-4.3754 	&	4.3754 	\\
5	&	Armenians	&	31	&	50	&	13	&	6	&	0.7012 	&	0.3027 	&	-1.6005 	&	1.6005 	\\
6	&	Asian(USA)	&	40	&	28	&	27	&	5	&	0.5646 	&	0.7694 	&	-2.5159 	&	2.5159 	\\
7	&	Bantus	&	46	&	30	&	19	&	5	&	0.4423 	&	0.5843 	&	0.0498 	&	-0.0498 	\\
8	&	Basques	&	51	&	44	&	4	&	1	&	0.3860 	&	0.1052 	&	-0.4424 	&	0.4424 	\\
9	&	Belgians	&	47	&	42	&	8	&	3	&	0.4285 	&	0.2140 	&	0.1153 	&	-0.1153 	\\
10	&	Blackfoot(N. Am. Indian)	&	17	&	82	&	0	&	1	&	0.9357 	&	0.0000 	&	1.0000 	&	-1.0000 	\\
11	&	Bororo	&	100	&	0	&	0	&	0	&	0.0000 	&		&	0.0000 	&	0.0000 	\\
12	&	Brazilians	&	47	&	41	&	9	&	3	&	0.4323 	&	0.2437 	&	-0.1799 	&	0.1799 	\\
13	&	Burmese	&	36	&	24	&	33	&	7	&	0.6087 	&	0.9224 	&	-1.2738 	&	1.2738 	\\
14	&	Buryats	&	33	&	21	&	38	&	8	&	0.6504 	&	1.0313 	&	-0.7470 	&	0.7470 	\\
15	&	Chinese-Canton	&	46	&	23	&	25	&	6	&	0.4290 	&	0.8232 	&	1.0825 	&	-1.0825 	\\
16	&	Danes	&	41	&	44	&	11	&	4	&	0.5206 	&	0.2798 	&	-0.6050 	&	0.6050 	\\
17	&	Dutch	&	45	&	43	&	9	&	3	&	0.4622 	&	0.2354 	&	-0.4505 	&	0.4505 	\\
18	&	Eskimos(Alaska)	&	38	&	44	&	13	&	5	&	0.5680 	&	0.3264 	&	-0.7347 	&	0.7347 	\\
19	&	Finns	&	34	&	41	&	18	&	7	&	0.6372 	&	0.4546 	&	-0.9803 	&	0.9803 	\\
20	&	French	&	43	&	47	&	7	&	3	&	0.4836 	&	0.1736 	&	-0.0098 	&	0.0098 	\\
21	&	Germans	&	41	&	43	&	11	&	5	&	0.5034 	&	0.2842 	&	0.5868 	&	-0.5868 	\\
22	&	Hindus(Bombay)	&	32	&	29	&	28	&	11	&	0.6127 	&	0.7701 	&	2.4826 	&	-2.4826 	\\
23	&	Hungarians	&	36	&	43	&	16	&	5	&	0.6267 	&	0.3990 	&	-2.3598 	&	2.3598 	\\
24	&	Indians(USA)	&	79	&	16	&	4	&	1	&	0.1201 	&	0.2532 	&	0.6212 	&	-0.6212 	\\
25	&	Irish	&	52	&	35	&	10	&	3	&	0.3667 	&	0.3035 	&	0.2035 	&	-0.2035 	\\
26	&	Japanese	&	30	&	38	&	22	&	10	&	0.6810 	&	0.5591 	&	0.5134 	&	-0.5134 	\\
27	&	Jews(Germany)	&	42	&	41	&	12	&	5	&	0.4907 	&	0.3184 	&	0.4772 	&	-0.4772 	\\
28	&	Jews(Poland)	&	33	&	41	&	18	&	8	&	0.6372 	&	0.4546 	&	0.0197 	&	-0.0197 	\\
29	&	Kikuyu(Kenya)	&	60	&	19	&	20	&	1	&	0.3015 	&	0.8093 	&	-1.8112 	&	1.8112 	\\
30	&	Lapps	&	29	&	63	&	4	&	4	&	0.6892 	&	0.0847 	&	1.0566 	&	-1.0566 	\\
31	&	Latvians	&	32	&	37	&	24	&	7	&	0.7180 	&	0.6056 	&	-3.5096 	&	3.5096 	\\
32	&	Lithuanians	&	40	&	34	&	20	&	6	&	0.5345 	&	0.5582 	&	-0.5444 	&	0.5444 	\\
33	&	Malasians	&	62	&	18	&	20	&	0	&	0.2894 	&	0.8346 	&	-2.6261 	&	2.6261 	\\
34	&	Moros	&	64	&	16	&	20	&	0	&	0.2663 	&	0.8895 	&	-2.2717 	&	2.2717 	\\
35	&	Navajo(N. Am. Indain)	&	73	&	27	&	0	&	0	&	0.1688 	&	0.0000 	&	0.0000 	&	0.0000 	\\
36	&	Papuas(New Guinea)	&	41	&	27	&	23	&	9	&	0.4631 	&	0.7135 	&	3.4828 	&	-3.4828 	\\
37	&	Persians	&	38	&	33	&	22	&	7	&	0.5594 	&	0.6103 	&	-0.1801 	&	0.1801 	\\
38	&	Peru(Indians)	&	100	&	0	&	0	&	0	&	0.0000 	&		&	0.0000 	&	0.0000 	\\
39	&	Philippinos	&	45	&	22	&	27	&	6	&	0.4452 	&	0.8774 	&	0.8265 	&	-0.8265 	\\
40	&	Portuguese	&	35	&	53	&	8	&	4	&	0.6123 	&	0.1834 	&	-0.4928 	&	0.4928 	\\
41	&	Rumanians	&	34	&	41	&	19	&	6	&	0.6673 	&	0.4755 	&	-2.7138 	&	2.7138 	\\
42	&	Russians	&	33	&	36	&	23	&	8	&	0.6572 	&	0.5969 	&	-1.1401 	&	1.1401 	\\
43	&	Sardinians	&	50	&	26	&	19	&	5	&	0.3808 	&	0.6435 	&	0.9971 	&	-0.9971 	\\
44	&	Scotts	&	51	&	34	&	12	&	3	&	0.3846 	&	0.3659 	&	-0.3254 	&	0.3254 	\\
45	&	Shompen(Nicobars)	&	100	&	0	&	0	&	0	&	0.0000 	&		&	0.0000 	&	0.0000 	\\
46	&	Slovaks	&	42	&	37	&	16	&	5	&	0.5049 	&	0.4409 	&	-0.5190 	&	0.5190 	\\
47	&	South Africans	&	45	&	40	&	11	&	4	&	0.4547 	&	0.2993 	&	0.1149 	&	-0.1149 	\\
48	&	Spanish	&	38	&	47	&	10	&	5	&	0.5510 	&	0.2447 	&	0.3680 	&	-0.3680 	\\
49	&	Swedes	&	38	&	47	&	10	&	5	&	0.5510 	&	0.2447 	&	0.3680 	&	-0.3680 	\\
50	&	Swiss	&	40	&	50	&	7	&	3	&	0.5321 	&	0.1666 	&	-0.3848 	&	0.3849 	\\
51	&	Tartars	&	28	&	30	&	29	&	13	&	0.6671 	&	0.7708 	&	3.2975 	&	-3.2975 	\\
52	&	Thais	&	37	&	22	&	33	&	8	&	0.5594 	&	0.9605 	&	0.8199 	&	-0.8199 	\\
53	&	United Kingdom(GB)	&	47	&	42	&	8	&	3	&	0.4285 	&	0.2140 	&	0.1153 	&	-0.1153 	\\
54	&	USA(blacks)	&	49	&	27	&	20	&	4	&	0.4114 	&	0.6504 	&	-0.5293 	&	0.5293 	\\
55	&	USA(whites)	&	45	&	40	&	11	&	4	&	0.4547 	&	0.2993 	&	0.1149 	&	-0.1149 	\\
\hline
\end{tabular} 
\end{ruledtabular}
\caption{ETHNIC GROUP.1 This table shows available {\bf experimental data}
of blood type ratios of O,A,B, and AB (denoted with a sub-indices ``$\exp $.'')
for 55 different ethnics (ethnic groups) around the world \cite{data}. The index of
certain ethnic corresponds to the specified number of a data point
distributed on the plots of FIG \ref{fig:2}, FIG \ref{fig:3}, FIG \ref{fig:10} and FIG \ref{fig:11}.
We fit blood type population percentage of
avalable experimental results to the 2 parameter spaces $\protect\theta _{1}$
and $\protect\theta _{2}$ of our equilibrium solution. The error bar is 
comparably small. The FIG \ref{fig:2}, \ref{fig:3}, \ref{fig:10} and \ref{fig:11} (the mappings of equilibrium
solutions) may be a good way of data organization for revealing the relations of different
ethnic groups. }
 \label{Table1}
\end{table*}

\begin{table*}[tbp] 
\begin{ruledtabular}
\begin{tabular}{cccccccccc} 
Index&Ethnic (People Group) &O$_{t}$(\%)&AA$_{t}$(\%)&Ai$_{t}$(\%)
&BB$_{t}$(\%)&Bi$_{t}$(\%)&AB$_{t}$(\%)&Error\%($\frac{\text{O}_{t}-\text{O}_{\exp }}{\text{O}_{\exp }}$)&Error\%($\frac{\text{AB}_{t}-\text{AB}_{\exp }}{\text{AB}_{\exp }}$) \\ \hline
1	&	Aborigines	&	61.00 	&	4.80 	&	34.21 	&	0.00 	&	0.00 	&	0.00 	&	0.00 	&		\\
2	&	Abyssinians	&	41.75 	&	3.35 	&	23.65 	&	2.92 	&	22.08 	&	6.25 	&	-2.92 	&	25.09 	\\
3	&	Albanians	&	38.52 	&	7.97 	&	35.03 	&	0.94 	&	12.06 	&	5.48 	&	1.36 	&	-8.62 	\\
4	&	Arabs	&	29.62 	&	5.49 	&	25.51 	&	4.90 	&	24.10 	&	10.38 	&	-12.87 	&	72.92 	\\
5	&	Armenians	&	29.40 	&	12.17 	&	37.83 	&	1.19 	&	11.81 	&	7.60 	&	-5.16 	&	26.67 	\\
6	&	Asian(USA)	&	37.48 	&	3.88 	&	24.12 	&	3.64 	&	23.36 	&	7.52 	&	-6.29 	&	50.32 	\\
7	&	Bantus	&	46.05 	&	3.74 	&	26.26 	&	1.64 	&	17.36 	&	4.95 	&	0.11 	&	-1.00 	\\
8	&	Basques	&	50.56 	&	6.83 	&	37.17 	&	0.08 	&	3.92 	&	1.44 	&	-0.87 	&	44.24 	\\
9	&	Belgians	&	47.12 	&	6.64 	&	35.36 	&	0.31 	&	7.69 	&	2.88 	&	0.25 	&	-3.84 	\\
10	&	Blackfoot(N. Am. Indian)	&	18.00 	&	33.15 	&	48.85 	&	0.00 	&	0.00 	&	0.00 	&	5.88 	&	-100.00 	\\
11	&	Bororo	&	100.00 	&	0.00 	&	0.00 	&	0.00 	&	0.00 	&	0.00 	&	0.00 	&		\\
12	&	Brazilians	&	46.82 	&	6.39 	&	34.61 	&	0.40 	&	8.60 	&	3.18 	&	-0.38 	&	6.00 	\\
13	&	Burmese	&	34.73 	&	3.13 	&	20.87 	&	5.46 	&	27.54 	&	8.27 	&	-3.54 	&	18.20 	\\
14	&	Buryats	&	32.25 	&	2.62 	&	18.38 	&	7.30 	&	30.70 	&	8.75 	&	-2.26 	&	9.34 	\\
15	&	Chinese-Canton	&	47.08 	&	2.28 	&	20.72 	&	2.65 	&	22.35 	&	4.92 	&	2.35 	&	-18.04 	\\
16	&	Danes	&	40.40 	&	8.01 	&	35.99 	&	0.66 	&	10.34 	&	4.60 	&	-1.48 	&	15.12 	\\
17	&	Dutch	&	44.55 	&	7.19 	&	35.81 	&	0.41 	&	8.59 	&	3.45 	&	-1.00 	&	15.02 	\\
18	&	Eskimos(Alaska)	&	37.27 	&	8.47 	&	35.53 	&	0.97 	&	12.03 	&	5.73 	&	-1.93 	&	14.69 	\\
19	&	Finns	&	33.02 	&	8.16 	&	32.84 	&	1.95 	&	16.05 	&	7.98 	&	-2.88 	&	14.00 	\\
20	&	French	&	42.99 	&	8.58 	&	38.42 	&	0.26 	&	6.74 	&	3.01 	&	-0.02 	&	0.33 	\\
21	&	Germans	&	41.59 	&	7.55 	&	35.45 	&	0.64 	&	10.36 	&	4.41 	&	1.43 	&	-11.74 	\\
22	&	Hindus(Bombay)	&	34.48 	&	4.39 	&	24.61 	&	4.13 	&	23.87 	&	8.52 	&	7.76 	&	-22.57 	\\
23	&	Hungarians	&	33.64 	&	8.73 	&	34.27 	&	1.55 	&	14.45 	&	7.36 	&	-6.56 	&	47.20 	\\
24	&	Indians(USA)	&	79.62 	&	0.73 	&	15.27 	&	0.05 	&	3.95 	&	0.38 	&	0.79 	&	-62.12 	\\
25	&	Irish	&	52.20 	&	4.47 	&	30.53 	&	0.44 	&	9.56 	&	2.80 	&	0.39 	&	-6.78 	\\
26	&	Japanese	&	30.51 	&	7.58 	&	30.42 	&	2.97 	&	19.03 	&	9.49 	&	1.71 	&	-5.13 	\\
27	&	Jews(Germany)	&	42.48 	&	6.86 	&	34.14 	&	0.75 	&	11.25 	&	4.52 	&	1.14 	&	-9.54 	\\
28	&	Jews(Poland)	&	33.02 	&	8.16 	&	32.84 	&	1.95 	&	16.05 	&	7.98 	&	0.06 	&	-0.25 	\\
29	&	Kikuyu(Kenya)	&	58.19 	&	1.34 	&	17.66 	&	1.47 	&	18.53 	&	2.81 	&	-3.02 	&	181.12 	\\
30	&	Lapps	&	30.06 	&	17.34 	&	45.66 	&	0.12 	&	3.88 	&	2.94 	&	3.64 	&	-26.42 	\\
31	&	Latvians	&	28.49 	&	7.59 	&	29.41 	&	3.64 	&	20.36 	&	10.51 	&	-10.97 	&	50.14 	\\
32	&	Lithuanians	&	39.46 	&	5.24 	&	28.76 	&	2.04 	&	17.96 	&	6.54 	&	-1.36 	&	9.07 	\\
33	&	Malasians	&	59.37 	&	1.19 	&	16.81 	&	1.45 	&	18.55 	&	2.63 	&	-4.24 	&		\\
34	&	Moros	&	61.73 	&	0.92 	&	15.08 	&	1.40 	&	18.60 	&	2.27 	&	-3.55 	&		\\
35	&	Navajo(N. Am. Indain)	&	73.00 	&	2.12 	&	24.88 	&	0.00 	&	0.00 	&	0.00 	&	0.00 	&		\\
36	&	Papuas(New Guinea)	&	44.48 	&	3.19 	&	23.81 	&	2.39 	&	20.61 	&	5.52 	&	8.49 	&	-38.70 	\\
37	&	Persians	&	37.82 	&	5.13 	&	27.87 	&	2.51 	&	19.49 	&	7.18 	&	-0.47 	&	2.57 	\\
38	&	Peru(Indians)	&	100.00 	&	0.00 	&	0.00 	&	0.00 	&	0.00 	&	0.00 	&	0.00 	&		\\
39	&	Philippinos	&	45.83 	&	2.15 	&	19.85 	&	3.11 	&	23.89 	&	5.17 	&	1.84 	&	-13.77 	\\
40	&	Portuguese	&	34.51 	&	12.11 	&	40.89 	&	0.42 	&	7.58 	&	4.49 	&	-1.41 	&	12.32 	\\
41	&	Rumanians	&	31.29 	&	8.46 	&	32.54 	&	2.24 	&	16.76 	&	8.71 	&	-7.98 	&	45.23 	\\
42	&	Russians	&	31.86 	&	6.72 	&	29.28 	&	3.11 	&	19.89 	&	9.14 	&	-3.45 	&	14.25 	\\
43	&	Sardinians	&	51.00 	&	2.67 	&	23.33 	&	1.50 	&	17.50 	&	4.00 	&	1.99 	&	-19.94 	\\
44	&	Scotts	&	50.67 	&	4.34 	&	29.66 	&	0.64 	&	11.36 	&	3.33 	&	-0.64 	&	10.85 	\\
45	&	Shompen(Nicobars)	&	100.00 	&	0.00 	&	0.00 	&	0.00 	&	0.00 	&	0.00 	&	0.00 	&		\\
46	&	Slovaks	&	41.48 	&	5.85 	&	31.15 	&	1.30 	&	14.70 	&	5.52 	&	-1.24 	&	10.38 	\\
47	&	South Africans	&	45.11 	&	6.30 	&	33.70 	&	0.60 	&	10.40 	&	3.89 	&	0.26 	&	-2.87 	\\
48	&	Spanish	&	38.37 	&	9.27 	&	37.73 	&	0.58 	&	9.42 	&	4.63 	&	0.97 	&	-7.36 	\\
49	&	Swedes	&	38.37 	&	9.27 	&	37.73 	&	0.58 	&	9.42 	&	4.63 	&	0.97 	&	-7.36 	\\
50	&	Swiss	&	39.62 	&	10.06 	&	39.94 	&	0.28 	&	6.72 	&	3.38 	&	-0.96 	&	12.83 	\\
51	&	Tartars	&	31.30 	&	4.99 	&	25.01 	&	4.71 	&	24.29 	&	9.70 	&	11.78 	&	-25.37 	\\
52	&	Thais	&	37.82 	&	2.51 	&	19.49 	&	5.13 	&	27.87 	&	7.18 	&	2.22 	&	-10.25 	\\
53	&	United Kingdom(GB)	&	47.12 	&	6.64 	&	35.36 	&	0.31 	&	7.69 	&	2.88 	&	0.25 	&	-3.84 	\\
54	&	USA(blacks)	&	48.47 	&	2.98 	&	24.02 	&	1.72 	&	18.28 	&	4.53 	&	-1.08 	&	13.23 	\\
55	&	USA(whites)	&	45.11 	&	6.30 	&	33.70 	&	0.60 	&	10.40 	&	3.89 	&	0.26 	&	-2.87 	\\
\hline 
\end{tabular} 
\end{ruledtabular}
\caption{ETHNIC GROUP.2 This table continues from the previous
TABLE \ref{Table1}. 
This TABLE \ref{Table2} 
shows our {\bf theoretical prediction} corresponding to the experimental data of TABLE \ref{Table1}. 
The experiment data fitting procedure is that we use two constraints
type A ($AA$ and $Ai$) and type B ($BB$ and $Bi$), comparing these with
2-dimensional equilibrium solution, and finding out total 6 different blood types by
our  {\bf theoretical stable fixed-point prediction}. In this table, we compute the error bar of O and AB.
In addition, we 
compute our theoretical prediction on the population ratio of $AA$ and $Ai$, $BB$ and $Bi$. %
For the usual blood type test, these data may not be easily determined.
However, we compile our theoretical prediction data here which may be testable data for future experiments. }
 \label{Table2}
\end{table*}

\begin{table*}[tbp]
\begin{ruledtabular}
\begin{tabular}{cccccccccc}
 Index&Country&O$_{exp}$(\%)&A$_{exp}$(\%)
&B$_{exp}$(\%)&AB$_{exp}$(\%)& $\theta_{1} $ & $\theta_{2} $ &Diff\%(O$_{t}$-O$_{exp}$)&Diff\%(AB$_{t}$-AB$_{exp}$) \\ \hline
1	&	Canada	&	41	&	45	&	10	&	4	&	0.5156 	&	0.2519 	&	-0.2600 	&	0.2600 	\\
2	&	USA	&	45	&	41	&	10	&	4	&	0.4510 	&	0.2691 	&	0.3930 	&	-0.3930 	\\
3	&	Mexico	&	84	&	11	&	4	&	1	&	0.0860 	&	0.3550 	&	0.7512 	&	-0.7512 	\\
4	&	Guatemala	&	95	&	3	&	2	&	0	&	0.0261 	&	0.5892 	&	-0.0312 	&	0.0312 	\\
5	&	El Salvador	&	60	&	26	&	11	&	3	&	0.2738 	&	0.4187 	&	0.9503 	&	-0.9503 	\\
6	&	Nicaragua	&	92	&	7	&	1	&	0	&	0.2738 	&	0.1441 	&	-0.0373 	&	0.0373 	\\
7	&	Costa Rica	&	53	&	31	&	13	&	3	&	0.3597 	&	0.4210 	&	-0.1914 	&	0.1914 	\\
8	&	Cuba	&	49	&	36	&	12	&	3	&	0.4128 	&	0.3503 	&	-0.6358 	&	0.6358 	\\
9	&	Jamaica	&	56	&	20	&	21	&	3	&	0.3266 	&	0.8080 	&	-0.1988 	&	0.1988 	\\
10	&	Dominican Rep	&	53	&	31	&	13	&	3	&	0.3597 	&	0.4210 	&	-0.1914 	&	0.1914 	\\
11	&	Haiti	&	51	&	27	&	18	&	4	&	0.3800 	&	0.6035 	&	0.0651 	&	-0.0651 	\\
12	&	Puerto Rico	&	47	&	30	&	13	&	10	&	0.3468 	&	0.4317 	&	6.9573 	&	-6.9573 	\\
13	&	Colombia	&	62	&	27	&	9	&	2	&	0.2610 	&	0.3401 	&	0.2843 	&	-0.2843 	\\
14	&	Ecuador	&	63	&	29	&	6	&	2	&	0.2469 	&	0.2195 	&	0.7914 	&	-0.7914 	\\
15	&	Peru	&	71	&	19	&	9	&	1	&	0.1855 	&	0.4544 	&	-0.1006 	&	0.1006 	\\
16	&	Chile	&	56	&	31	&	10	&	3	&	0.3174 	&	0.3342 	&	0.6607 	&	-0.6607 	\\
17	&	Bolivea	&	93	&	5	&	2	&	0	&	0.0372 	&	0.3832 	&	-0.0528 	&	0.0528 	\\
18	&	Venezuela	&	44	&	37	&	14	&	5	&	0.4645 	&	0.3933 	&	0.3814 	&	-0.3814 	\\
19	&	Brazil	&	48	&	39	&	10	&	3	&	0.4213 	&	0.2793 	&	-0.3173 	&	0.3173 	\\
20	&	Argentina	&	48	&	39	&	10	&	3	&	0.4213 	&	0.2793 	&	-0.3173 	&	0.3173 	\\
21	&	Paraguay	&	56	&	34	&	9	&	1	&	0.3393 	&	0.2818 	&	-1.3696 	&	1.3696 	\\
22	&	Uruguay	&	45	&	42	&	10	&	3	&	0.4664 	&	0.2644 	&	-0.7603 	&	0.7603 	\\
23	&	Australia	&	47	&	40	&	10	&	3	&	0.4359 	&	0.2740 	&	-0.4594 	&	0.4594 	\\
24	&	New Zealand	&	48	&	40	&	9	&	3	&	0.4180 	&	0.2483 	&	-0.0520 	&	0.0520 	\\
25	&	Iceland	&	56	&	32	&	10	&	2	&	0.3293 	&	0.3258 	&	-0.4484 	&	0.4484 	\\
26	&	Ireland	&	54	&	33	&	10	&	3	&	0.3414 	&	0.3179 	&	0.4391 	&	-0.4391 	\\
27	&	United Kingdom	&	47	&	41	&	9	&	3	&	0.4323 	&	0.2437 	&	-0.1799 	&	0.1799 	\\
28	&	Scotland	&	51	&	35	&	11	&	3	&	0.3823 	&	0.3310 	&	-0.1307 	&	0.1307 	\\
29	&	England	&	46	&	43	&	8	&	3	&	0.4430 	&	0.2103 	&	-0.0025 	&	0.0025 	\\
30	&	Norway	&	39	&	49	&	8	&	4	&	0.5386 	&	0.1922 	&	0.1844 	&	-0.1844 	\\
31	&	Sweden	&	38	&	47	&	10	&	5	&	0.5510 	&	0.2447 	&	0.3680 	&	-0.3680 	\\
32	&	Finland	&	33	&	42	&	18	&	7	&	0.6619 	&	0.4480 	&	-1.4095 	&	1.4095 	\\
33	&	Denmark	&	42	&	44	&	10	&	4	&	0.4987 	&	0.2558 	&	-0.0863 	&	0.0863 	\\
34	&	Netherland	&	45	&	43	&	9	&	3	&	0.4622 	&	0.2354 	&	-0.4505 	&	0.4505 	\\
35	&	Belgium	&	46	&	43	&	8	&	3	&	0.4430 	&	0.2103 	&	-0.0025 	&	0.0025 	\\
36	&	France	&	43	&	45	&	9	&	3	&	0.4938 	&	0.2279 	&	-0.7431 	&	0.7431 	\\
37	&	Spain	&	42	&	45	&	9	&	4	&	0.4938 	&	0.2279 	&	0.2569 	&	-0.2569 	\\
38	&	Portugal	&	41	&	48	&	8	&	3	&	0.5215 	&	0.1948 	&	-0.6655 	&	0.6655 	\\
39	&	Germany	&	38	&	43	&	13	&	6	&	0.5487 	&	0.3314 	&	0.5169 	&	-0.5169 	\\
40	&	Austria	&	39	&	44	&	13	&	4	&	0.5680 	&	0.3264 	&	-1.7347 	&	1.7347 	\\
41	&	Swiss	&	42	&	47	&	8	&	3	&	0.5049 	&	0.1976 	&	-0.5218 	&	0.5218 	\\
42	&	Italy	&	45	&	40	&	11	&	4	&	0.4547 	&	0.2993 	&	0.1149 	&	-0.1149 	\\
43	&	North Italy	&	42	&	43	&	11	&	4	&	0.5034 	&	0.2842 	&	-0.4132 	&	0.4132 	\\
44	&	South Italy	&	46	&	35	&	15	&	4	&	0.4510 	&	0.4343 	&	-0.6086 	&	0.6086 	\\
45	&	Czech	&	33	&	41	&	18	&	8	&	0.6372 	&	0.4546 	&	0.0197 	&	-0.0197 	\\
46	&	Hungary	&	30	&	43	&	19	&	8	&	0.7223 	&	0.4626 	&	-1.7437 	&	1.7437 	\\
47	&	*Gypsies	&	30	&	25	&	36	&	9	&	0.7213 	&	0.9376 	&	-1.6902 	&	1.6902 	\\
48	&	Poland	&	35	&	38	&	19	&	8	&	0.5951 	&	0.4987 	&	0.5608 	&	-0.5608 	\\
49	&	Yogoslavia	&	37	&	42	&	14	&	7	&	0.5534 	&	0.3600 	&	1.2048 	&	-1.2048 	\\
50	&	Romania	&	34	&	45	&	17	&	7	&	0.7083 	&	0.4097 	&	-4.9869 	&	1.9869 	\\
51	&	Bulgalia	&	32	&	44	&	16	&	8	&	0.6502 	&	0.3934 	&	0.2636 	&	-0.2636 	\\
52	&	Albania	&	42	&	36	&	17	&	5	&	0.5078 	&	0.4725 	&	-0.7220 	&	0.7220 	\\
53	&	Greece	&	43	&	39	&	13	&	5	&	0.4778 	&	0.3547 	&	0.4094 	&	-0.4094 	\\
54	&	Cyprus	&	35	&	46	&	13	&	6	&	0.6088 	&	0.3174 	&	-0.2807 	&	0.2807 	\\
55	&	Turkey	&	33	&	45	&	15	&	7	&	0.6439 	&	0.3665 	&	-0.3693 	&	0.3693 	\\
56	&	Iran(Westnorth)	&	36	&	33	&	24	&	7	&	0.6087 	&	0.6484 	&	-1.2738 	&	1.2738 	\\
57	&	Iran(Eastsouth)	&	42	&	26	&	25	&	7	&	0.4820 	&	0.7678 	&	1.1033 	&	-1.1033 	\\
58	&	Iran(Middleeast)	&	31	&	25	&	34	&	10	&	0.6623 	&	0.9162 	&	0.5671 	&	-0.5671 	\\
59	&	Iraq	&	35	&	31	&	26	&	8	&	0.6115 	&	0.7091 	&	-0.4441 	&	0.4441 	\\
60	&	Kuwait	&	47	&	24	&	24	&	5	&	0.4291 	&	0.7854 	&	0.0733 	&	-0.0733 	\\
61	&	Lebanon	&	36	&	47	&	12	&	5	&	0.6023 	&	0.2907 	&	-0.8845 	&	0.8845 	\\
62	&	Israel	&	36	&	41	&	17	&	6	&	0.6092 	&	0.4332 	&	-1.3042 	&	1.3042 	\\
\hline
\end{tabular}
\end{ruledtabular}
\caption{COUNTRY.1-1 This table shows available {\bf experimental data} of blood
type ratios for 62 different countries around the world, where totally 114 different countries' data are available \cite{data}. To be continued to
TABLE \ref{Table4}. 
} \label{Table3}
\end{table*}


\begin{table*}[tbp]
\begin{ruledtabular}
\begin{tabular}{cccccccccc}
 Index&Country &O$_{exp}$(\%)&A$_{exp}$(\%)
&B$_{exp}$(\%)&AB$_{exp}$(\%)& $\theta_{1} $ & $\theta_{2} $ &Diff\%(O$_{t}$-O$_{exp}$)&Diff\%(AB$_{t}$-AB$_{exp}$) \\ \hline
63	&	Pakistan	&	31	&	25	&	34	&	10	&	0.6623 	&	0.9162 	&	0.5671 	&	-0.5671 	\\
64	&	Afgjanistan	&	29	&	25	&	36	&	10	&	0.7213 	&	0.9376 	&	-0.6902 	&	0.6902 	\\
65	&	India(Westnorth)	&	29	&	21	&	41	&	9	&	0.7349 	&	1.0549 	&	-1.4084 	&	1.4084 	\\
66	&	India(Eastnorth)	&	32	&	25	&	36	&	7	&	0.7213 	&	0.9376 	&	-3.6902 	&	3.6902 	\\
67	&	India(Middlesouth)	&	37	&	22	&	34	&	7	&	0.5813 	&	0.9719 	&	-0.5815 	&	0.5815 	\\
68	&	Srilanka	&	47	&	22	&	26	&	5	&	0.4286 	&	0.8609 	&	0.1098 	&	-0.1098 	\\
69	&	Myanmar	&	35	&	25	&	32	&	8	&	0.6103 	&	0.8921 	&	-0.3710 	&	0.3710 	\\
70	&	Thailand	&	39	&	22	&	33	&	6	&	0.5594 	&	0.9605 	&	-1.1801 	&	1.1801 	\\
71	&	Malaysia	&	38	&	25	&	29	&	8	&	0.5418 	&	0.8508 	&	0.9718 	&	-0.9718 	\\
72	&	Philippines	&	45	&	26	&	24	&	5	&	0.4635 	&	0.7494 	&	-0.5460 	&	0.5460 	\\
73	&	Vietnam	&	42	&	22	&	31	&	5	&	0.5183 	&	0.9357 	&	-1.4429 	&	1.4429 	\\
74	&	Cambodia	&	39	&	23	&	35	&	3	&	0.6313 	&	0.9636 	&	-5.6284 	&	5.6284 	\\
75	&	South China	&	44	&	27	&	23	&	6	&	0.4631 	&	0.7135 	&	0.4828 	&	-0.4828 	\\
76	&	North China	&	33	&	30	&	29	&	8	&	0.6671 	&	0.7708 	&	-1.7025 	&	1.7025 	\\
77	&	Middle China	&	35	&	32	&	26	&	7	&	0.6371 	&	0.6960 	&	-1.9711 	&	1.9711 	\\
78	&	Hong Kong	&	44	&	24	&	26	&	6	&	0.4635 	&	0.8214 	&	0.4540 	&	-0.4540 	\\
79	&	Macao	&	40	&	25	&	28	&	7	&	0.5210 	&	0.8357 	&	0.3689 	&	-0.3689 	\\
80	&	Tibet	&	30	&	37	&	24	&	9	&	0.7180 	&	0.6056 	&	-1.5096 	&	1.5096 	\\
81	&	Taiwan	&	44	&	27	&	23	&	6	&	0.4631 	&	0.7135 	&	0.4828 	&	-0.4828 	\\
82	&	Nepal	&	30	&	37	&	24	&	9	&	0.7180 	&	0.6056 	&	-1.5096 	&	1.5096 	\\
83	&	Mongolia	&	37	&	22	&	34	&	7	&	0.5813 	&	0.9719 	&	-0.5815 	&	0.5815 	\\
84	&	South Korea	&	27	&	32	&	30	&	11	&	0.7656 	&	0.7584 	&	-1.0020 	&	1.0020 	\\
85	&	Japan	&	31	&	38	&	22	&	9	&	0.6810 	&	0.5591 	&	-0.4866 	&	0.4866 	\\
86	&	Sumatra	&	41	&	23	&	30	&	6	&	0.5195 	&	0.9025 	&	-0.5264 	&	0.5264 	\\
87	&	Java	&	37	&	26	&	29	&	8	&	0.5642 	&	0.8334 	&	0.5067 	&	-0.5067 	\\
88	&	Middle Asia(Russia)	&	32	&	28	&	31	&	9	&	0.6666 	&	0.8291 	&	-0.6754 	&	0.6754 	\\
89	&	Sibereria(Russia)	&	39	&	25	&	29	&	7	&	0.5418 	&	0.8508 	&	-0.0282 	&	0.0282 	\\
90	&	Russia(Europe)	&	34	&	35	&	23	&	8	&	0.6313 	&	0.6072 	&	-0.6284 	&	0.6284 	\\
91	&	Caucasus	&	44	&	32	&	20	&	4	&	0.4961 	&	0.5810 	&	-1.9009 	&	1.9009 	\\
92	&	Georgia	&	50	&	36	&	11	&	4	&	0.3960 	&	0.3239 	&	-0.2710 	&	-0.7290 	\\
93	&	Armenia	&	29	&	50	&	13	&	8	&	0.7012 	&	0.3027 	&	0.3995 	&	-0.3995 	\\
94	&	Egypt	&	34	&	34	&	24	&	8	&	0.6336 	&	0.6367 	&	-0.7674 	&	0.7674 	\\
95	&	Libya	&	40	&	36	&	19	&	5	&	0.5522 	&	0.5166 	&	-1.7167 	&	1.7167 	\\
96	&	Tunisia	&	46	&	31	&	18	&	5	&	0.4409 	&	0.5485 	&	0.1610 	&	-0.1610 	\\
97	&	Algelia	&	44	&	39	&	13	&	4	&	0.4778 	&	0.3547 	&	-0.5906 	&	0.5906 	\\
98	&	Morocco	&	37	&	34	&	23	&	6	&	0.6067 	&	0.6181 	&	-2.1529 	&	2.1529 	\\
99	&	Ethiopia	&	47	&	28	&	21	&	4	&	0.4445 	&	0.6568 	&	-1.1176 	&	1.1176 	\\
100	&	Somalia	&	58	&	26	&	13	&	3	&	0.2982 	&	0.4818 	&	0.5078 	&	-0.5078 	\\
101	&	Chad	&	45	&	27	&	26	&	2	&	0.5213 	&	0.7686 	&	-4.6521 	&	4.6521 	\\
102	&	Senegal	&	50	&	25	&	21	&	4	&	0.3967 	&	0.7061 	&	-0.3307 	&	0.3307 	\\
103	&	Libelia	&	46	&	25	&	24	&	5	&	0.4460 	&	0.7670 	&	-0.2296 	&	0.2296 	\\
104	&	Ghana	&	49	&	22	&	25	&	4	&	0.4126 	&	0.8435 	&	-0.6182 	&	0.6182 	\\
105	&	Cameroon	&	54	&	22	&	21	&	3	&	0.3533 	&	0.7640 	&	-0.6282 	&	0.6282 	\\
106	&	Congo	&	49	&	27	&	20	&	4	&	0.4114 	&	0.6504 	&	-0.5293 	&	0.5293 	\\
107	&	Guinea	&	64	&	17	&	17	&	2	&	0.2450 	&	0.7854 	&	0.0000 	&	0.0000 	\\
108	&	Angola	&	49	&	25	&	22	&	4	&	0.4126 	&	0.7273 	&	-0.6182 	&	0.6182 	\\
109	&	Uganda	&	49	&	23	&	23	&	5	&	0.3971 	&	0.7854 	&	0.6345 	&	-0.6345 	\\
110	&	Mozambique	&	57	&	23	&	17	&	3	&	0.3131 	&	0.6465 	&	0.0677 	&	-0.0677 	\\
111	&	Madagascar	&	43	&	23	&	28	&	6	&	0.4812 	&	0.8731 	&	0.1624 	&	-0.1624 	\\
112	&	Sorth Africa	&	45	&	35	&	16	&	4	&	0.4699 	&	0.4584 	&	-1.0201 	&	1.0201 	\\
113	&	Kenya	&	49	&	26	&	22	&	3	&	0.4286 	&	0.7099 	&	-1.8902 	&	1.8902 	\\
114	&	Zambia	&	45	&	25	&	27	&	3	&	0.5011 	&	0.8197 	&	-3.2547 	&	3.2547 	\\
%
\hline
\end{tabular}
\end{ruledtabular}
\caption{COUNTRY.1-2. This table shows available {\bf experimental data} of blood
type ratios for another 52 different countries around the world, where totally 114 different countries' data are available \cite{data},
continued from TABLE \ref{Table3}. 
Be aware that each country may consist of many
different ethnic groups, so these fittings are some approximate examples. 
For a country with multi-ethnicities such as USA, the theoretical prediction does not work 
as well as a country with a pure ethnic group.} \label{Table4}
\end{table*}

\begin{table*}[tbp]
\begin{ruledtabular}
\begin{tabular}{cccccccccc}
 Index&Country &O$_{t}$(\%)&AA$_{t}$(\%)&Ai$_{t}$(\%)
&BB$_{t}$(\%)&Bi$_{t}$(\%)&AB$_{t}$(\%)&Error\%($\frac{\text{O}_{t}-\text{O}_{\exp }}{\text{O}_{\exp }}$)&Error\%($\frac{\text{AB}_{t}-\text{AB}_{\exp }}{\text{AB}_{\exp }}$) \\ \hline
1	&	Canada	&	40.74 	&	8.28 	&	36.72 	&	0.55 	&	9.45 	&	4.26 	&	-0.63 	&	6.50 	\\
2	&	USA	&	45.39 	&	6.54 	&	34.46 	&	0.50 	&	9.50 	&	3.61 	&	0.87 	&	-9.83 	\\
3	&	Mexico	&	84.75 	&	0.34 	&	10.66 	&	0.05 	&	3.95 	&	0.25 	&	0.89 	&	-75.12 	\\
4	&	Guatemala	&	94.97 	&	0.02 	&	2.98 	&	0.01 	&	1.99 	&	0.03 	&	-0.03 	&		\\
5	&	El Salvador	&	60.95 	&	2.30 	&	23.70 	&	0.46 	&	10.54 	&	2.05 	&	1.58 	&	-31.68 	\\
6	&	Nicaragua	&	91.96 	&	0.13 	&	6.87 	&	0.00 	&	1.00 	&	0.04 	&	-0.04 	&		\\
7	&	Costa Rica	&	52.81 	&	3.56 	&	27.44 	&	0.71 	&	12.29 	&	3.19 	&	-0.36 	&	6.38 	\\
8	&	Cuba	&	48.36 	&	4.98 	&	31.02 	&	0.66 	&	11.34 	&	3.64 	&	-1.30 	&	21.19 	\\
9	&	Jamaica	&	55.80 	&	1.53 	&	18.47 	&	1.67 	&	19.33 	&	3.20 	&	-0.35 	&	6.63 	\\
10	&	Dominican Rep	&	52.81 	&	3.56 	&	27.44 	&	0.71 	&	12.29 	&	3.19 	&	-0.36 	&	6.38 	\\
11	&	Haiti	&	51.07 	&	2.85 	&	24.15 	&	1.36 	&	16.64 	&	3.93 	&	0.13 	&	-1.63 	\\
12	&	Puerto Rico	&	53.96 	&	3.30 	&	26.70 	&	0.70 	&	12.30 	&	3.04 	&	14.80 	&	-69.57 	\\
13	&	Colombia	&	62.28 	&	2.42 	&	24.58 	&	0.30 	&	8.70 	&	1.72 	&	0.46 	&	-14.22 	\\
14	&	Ecuador	&	63.79 	&	2.71 	&	26.29 	&	0.13 	&	5.87 	&	1.21 	&	1.26 	&	-39.57 	\\
15	&	Peru	&	70.90 	&	1.13 	&	17.87 	&	0.27 	&	8.73 	&	1.10 	&	-0.14 	&	10.06 	\\
16	&	Chile	&	56.66 	&	3.37 	&	27.63 	&	0.41 	&	9.59 	&	2.34 	&	1.18 	&	-22.02 	\\
17	&	Bolivea	&	92.95 	&	0.07 	&	4.93 	&	0.01 	&	1.99 	&	0.05 	&	-0.06 	&		\\
18	&	Venezuela	&	44.38 	&	5.57 	&	31.43 	&	0.96 	&	13.04 	&	4.62 	&	0.87 	&	-7.63 	\\
19	&	Brazil	&	47.68 	&	5.78 	&	33.22 	&	0.48 	&	9.52 	&	3.32 	&	-0.66 	&	10.58 	\\
20	&	Argentina	&	47.68 	&	5.78 	&	33.22 	&	0.48 	&	9.52 	&	3.32 	&	-0.66 	&	10.58 	\\
21	&	Paraguay	&	54.63 	&	4.09 	&	29.91 	&	0.34 	&	8.66 	&	2.37 	&	-2.45 	&	136.96 	\\
22	&	Uruguay	&	44.24 	&	6.94 	&	35.06 	&	0.51 	&	9.49 	&	3.76 	&	-1.69 	&	25.34 	\\
23	&	Australia	&	46.54 	&	6.15 	&	33.85 	&	0.49 	&	9.51 	&	3.46 	&	-0.98 	&	15.31 	\\
24	&	New Zealand	&	47.95 	&	6.02 	&	33.98 	&	0.39 	&	8.61 	&	3.05 	&	-0.11 	&	1.73 	\\
25	&	Iceland	&	55.55 	&	3.62 	&	28.38 	&	0.41 	&	9.59 	&	2.45 	&	-0.80 	&	22.42 	\\
26	&	Ireland	&	54.44 	&	3.89 	&	29.11 	&	0.42 	&	9.58 	&	2.56 	&	0.81 	&	-14.64 	\\
27	&	United Kingdom	&	46.82 	&	6.39 	&	34.61 	&	0.40 	&	8.60 	&	3.18 	&	-0.38 	&	6.00 	\\
28	&	Scotland	&	50.87 	&	4.56 	&	30.44 	&	0.54 	&	10.46 	&	3.13 	&	-0.26 	&	4.36 	\\
29	&	England	&	46.00 	&	7.03 	&	35.97 	&	0.32 	&	7.68 	&	3.00 	&	-0.01 	&	0.08 	\\
30	&	Norway	&	39.18 	&	9.80 	&	39.20 	&	0.37 	&	7.63 	&	3.82 	&	0.47 	&	-4.61 	\\
31	&	Sweden	&	38.37 	&	9.27 	&	37.73 	&	0.58 	&	9.42 	&	4.63 	&	0.97 	&	-7.36 	\\
32	&	Finland	&	31.59 	&	8.75 	&	33.25 	&	2.02 	&	15.98 	&	8.41 	&	-4.27 	&	20.14 	\\
33	&	Denmark	&	41.91 	&	7.81 	&	36.19 	&	0.53 	&	9.47 	&	4.09 	&	-0.21 	&	2.16 	\\
34	&	Netherland	&	44.55 	&	7.19 	&	35.81 	&	0.41 	&	8.59 	&	3.45 	&	-1.00 	&	15.02 	\\
35	&	Belgium	&	46.00 	&	7.03 	&	35.97 	&	0.32 	&	7.68 	&	3.00 	&	-0.01 	&	0.08 	\\
36	&	France	&	42.26 	&	8.07 	&	36.93 	&	0.43 	&	8.57 	&	3.74 	&	-1.73 	&	24.77 	\\
37	&	Spain	&	42.26 	&	8.07 	&	36.93 	&	0.43 	&	8.57 	&	3.74 	&	0.61 	&	-6.42 	\\
38	&	Portugal	&	40.33 	&	9.29 	&	38.71 	&	0.36 	&	7.64 	&	3.67 	&	-1.62 	&	22.18 	\\
39	&	Germany	&	38.52 	&	7.97 	&	35.03 	&	0.94 	&	12.06 	&	5.48 	&	1.36 	&	-8.62 	\\
40	&	Austria	&	37.27 	&	8.47 	&	35.53 	&	0.97 	&	12.03 	&	5.73 	&	-4.45 	&	43.37 	\\
41	&	Swiss	&	41.48 	&	8.80 	&	38.20 	&	0.35 	&	7.65 	&	3.52 	&	-1.24 	&	17.39 	\\
42	&	Italy	&	45.11 	&	6.30 	&	33.70 	&	0.60 	&	10.40 	&	3.89 	&	0.26 	&	-2.87 	\\
43	&	North Italy	&	41.59 	&	7.55 	&	35.45 	&	0.64 	&	10.36 	&	4.41 	&	-0.98 	&	10.33 	\\
44	&	South Italy	&	45.39 	&	4.97 	&	30.03 	&	1.07 	&	13.93 	&	4.61 	&	-1.32 	&	15.22 	\\
45	&	Czech	&	33.02 	&	8.16 	&	32.84 	&	1.95 	&	16.05 	&	7.98 	&	0.06 	&	-0.25 	\\
46	&	Hungary	&	28.26 	&	9.77 	&	33.23 	&	2.43 	&	16.57 	&	9.74 	&	-5.81 	&	21.80 	\\
47	&	*Gypsies	&	28.31 	&	3.92 	&	21.08 	&	7.28 	&	28.72 	&	10.69 	&	-5.63 	&	18.78 	\\
48	&	Poland	&	35.56 	&	6.83 	&	31.17 	&	2.03 	&	16.97 	&	7.44 	&	1.60 	&	-7.01 	\\
49	&	Yogoslavia	&	38.20 	&	7.70 	&	34.30 	&	1.09 	&	12.91 	&	5.80 	&	3.26 	&	-17.21 	\\
50	&	Romania	&	29.01 	&	10.35 	&	34.65 	&	1.95 	&	15.05 	&	8.99 	&	-14.67 	&	28.38 	\\
51	&	Bulgalia	&	32.26 	&	9.32 	&	34.68 	&	1.61 	&	14.39 	&	7.74 	&	0.82 	&	-3.30 	\\
52	&	Albania	&	41.28 	&	5.60 	&	30.40 	&	1.46 	&	15.54 	&	5.72 	&	-1.72 	&	14.44 	\\
53	&	Greece	&	43.41 	&	6.20 	&	32.80 	&	0.85 	&	12.15 	&	4.59 	&	0.95 	&	-8.19 	\\
54	&	Cyprus	&	34.72 	&	9.56 	&	36.44 	&	1.03 	&	11.97 	&	6.28 	&	-0.80 	&	4.68 	\\
55	&	Turkey	&	32.63 	&	9.60 	&	35.40 	&	1.41 	&	13.59 	&	7.37 	&	-1.12 	&	5.28 	\\
56	&	Iran(Westnorth)	&	34.73 	&	5.46 	&	27.54 	&	3.13 	&	20.87 	&	8.27 	&	-3.54 	&	18.20 	\\
57	&	Iran(Eastsouth)	&	43.10 	&	3.05 	&	22.95 	&	2.85 	&	22.15 	&	5.90 	&	2.63 	&	-15.76 	\\
58	&	Iran(Middleeast)	&	31.57 	&	3.62 	&	21.38 	&	6.14 	&	27.86 	&	9.43 	&	1.83 	&	-5.67 	\\
59	&	Iraq	&	34.56 	&	4.92 	&	26.08 	&	3.62 	&	22.38 	&	8.44 	&	-1.27 	&	5.55 	\\
60	&	Kuwait	&	47.07 	&	2.46 	&	21.54 	&	2.46 	&	21.54 	&	4.93 	&	0.16 	&	-1.47 	\\
61	&	Lebanon	&	35.12 	&	9.83 	&	37.17 	&	0.88 	&	11.12 	&	5.88 	&	-2.46 	&	17.69 	\\
62	&	Israel	&	34.70 	&	7.90 	&	33.10 	&	1.69 	&	15.31 	&	7.30 	&	-3.62 	&	21.74 	\\
\hline
\end{tabular}
\end{ruledtabular}
\caption{COUNTRY.2-1 This table shows our {\bf theoretical prediction} of blood
type ratios for 62 different countries around the world. 
The theoretical prediction here in TABLE \ref{Table5} is compared to the experimental data in TABLE \ref{Table3}.
To be continued to
TABLE \ref{Table6} 
} \label{Table5}
\end{table*}

\begin{table*}[tbp]
\begin{ruledtabular}
\begin{tabular}{cccccccccc}
 Index&Country &O$_{t}$(\%)&AA$_{t}$(\%)&Ai$_{t}$(\%)
&BB$_{t}$(\%)&Bi$_{t}$(\%)&AB$_{t}$(\%)&Error\%($\frac{\text{O}_{t}-\text{O}_{\exp }}{\text{O}_{\exp }}$)&Error\%($\frac{\text{AB}_{t}-\text{AB}_{\exp }}{\text{AB}_{\exp }}$) \\ \hline
63	&	Pakistan	&	31.57 	&	3.62 	&	21.38 	&	6.14 	&	27.86 	&	9.43 	&	1.83 	&	-5.67 	\\
64	&	Afgjanistan	&	28.31 	&	3.92 	&	21.08 	&	7.28 	&	28.72 	&	10.69 	&	-2.38 	&	6.90 	\\
65	&	India(Westnorth)	&	27.59 	&	2.95 	&	18.05 	&	9.18 	&	31.82 	&	10.41 	&	-4.86 	&	15.65 	\\
66	&	India(Eastnorth)	&	28.31 	&	3.92 	&	21.08 	&	7.28 	&	28.72 	&	10.69 	&	-11.53 	&	52.72 	\\
67	&	India(Middlesouth)	&	36.42 	&	2.59 	&	19.41 	&	5.55 	&	28.45 	&	7.58 	&	-1.57 	&	8.31 	\\
68	&	Srilanka	&	47.11 	&	2.10 	&	19.90 	&	2.85 	&	23.15 	&	4.89 	&	0.23 	&	-2.20 	\\
69	&	Myanmar	&	34.63 	&	3.38 	&	21.62 	&	5.19 	&	26.81 	&	8.37 	&	-1.06 	&	4.64 	\\
70	&	Thailand	&	37.82 	&	2.51 	&	19.49 	&	5.13 	&	27.87 	&	7.18 	&	-3.03 	&	19.67 	\\
71	&	Malaysia	&	38.97 	&	3.08 	&	21.92 	&	4.01 	&	24.99 	&	7.03 	&	2.56 	&	-12.15 	\\
72	&	Philippines	&	44.45 	&	2.98 	&	23.02 	&	2.58 	&	21.42 	&	5.55 	&	-1.21 	&	10.92 	\\
73	&	Vietnam	&	40.56 	&	2.37 	&	19.63 	&	4.37 	&	26.63 	&	6.44 	&	-3.44 	&	28.86 	\\
74	&	Cambodia	&	33.37 	&	3.00 	&	20.00 	&	6.21 	&	28.79 	&	8.63 	&	-14.43 	&	187.61 	\\
75	&	South China	&	44.48 	&	3.19 	&	23.81 	&	2.39 	&	20.61 	&	5.52 	&	1.10 	&	-8.05 	\\
76	&	North China	&	31.30 	&	4.99 	&	25.01 	&	4.71 	&	24.29 	&	9.70 	&	-5.16 	&	21.28 	\\
77	&	Middle China	&	33.03 	&	5.37 	&	26.63 	&	3.75 	&	22.25 	&	8.97 	&	-5.63 	&	28.16 	\\
78	&	Hong Kong	&	44.45 	&	2.58 	&	21.42 	&	2.98 	&	23.02 	&	5.55 	&	1.03 	&	-7.57 	\\
79	&	Macao	&	40.37 	&	3.00 	&	22.00 	&	3.67 	&	24.33 	&	6.63 	&	0.92 	&	-5.27 	\\
80	&	Tibet	&	28.49 	&	7.59 	&	29.41 	&	3.64 	&	20.36 	&	10.51 	&	-5.03 	&	16.77 	\\
81	&	Taiwan	&	44.48 	&	3.19 	&	23.81 	&	2.39 	&	20.61 	&	5.52 	&	1.10 	&	-8.05 	\\
82	&	Nepal	&	28.49 	&	7.59 	&	29.41 	&	3.64 	&	20.36 	&	10.51 	&	-5.03 	&	16.77 	\\
83	&	Mongolia	&	36.42 	&	2.59 	&	19.41 	&	5.55 	&	28.45 	&	7.58 	&	-1.57 	&	8.31 	\\
84	&	South Korea	&	26.00 	&	6.33 	&	25.67 	&	5.69 	&	24.31 	&	12.00 	&	-3.71 	&	9.11 	\\
85	&	Japan	&	30.51 	&	7.58 	&	30.42 	&	2.97 	&	19.03 	&	9.49 	&	-1.57 	&	5.41 	\\
86	&	Sumatra	&	40.47 	&	2.58 	&	20.42 	&	4.13 	&	25.87 	&	6.53 	&	-1.28 	&	8.77 	\\
87	&	Java	&	37.51 	&	3.40 	&	22.60 	&	4.12 	&	24.88 	&	7.49 	&	1.37 	&	-6.33 	\\
88	&	Middle Asia(Russia)	&	31.32 	&	4.43 	&	23.57 	&	5.28 	&	25.72 	&	9.68 	&	-2.11 	&	7.50 	\\
89	&	Sibereria(Russia)	&	38.97 	&	3.08 	&	21.92 	&	4.01 	&	24.99 	&	7.03 	&	-0.07 	&	0.40 	\\
90	&	Russia(Europe)	&	33.37 	&	6.21 	&	28.79 	&	3.00 	&	20.00 	&	8.63 	&	-1.85 	&	7.85 	\\
91	&	Caucasus	&	42.10 	&	4.49 	&	27.51 	&	1.94 	&	18.06 	&	5.90 	&	-4.32 	&	47.52 	\\
92	&	Georgia	&	49.73 	&	4.87 	&	31.13 	&	0.55 	&	10.45 	&	3.27 	&	-0.54 	&	-18.23 	\\
93	&	Armenia	&	29.40 	&	12.17 	&	37.83 	&	1.19 	&	11.81 	&	7.60 	&	1.38 	&	-4.99 	\\
94	&	Egypt	&	33.23 	&	5.93 	&	28.07 	&	3.24 	&	20.76 	&	8.77 	&	-2.26 	&	9.59 	\\
95	&	Libya	&	38.28 	&	5.91 	&	30.09 	&	1.91 	&	17.09 	&	6.72 	&	-4.29 	&	34.33 	\\
96	&	Tunisia	&	46.16 	&	3.96 	&	27.04 	&	1.48 	&	16.52 	&	4.84 	&	0.35 	&	-3.22 	\\
97	&	Algelia	&	43.41 	&	6.20 	&	32.80 	&	0.85 	&	12.15 	&	4.59 	&	-1.34 	&	14.77 	\\
98	&	Morocco	&	34.85 	&	5.73 	&	28.27 	&	2.90 	&	20.10 	&	8.15 	&	-5.82 	&	35.88 	\\
99	&	Ethiopia	&	45.88 	&	3.32 	&	24.68 	&	1.97 	&	19.03 	&	5.12 	&	-2.38 	&	27.94 	\\
100	&	Somalia	&	58.51 	&	2.38 	&	23.62 	&	0.65 	&	12.35 	&	2.49 	&	0.88 	&	-16.93 	\\
101	&	Chad	&	40.35 	&	3.44 	&	23.56 	&	3.22 	&	22.78 	&	6.65 	&	-10.34 	&	232.60 	\\
102	&	Senegal	&	49.67 	&	2.54 	&	22.46 	&	1.85 	&	19.15 	&	4.33 	&	-0.66 	&	8.27 	\\
103	&	Libelia	&	45.77 	&	2.71 	&	22.29 	&	2.52 	&	21.48 	&	5.23 	&	-0.50 	&	4.59 	\\
104	&	Ghana	&	48.38 	&	2.06 	&	19.94 	&	2.59 	&	22.41 	&	4.62 	&	-1.26 	&	15.46 	\\
105	&	Cameroon	&	53.37 	&	1.89 	&	20.11 	&	1.74 	&	19.26 	&	3.63 	&	-1.16 	&	20.94 	\\
106	&	Congo	&	48.47 	&	2.98 	&	24.02 	&	1.72 	&	18.28 	&	4.53 	&	-1.08 	&	13.23 	\\
107	&	Guinea	&	64.00 	&	1.00 	&	16.00 	&	1.00 	&	16.00 	&	2.00 	&	0.00 	&	0.00 	\\
108	&	Angola	&	48.38 	&	2.59 	&	22.41 	&	2.06 	&	19.94 	&	4.62 	&	-1.26 	&	15.46 	\\
109	&	Uganda	&	49.63 	&	2.18 	&	20.82 	&	2.18 	&	20.82 	&	4.37 	&	1.29 	&	-12.69 	\\
110	&	Mozambique	&	57.07 	&	1.94 	&	21.06 	&	1.11 	&	15.89 	&	2.93 	&	0.12 	&	-2.26 	\\
111	&	Madagascar	&	43.16 	&	2.45 	&	20.55 	&	3.48 	&	24.52 	&	5.84 	&	0.38 	&	-2.71 	\\
112	&	Sorth Africa	&	43.98 	&	5.09 	&	29.91 	&	1.24 	&	14.76 	&	5.02 	&	-2.27 	&	25.50 	\\
113	&	Kenya	&	47.11 	&	2.85 	&	23.15 	&	2.10 	&	19.90 	&	4.89 	&	-3.86 	&	63.01 	\\
114	&	Zambia	&	41.75 	&	2.92 	&	22.08 	&	3.35 	&	23.65 	&	6.25 	&	-7.23 	&	108.49 	\\
\hline
\end{tabular}
\end{ruledtabular}
\caption{COUNTRY.2-2. This table shows our {\bf theoretical prediction} of blood
type ratios for another 52 different countries around the world,
continued from TABLE \ref{Table5}. 
The theoretical prediction here in TABLE \ref{Table6} is compared to the experimental data in TABLE \ref{Table4}
} \label{Table6}
\end{table*}








\section{\label{sec:level4}General Gene-Mating Evolution
Model and Theory}
\subsection{Governing equations, stable equilibrium solutions and exact analytic solutions}

We can generalize our model and theory of Sec.\ref{sec:level3} to $n+1$ alleles. 
Alternatively, one may say $n$ dominant alleles and 1 recessive allele in a single locus.
For dominant gene with a label $\alpha $, we denote it as $%
D_{\alpha }$; 
and for the only recessive gene we denote as $r$.  
We denote biological traits 
of $D_{\alpha }D_{\alpha }$ as $G_{\alpha
\alpha }$, $D_{\alpha }D_{\beta }$ as $G_{\alpha \beta }$ ($\equiv G_{\beta
\alpha }$) for $\alpha \neq \beta $, $D_{\alpha }r$ as $G_{\alpha 0}$ ($%
\equiv G_{0\alpha }$), $rr$ as $G_{00}$. 
Indeed, whether the genotypes we studied have dominant or recessive alleles do not matter, we can simply label
them as $G_{\alpha \beta }$ generically.
We derive the governing
equations for \emph{population parameters} as: 
\begin{equation} \label{eq:popGenen-1}
\left\{ 
\begin{array}{l}
\frac{d}{dt}{\small G}_{\alpha \alpha }{\small =k}_{b}\frac{1}{P}{\small (G}%
_{\alpha \alpha }\overset{n}{\underset{j=0}{\sum }}{\small G}_{\alpha j}%
{\small +}\frac{1}{4}\underset{i\neq \alpha }{\underset{i=0}{\overset{n}{%
\sum }}}\underset{j\neq \alpha }{\overset{n}{\underset{j=0}{\sum }}}{\small G%
}_{\alpha i}{\small G}_{\alpha j}{\small )-k}_{d}{\small G}_{\alpha \alpha },
\\ 
\frac{d}{dt}{\small G}_{\alpha \beta }{\small =k}_{b}\frac{1}{P}{\small (2G}%
_{\alpha \alpha }{\small G}_{\beta \beta }{\small +}\underset{i\neq \alpha }{%
\overset{n}{\underset{i=0}{\sum }}}{\small G}_{\alpha i}{\small G}_{\beta
\beta } \\ 
\text{ \ \ \ \ \ \ \ \ \ \ \ \ \ \ \ \ \ \ }{\small +}\underset{j\neq \beta }%
{\overset{n}{\underset{j=0}{\sum }}}{\small G}_{\alpha \alpha }{\small G}%
_{\beta j}{\small +}\frac{1}{2}\underset{i\neq \alpha }{\underset{i=0}{%
\overset{n}{\sum }}}\underset{j\neq \beta }{\overset{n}{\underset{j=0}{\sum }%
}}{\small G}_{\alpha i}{\small G}_{\beta j}{\small )-k}_{d}{\small G}%
_{\alpha \beta }.%
\end{array}%
\right.
\end{equation}
%
By redefining a \emph{population parameters} $G$ to the \emph{percentages parameter} $\frac{G}{P} \to G$,
we derive the governing equations for \emph{percentages parameters} as: 
\begin{equation} \label{eq:Genen-1}
\left\{ 
\begin{array}{l}
\frac{d}{dt}{\small G}_{\alpha \alpha }{\small =k}_{b}{\small (G}_{\alpha
\alpha }\overset{n}{\underset{j=0}{\sum }}{\small G}_{\alpha j}{\small +}%
\frac{1}{4}\underset{i\neq \alpha }{\underset{i=0}{\overset{n}{\sum }}}%
\underset{j\neq \alpha }{\overset{n}{\underset{j=0}{\sum }}}{\small G}%
_{\alpha i}{\small G}_{\alpha j}{\small -G}_{\alpha \alpha }{\small )}, \\ 
\frac{d}{dt}{\small G}_{\alpha \beta }{\small =k}_{b}{\small (2G}_{\alpha
\alpha }{\small G}_{\beta \beta }{\small +}\underset{i\neq \alpha }{\overset{%
n}{\underset{i=0}{\sum }}}{\small G}_{\alpha i}{\small G}_{\beta \beta } \\ 
\text{ \ \ \ \ \ \ \ \ \ \ \ \ \ \ }{\small +}\underset{j\neq \beta }{%
\overset{n}{\underset{j=0}{\sum }}}{\small G}_{\alpha \alpha }{\small G}%
_{\beta j}{\small +}\frac{1}{2}\underset{i\neq \alpha }{\underset{i=0}{%
\overset{n}{\sum }}}\underset{j\neq \beta }{\overset{n}{\underset{j=0}{\sum }%
}}{\small G}_{\alpha i}{\small G}_{\beta j}{\small -G}_{\alpha \beta }%
{\small )}. %
\end{array}%
\right.
\end{equation}
%
The equilibrium solutions consist of a continuous $n$-dimensional manifold as a continuous set of 
fixed-points:\footnote{It may be interesting to compare this fixed-point manifold (also known as the Hardy-Weinberg manifold) 
with the Wright manifold (the attractor manifold in the case of multi loci) \cite{Wright}.}
\begin{equation}  \label{eq:Equiln}
\left\{ 
\begin{array}{l}
{\small G}_{\alpha \alpha ,eq}\left( \theta _{k}\right) {\small =[}\underset{%
i=1}{\overset{\alpha }{\Pi }}\frac{\sin ^{2}(\theta _{i})}{(\cos (\theta
_{i})+\sin (\theta _{i}))^{2}}{\small ]}\frac{\cos ^{2}(\theta _{\alpha +1})%
}{(\cos (\theta _{\alpha +1})+\sin (\theta _{\alpha +1}))^{2}}, \\ 
{\small G}_{\alpha \beta ,eq}\left( \theta _{k}\right) {\small =[}\underset{%
i=1}{\overset{\alpha }{\Pi }}\frac{\sin ^{2}(\theta _{i})}{(\cos (\theta
_{i})+\sin (\theta _{i}))^{2}}{\small ]}\frac{2\sin (\theta _{\alpha
+1})\cos (\theta _{\alpha +1})}{(\cos (\theta _{\alpha +1})+\sin (\theta
_{\alpha +1}))^{2}}{\small  } \\ 
\text{ \ \ \ \ \   }
\cdot {\small [}\underset{j=1}{\overset{%
\beta -\alpha -1}{\Pi }}\frac{\sin (\theta _{\alpha +1+j})}{\cos (\theta
_{\alpha +1+j})+\sin (\theta _{\alpha +1+j})}{\small ]}\frac{\cos (\theta
_{\beta +1})}{\cos (\theta _{\beta +1})+\sin (\theta _{\beta +1})}.%
\end{array}%
\right.
\end{equation}
We are confident to anticipate that the linear stability analysis of the system would
give $n$ eigenvalues $0$, and $\binom{n+1}{2}$ eigenvalues $-1$ \cite{RG}. 
Note that $\widehat{G}_{\alpha \alpha }+\widehat{G}_{\beta \beta }-2\widehat{G}_{\alpha \beta
}\equiv \widehat{s}_{\alpha \beta }$ (while symmetrically we define $s_{ij}\equiv s_{ji}$) would be the
eigenvectors corresponding to $-1$ eigenvalues. We can parameterize the
whole space of $G_{ij}$ ($n+1+\binom{n+1}{2}$-dimensions with one constraint) by
a new set of $n+\binom{n+1}{2}$ coordinates $\theta _{k},$ $s_{ij}$:
\begin{equation} \label{eq:Gn-1para} 
\left\{ 
\begin{array}{l}
G_{\alpha \alpha }=G_{\alpha \alpha ,eq}\left( \theta _{k}\right) +\overset{n%
}{\underset{i=0,i\neq \alpha }{\sum }}s_{\alpha i}, \\ 
G_{\alpha \beta }=G_{\alpha \beta ,eq}\left( \theta _{k}\right) -2s_{\alpha
\beta }.%
\end{array}%
\right.
\end{equation}
%
\begin{widetext}
The inverse transformation is: 
\begin{equation} \label{eq:Gn-1inverse}
\left\{ 
\begin{array}{l}
{\small \theta }_{k}{\small =}\tan ^{-1}{\small \{[}\underset{i=k}{\overset{n%
}{\sum }}{\small (2G}_{ii}{\small +}\underset{j\neq i}{\underset{j=0}{%
\overset{n}{\sum }}}{\small G}_{ij}{\small )]/(2G}_{k-1,k-1}{\small +}%
\underset{j\neq k-1}{\underset{j=0}{\overset{n}{\sum }}}{\small 2G}_{k-1,%
\text{ }j}{\small )\}}\text{, } \\ 
\text{ \ \ \ \ \ \ \ \ \ \ \ \ \ \ \ \ \ \ \ \ \ \ \ \ \ \ \ \ \ \ \ \ \ \ \
\ \ \ \ \ \ \ \ }{\small 1\leq k\leq n}\text{, totally n equations}. \\ 
\\ 
{\small s}_{ij}\text{ can be solved from the following equations:} \\ 
\text{ }\left\{ 
\begin{array}{l}
\tan ^{2}{\small (\theta }_{k}{\small )=}\frac{\underset{k\leq i\leq j\leq n}%
{\overset{n}{\sum }}G_{ij}-\underset{i=k}{\overset{n}{\sum }}\underset{j=0}{%
\overset{k-1}{\sum }}s_{ij}}{(G_{k-1,k-1}-\underset{i\neq k-1}{\underset{i=0}%
{\overset{n}{\sum }}}s_{k-1,\text{ }i})}\text{,} \\ 
\text{ \ \ \ \ \ \ \ \ \ \ \ \ \ \ \ \ \ \ \ \ \ \ }{\small 1\leq k\leq n}%
\text{, totally n equations.} \\ 
\tan {\small (\theta }_{k}{\small )=[}\underset{i=k}{\overset{n}{\sum }}%
{\small (G}_{mi}{\small +2s}_{mi}{\small )]/(G}_{m,k-1}{\small +2s}_{m,k-1}%
{\small ),} \\ 
\text{\ \ \ \ }{\small 1\leq k\leq n}\text{, }{\small 0\leq m\leq k-2}\text{%
, totally }\binom{n}{2}\text{ equations}.%
\end{array}%
\right.%
\end{array}%
\right.
\end{equation}
\end{widetext}

Substitute Eq.(\ref{eq:Gn-1para}) into Eq.(\ref{eq:Genen-1}), 
we have the de-coupled governing
linear equations in the new coordinates:
\begin{equation} \label{eq:Gn-1linear}
\left\{ 
\begin{array}{l}
\frac{ds_{ij}}{dt}=-k_{b}s_{ij},\text{ }0\leq i < j\leq n. \\ 
\frac{d\theta _{k}}{dt}=0,\text{ \ \ \ \ \ \ \ \ \ }1\leq k\leq n.%
\end{array}%
\right.
\end{equation}
%
Given initial values $\widetilde{G}_{ij}$ in the percentage parameter space,
the exact analytic solution under time evolution is
%
\begin{eqnarray}  \label{eq:exactGn-1}
\overset{n}{\underset{0\leq i\leq j\leq n}{\sum }}{\small G}_{ij}{\small (t)}%
\widehat{G}_{ij} &{\small =}& 
\underset{0\leq i\leq j\leq n}{\overset{n}{\sum }}{\small G}_{ij,eq}{\small %
(\theta }_{i} {\small =}\widetilde{\theta }_{i}{\small )}\widehat{G}_{ij} \\%
&{\small +}&\overset{n}{\underset{0\leq \alpha <\beta \leq n}{\sum }}%
\widetilde{s}_{\alpha \beta }{\small e}^{-k_{b}t}{\small (}\widehat{G}%
_{\alpha \alpha }{\small +}\widehat{G}_{\beta \beta }{\small -2}\widehat{G}%
_{\alpha \beta }{\small )}.    \notag
\end{eqnarray}
where $\widetilde{\theta }_{i}$ and $\widetilde{s}_{\alpha \beta }$ can be
solved in the form of $\widetilde{G}_{ij}$ from Eq.(\ref{eq:Gn-1inverse}). 
{\bf The key to obtain our exact analytic solution is to transform
the nonlinear coupled Eq.(\ref{eq:Genen-1}) to the decoupled linear Eq.(\ref{eq:Gn-1linear}).} \\

\noindent
{\bf The dimensionality}:
We explain again the physical meaning on the dimensionality of the fibers and the stable base manifold, shown in Fig.\ref{fig:01}.
For Eq.(\ref{eq:Genen-1}), there is a permutation symmetric group S${}_{n+1}$ symmetry by permuting $G_{\alpha\beta}$.
The S${}_{n+1}$ symmetry also results in time-independent ${n+1}$ \emph{conserved quantities}, 
spanned by  
$(2G_{k-1,k-1}{+}\underset{j\neq k-1}{\underset{j=0}{\overset{n}{\sum }}}{\small 2G}_{k-1,j} )$ with $k=1,\dots,n+1$.
Since there is a 1-dimensional constraint  $\sum G_{jk}=1$, overall there is $n$ independent degrees of freedom.
We can say the S${}_{n+1}$ symmetry results in  
the dimensionality of fixed-point solutions is $n$, here parametrized by $\theta_1,\dots, \theta_n$.
The stable equilibrium base manifold is $n$-dimensional, the fibers are ${n+1 \choose 2}$-dimensional.\\


\subsection{Geodesic distance and genetic distance}
\label{subsec:distance}

\noindent
{\bf The geodesic distance as the genetic distance}:
Again, as stated in Sec.\ref{sec:G2-1geo}. we can solve the geodesic distance of of two populations 
on the manifold Eq.(\ref{eq:Equiln}) in the genotype frequency space to define the genetic distance of two populations. Note that 
the intrinsic metric $g_{\theta_\mu,\theta_\nu}$ can be derived from
\bea
ds^2 &=&\sum_{0 \leq i \leq n; i \leq j \leq n} (dG_{ij}(\theta_1, \dots, \theta_n)){}^2 \nonumber \\
&\equiv& \sum_{\mu,\nu=1,\dots,n}g_{\theta_\mu,\theta_\nu} \;d\theta_\mu \; d\theta_\nu. \;\;\;\;\;\;\;
\eea
The geodesic is solved from the geodesic equation on the manifold Eq.(\ref{eq:Equiln}):
\bea
\frac{d^2 \theta_\mu }{ds^2}+\Gamma^{\mu}_{\nu \lambda} \frac{d^2 \theta_\nu }{ds} \frac{d^2 \theta_\lambda }{ds}=0
\eea
with $\Gamma^{\mu}_{\nu \lambda} \equiv \frac{1}{2} g^{\mu \rho} (\partial_\nu g_{\lambda \rho}+\partial_\lambda g_{\nu\rho} - \partial_\rho  g_{\nu \lambda})$ is the Christoffel symbol.

\subsection{1-1 mapping and the well-defined manifold} 

Recall that in Sec.\ref{sec:2-1G}, 
we are aware that in order to have an one-to-one (denoted as 1-1) mapping,  
we have to specify the valid domain of $(\theta_1,\theta_2)$. 
From Eq.(\ref{eq:Equil2}), 
$(\theta _{1}=0,\forall $ $%
\theta _{2})$ would map to $o=1$ case, this is many-to-one mapping.
Topologically we could shrink $(\theta _{1}=0,\forall $ $\theta _{2})$ as a
point to make it a well-defined 1-1 mapping 2-dimensional manifold. Similarly, we
should perform an analogous operation on the general case of Eq.(\ref{eq:Equiln}):  
\begin{equation} \label{eq:G-one-to-one}
\left\{ 
\begin{array}{l}
{\small \theta }_{1}{\small =0,\forall (\theta }_{2}{\small ,\theta }_{3}%
{\small ,\ldots ,\theta }_{n}{\small )}\text{ shrink into a point}. \\ 
{\small \theta }_{1}{\small \neq 0,\theta }_{2}{\small =0,\forall (\theta }%
_{3}{\small ,\ldots ,\theta }_{n}{\small )}\text{ shrink into a point}. \\ 
{\small \theta }_{1}{\small \neq 0,\theta }_{2}{\small \neq 0,\theta }_{3}%
{\small =0,\forall (\theta }_{4}{\small ,\ldots ,\theta }_{n}{\small )}\text{
shrink into a point}. \\ 
{\small \vdots } \\ 
{\small \theta }_{1}{\small \neq 0,\theta }_{2}{\small \neq 0,\ldots ,\theta 
}_{n-2}{\small \neq 0,\theta }_{n-1}{\small =0,}\\
{\forall \theta }_{n}\text{
shrink into a point}.%
\end{array}%
\right. {\tiny \ }
\end{equation}

For the same reason, the inverse functions Eq.(\ref{eq:G2-1inverse}) and Eq.(\ref{eq:Gn-1inverse}) are not
one-to-one defined at those few points. Nonetheless, we could follow the rule of Eq.(\ref{eq:G-one-to-one})
to make them one-to-one well-defined. 


\subsection{Equilibrium solutions as a global attractor, monotonic behavior,
non-extinction and the Hardy-Weinberg Law}


Now we can prove several fundamental common properties of macroscopic gene-mating dynamical evolutionary systems.
Our proofs are straightforward 
since we have the time-evolutionary analytic solution in Eq.(\ref{eq:exactGn-1}).\\

\noindent 
(1) {\bf The global stability} of the system: as the time approaches
infinity (approximately), the LHS parameters of Eq.(\ref{eq:exactGn-1}) will evolve to an 
equilibrium solution
through a 1-dimensional line direction{\small \ }$\underset{0\leq \alpha
<\beta \leq n}{\overset{n}{\sum }}s_{\alpha \beta }${\tiny \ }$\widehat{s}%
_{\alpha \beta }$. \\

\noindent
(2) {\bf Monotonic}: Time-evolution of parameters in Eq.(\ref{eq:exactGn-1}) is also monotonic 
through the exponential decay along the $\hat{s}_{\alpha \beta}$ direction to the corresponding equilibrium
fixed-point on the stable manifold Eq.(\ref{eq:Equiln}).\\

\noindent
(3) {\bf Non-extinction and the Hardy-Weinberg law}:
Based on our model, here we further claim that our following non-extinction
statement shows a proof of the Hardy-Weinberg law \cite{{Hardy},{Weinberg}} for a gene-mating system.

Our non-extinction
statement states that \emph{any genotype or an allele that ever appears 
in the population will never become extinct}.
The Hardy-Weinberg law states that  in the absence of other evolutionary influences, 
the population
genetics will obtain an equilibrium. 

We prove that our model 
not only have the stable fixed points as a continuous manifold in Eq.(\ref{eq:Equiln}), 
but also verify a stronger claim from a dynamical viewpoint --- 
if there ever exists a certain genotype, no matter how tiny a portion
it is in the total population, it will never become extinct under time-evolution. 
Here is our proof.
The extinction evolution approaches zeros for certain genotypes. 
Zeros are
a final state, which must be located at some equilibrium point. This is easy to
verify by moving an equilibrium point (those points with a certain genotype
percentage equal to zero) away from the equilibrium manifold through the
time-reversal evolution direction (the line direction of Eq.(\ref{eq:Gn-1para})).  
We find that it
is impossible because there must be another genotype 
going to a value less
than $0$, but the population percentage  cannot be smaller than 0 at any moment.  
The contradiction shows time-evolution never brings genotype frequencies 
to 
approach extinction points. 
We have proven the non-extinction statement and the Hardy-Weinberg law for the genotype frequency.

We can prove 
the non-extinction for
the allele frequency from the fact that \emph{the time-evolution direction in the genotype frequency space
always conserves the allele frequency}.  Namely, in Eq.(\ref{eq:Gn-1para}), along the 
$(\widehat{G}_{\alpha \alpha }+\widehat{G}_{\beta \beta }-2\widehat{G}_{\alpha \beta
})\equiv \widehat{s}_{\alpha \beta }$ direction on the fiber, the allele frequency is conserved independently of time.
We have thus proven the non-extinction statement for both genotype frequencies and  allele frequencies.

Based on these three proven facts above, we know the global properties of \emph{genotype frequency} 
space of the evolutionary system; and we also know the
topological properties of \emph{genotype frequency} space from the fiber bundle picture, shown in Fig.\ref{fig:01}.
\bigskip

\section{\label{sec:level5}Mutation and Natural Selection}

Let us consider a more generic model beyond Eq.(\ref{eq:popGenen-1}) and Eq.(\ref{eq:Genen-1}) to include the process of mutation and natural selection.

Mutation, for example, corresponds to enlarging the parameter space through adding a new gene type (genotypes or alleles). 
%
Natural selection, for example, correponds to perturbe the birth rate, the death rate, and the inherited factor of our governing equations.
The population governing equations with various independent
birth rates ${\small k_{bij}}$ and various death rates ${\small k_{dij}}$ can be:
\begin{equation} \label{eq:naturalselect}
\left\{ 
\begin{array}{l}
\frac{d}{dt}{\small G}_{\alpha \alpha }{\small =}\frac{1}{P}{\small (}\sqrt{%
{\small k_{b\alpha \alpha }}}{\small G}_{\alpha \alpha }\overset{n}{\underset%
{j=0}{\sum }}\sqrt{{\small k_{b\alpha j}}}{\small G}_{\alpha j}+ \\ 
\text{ \ }\frac{1}{4}\underset{i\neq \alpha }{\underset{i=0}{\overset{n}{%
\sum }}}\underset{j\neq \alpha }{\overset{n}{\underset{j=0}{\sum }}}\sqrt{%
{\small k_{b\alpha i}}}{\small G}_{\alpha i}\sqrt{{\small k_{b\alpha j}}}%
{\small G}_{\alpha j}{\small )-k}_{d\alpha \alpha }{\small G}_{\alpha \alpha
}, \\ 
\frac{d}{dt}{\small G}_{\alpha \beta }{\small =}\frac{1}{P}{\small (2\sqrt{%
k_{b\alpha \alpha }}G}_{\alpha \alpha }{\small \sqrt{k_{b\beta \beta }}G}%
_{\beta \beta }{\small +} \\ 
\text{ }\underset{i\neq \alpha }{\overset{n}{\underset{i=0}{\sum }}}{\small 
\sqrt{k_{b\alpha i}}G}_{\alpha i}{\small \sqrt{k_{b\beta \beta }}G}_{\beta
\beta }+\underset{j\neq \beta }{\overset{n}{\underset{j=0}{\sum }}}{\small 
\sqrt{k_{b\alpha \alpha }}G}_{\alpha \alpha }{\small \sqrt{k_{b\beta j}}G}%
_{\beta j}{\small +} \\ 
\text{ \ \ }\frac{1}{2}\underset{i\neq \alpha }{\underset{i=0}{\overset{n}{%
\sum }}}\underset{j\neq \beta }{\overset{n}{\underset{j=0}{\sum }}}{\small 
\sqrt{k_{b\alpha i}}G}_{\alpha i}{\small \sqrt{k_{b\beta j}}G}_{\beta j}%
{\small )}\text{\ }{\small -k}_{d\alpha \beta }{\small G}_{\alpha \beta }.%
\end{array}%
\right.
\end{equation}

Here we apply the square root of birth rate or death rate, $\sqrt{k_{b\alpha \beta }}, \sqrt{k_{d\alpha \beta }}$,  to distribute the
birth and the death contributions from two genders carrying two independent sets of genes. 
On one hand, for a consistency check, this could reduce to the
original standard equations if both birth and death rates are universal ($\sqrt{k_{b\alpha \beta }}=\sqrt{k_{b}}, \sqrt{k_{d\alpha \beta }}=\sqrt{k_{d}}$) regardless
different of genetic traits. 
On the other hand,
the separated square roots from two genders show the natural selection effect, i.e., 
the preferred 
gene or genotype 
has 
a strong tendency to bear more offspring. The amount of
offspring depends on the population 
of both genders, and also on the multiplication product 
of their square root birth rates.

Next we rewrite Eq.(\ref{eq:naturalselect}) in terms of variables $\frac{G}{P}$, where $G$ is any of the population
parameters,\ and then redefine $\frac{G}{P} \to G$. The governing equations for percentage parameters
as \emph{genotype frequencies} for the redefined percentage $G$ 
are:
\begin{equation} \label{eq:naturalselectPerc}
\left\{ 
\begin{array}{l}
\frac{d}{dt}{\small G}_{\alpha \alpha }{\small =(\sqrt{k_{b\alpha \alpha }}G}%
_{\alpha \alpha }\overset{n}{\underset{j=0}{\sum }}\sqrt{{\small k_{b\alpha
j}}}{\small G}_{\alpha j}{\small +} \\ 
\text{ \ \ \ \ \ }\frac{1}{4}\underset{i\neq \alpha }{\underset{i=0}{\overset%
{n}{\sum }}}\underset{j\neq \alpha }{\overset{n}{\underset{j=0}{\sum }}}%
\sqrt{{\small k_{b\alpha i}}}{\small G}_{\alpha i}\sqrt{{\small k_{b\alpha j}%
}}{\small G}_{\alpha j}{\small -k_{d\alpha \alpha }G}_{\alpha \alpha }%
{\small )-} \\ 
\text{ \ \ \ \ \ }{\small G}_{\alpha \alpha }{\small (}\overset{n}{\underset{%
0\leq i\leq j}{\sum }}\sqrt{{\small k_{bij}}}{\small G}_{ij}{\small )}^{2}%
{\small +G}_{\alpha \alpha }{\small (}\overset{n}{\underset{0\leq i\leq j}{%
\sum }}{\small k}_{dij}{\small G}_{ij}{\small )}, \\ 
\frac{d}{dt}{\small G}_{\alpha \beta }{\small =(2\sqrt{k_{b\alpha \alpha }}G}%
_{\alpha \alpha }{\small \sqrt{k_{b\beta \beta }}G}_{\beta \beta }{\small +}
\\ 
\text{ }\underset{i\neq \alpha }{\overset{n}{\underset{i=0}{\sum }}}{\small 
\sqrt{k_{b\alpha i}}G}_{\alpha i}{\small \sqrt{k_{b\beta \beta }}G}_{\beta
\beta }+\underset{j\neq \beta }{\overset{n}{\underset{j=0}{\sum }}}{\small 
\sqrt{k_{b\alpha \alpha }}G}_{\alpha \alpha }{\small \sqrt{k_{b\beta j}}G}%
_{\beta j}{\small +} \\ 
\text{ \ \ \ }\frac{1}{2}\underset{i\neq \alpha }{\underset{i=0}{\overset{n}{%
\sum }}}\underset{j\neq \beta }{\overset{n}{\underset{j=0}{\sum }}}{\small 
\sqrt{k_{b\alpha i}}G}_{\alpha i}{\small \sqrt{k_{b\beta j}}G}_{\beta j}%
{\small -k_{d\alpha \beta }G}_{\alpha \beta }{\small )} \\ 
{\small \ \ \ \ \ }\text{\ }{\small -G}_{\alpha \beta }{\small (}\overset{n}{%
\underset{0\leq i\leq j}{\sum }}\sqrt{{\small k_{bij}}}{\small G}_{ij}%
{\small )}^{2}{\small +G}_{\alpha \beta }{\small (}\overset{n}{\underset{%
0\leq i\leq j}{\sum }}{\small k}_{dij}{\small G}_{ij}{\small )}.%
\end{array}%
\right.
\end{equation}

Neither Eq.(\ref{eq:naturalselect}) nor Eq.(\ref{eq:naturalselectPerc}) are exactly solvable.
Further analysis on the phase diagram of the time-evolving parameter space shows that 
the patterns of time-evolutions and fixed-points 
vary through tuning birth rates or death rates. 
The original
continuous manifold of stable equilibrium solutions Eq.(\ref{eq:Equiln}) will reduce to discrete points once
we tune any birth or death rate. This shows that the \emph{symmetry breaking} of
governing equations results in the discrete degeneracy of stable solutions. 


The discrete
degenerated stable solution will become the \emph{sink, source, or saddle point} of the 
phase diagram (rather than just an \emph{attractor of stable fixed-points}, as in the previous case in Sec.\ref{sec:level4}). These bring up
more interesting and complicated phenomena, 
especially for the case with more genes with a larger parameter space.


\section{Conclusion}

We have proposed a set of time-dependent coupled nonlinear differential equations as the governing equations to
describe a 
class of gene-mating dynamical evolutionary systems within the disciplines of population genetics and evolutionary biology.
Our model consists of the set of governing equations derived from the fundamental assumptions in Sec.\ref{sec:level2}.
The specific models for 
2 alleles and 3 alleles (blood type)
evolutionary systems are derived in Sec.\ref{sec:level3}, 
and the more generic systems with arbitrary $(n+1)$-alleles 
with $((n+1)+{{(n+1)} \choose 2})$-genotypes are studied in Sec.\ref{sec:level4}.
We find the exact analytic solutions for our models, 
where the solutions show their most generic forms in Sec.\ref{sec:level4}. \\ 

Based on the exact analytic solutions, 
we have proved the common properties of gene-mating evolutionary systems: 
(1) global stability, (2) monotonic evolution, and (3) non-extinction and the Hardy-Weinberg law,
under the assumption of no mutation and no natural selection.\\

More generally, our model in Sec.\ref{sec:level4} describes the phenomena of the many-body reaction or the many-body collision process.
In our work, we interpret the governing equations (\ref{eq:Genen-1}) as the population evolution within a given gene pool. 
It may be possible that one can also extend Eq.(\ref{eq:Genen-1}) as certain chemical compounds reactions in the chemical reaction pools ---
such as reactions involving enzymes. 
Another possible interpretation is to regard Eq.(\ref{eq:Genen-1}) as a discretized variant of the Boltzmann equations. 
Further extensions of our work will be reported elsewhere.\\ 

After the completion of our work, 
by searching for similar studies in the literature, we became aware of
a closely relevant model studied in a pioneer work Ref.\cite{Nagylaki} where they obtain the exact
solutions in a different parameterization. 
There is also a theoretical proof in \cite{Akin_Szucs} of a generalized model of Ref.\cite{Nagylaki}, where the 
birth rates (fertilities) and the death rates of different genotypes need not to be the same.\\

There is another study in Ref.\cite{Bloodgroup} concerning the blood groups (blood type) and their Hardy-Weinberg law, nonetheless 
their perspective is rather different from ours.
Ref.\cite{Bloodgroup}'s model is parameterized by phenotype (instead of our genotype) and its model is discrete model (instead of our continuous model).\\

The new ingredients 
for our study are the emphasis on stable fixed-point manifold. 
We believe that our findings of 
(1) a unified parameterization of the stable manifold together with the time-dependent evolution 
and  (2) the Euclidean fiber bundle\footnote{The fiber is bounded by the whole simplex of genotype frequency space, 
which has all coordinates of genotype frequency bounded from 0 to 1. 
Also, the sum of all genotype frequencies is 1. 
The Euclidean fiber is meant to emphasize that the fiber is straight as a Euclidean submanifold with Cartesian coordinates, 
instead of a curved Riemannian curved fiber/submanifold. When we refer to 
Euclidean fiber, we always mean the fiber bounded within the constrained genotype frequency space (within the constrained simplex).
} 
description of exact analytic solutions 
are new to the literature. We hope that 
our work 
can contribute to the population genetics and mathematical biology study.

For future directions, it can be illuminating to generalize our
explicit solutions from without natural selection rules,
to other situations of selection rules, for example, 
when the birth rates or death rates can be varied, 
 in Hofbauer and Sigmund (1988) \cite{HofbauerandSigmund1988}, 
or further generalizations similar to Akin and Szucs (1994) \cite{AkinandSzucs1994},  
Nagylaki and Crow (1974) \cite{Nagylaki} or Jost and Pepper (2008) \cite{JostPepper2008}.

This is the first paper for a sequence of three related studies. 
The second \cite{gene2} and the third paper \cite{gene3} will be reported elsewhere.

\section{Acknowledgements}

JW would like to thank Mehran Kardar and Leonid Mirny for showing great interests and encouragements, and giving comments. 
We thank Jeremy England, Hsien-Ching Kao and Matthew Pinson for comments on the manuscript.
JW wishes to thank Yih-Yuh Chen and Ning-Ning Pang for comments in 2006, and
thank Sze-Bi Hsu for introducing the reference \cite{book1} in 2007 and mentioning the alternative approach: Hirsch's monotone flow, 
for proving the global stability (our independent proof is in Sec.\ref{sec:level4} (1)).
JW thanks Mehran Kardar, 
Patrick Lee and Xiao-Gang Wen for 
encouraging posting the paper.

JW is supported by NSF Grant No.\;DMR-1005541, NSFC 11074140, NSFC
11274192,
the BMO Financial Group and
the John Templeton Foundation. Research at Perimeter Institute is supported by the Government of Canada
through Industry Canada and by the Province of Ontario
through the Ministry of Research.
JW  acknowledges the NSF Grant PHY-1606531. 
This work is also supported by NSF Grant DMS-1607871 ``Analysis, Geometry and Mathematical
Physics'' and Center for Mathematical Sciences and Applications at Harvard University.
JWC is supported in part by the MOST, NTU-CTS, and the NTU-CASTS of R.O.C. 
JWC is partly supported by the Ministry of Science and Technology, Taiwan, under Grant No. 108-2112-M-002-003-MY3 and the Kenda Foundation.


\onecolumngrid

\newpage
\section{Figures \label{sec:fig}}

\begin{figure}[!h] 
\begin{flushleft}
\includegraphics[scale=1.]{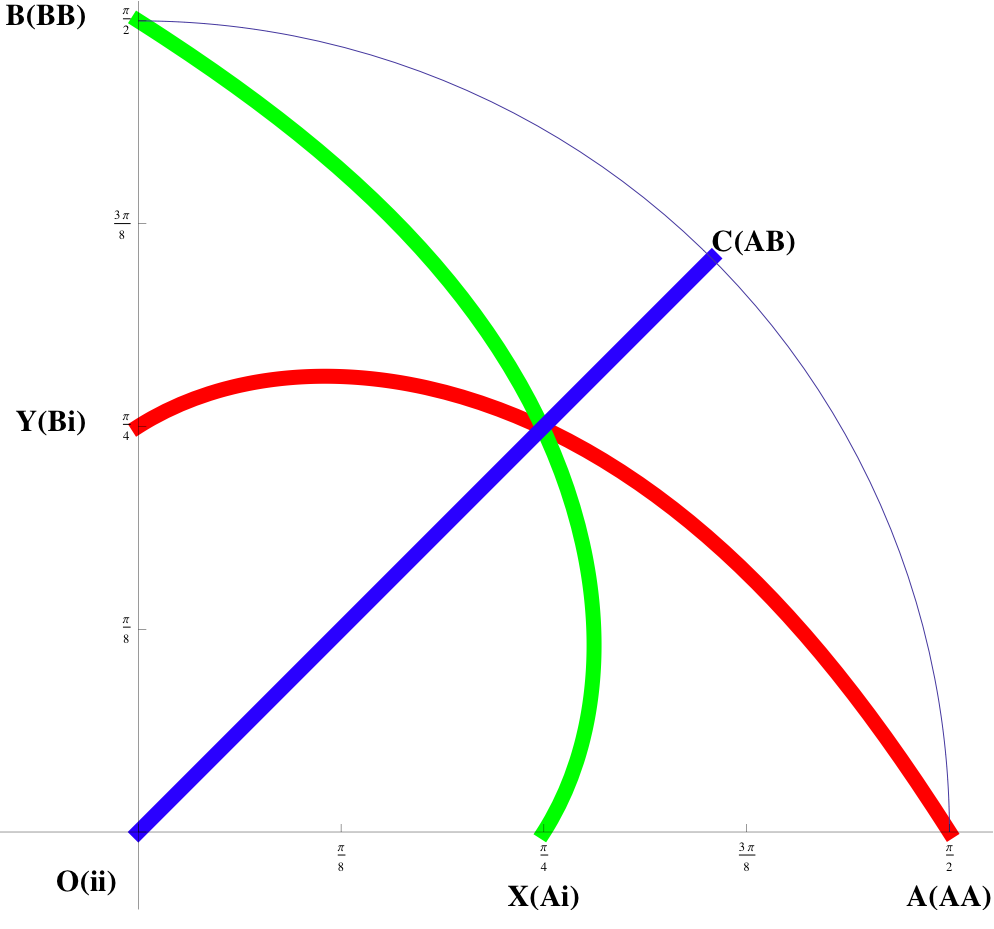} 
\end{flushleft}
\caption{ The quadrant presented here is a 2-dimensional $(\theta_1,\theta_2)$ parameterization of the 2-dimensional stable equilibrium manifold in Eq.(\ref{eq:Equil2})
following the description in Sec.\ref{sec:chart}. 
$\theta _{1}$ represents the radial
direction $(0\leq \theta _{1}\leq \frac{\pi }{2})$, $\theta _{2}$ represents
the angle direction $(0\leq \theta _{2}\leq \frac{\pi }{2})$ of the quadrant. 
The blue line is $\theta _{2}=\frac{\pi }{4}$,
reflecting the symmetry  
invariant under switching $x,a$ to $y,b$.
The green curve is parametrized by $\theta _{1}(\theta _{2})=\frac{(\sqrt{3+\cos (2\theta _{2})+2\sin (2\theta
_{2})}(1+\tan (\theta _{2})))}{\sqrt{2+2\sin (2\theta _{2})}}$,
reflecting the symmetry 
invariant under switching $a,c$ to $o,y$.
The red curve is parametrized by $\theta _{1}(\theta _{2})=\frac{(\sqrt{3-\cos (2\theta _{2})+2\sin (2\theta
_{2})}(1+\cot (\theta _{2})))}{\sqrt{2+2\sin (2\theta _{2})}}$, reflecting the symmetry 
invariant under switching $b,c$ to $o,x$.
The intersection of three color lines, is $(\frac{2}{9},\frac{2}{9},\frac{2}{%
9},\frac{1}{9},\frac{1}{9},\frac{1}{9})$, which is the most symmetric mid point.
} 
\label{fig:1} 
\end{figure}

\center{{\bf FIG.\ref{fig:1}.} Quadrant $(\theta _{1} \cos(\theta _{2}), \theta _{1} \sin(\theta _{2}))$ for the Blood Type Population Ratio Distribution.}

\newpage

\begin{figure}[!h] 
\begin{flushleft}
\includegraphics[scale=1.]{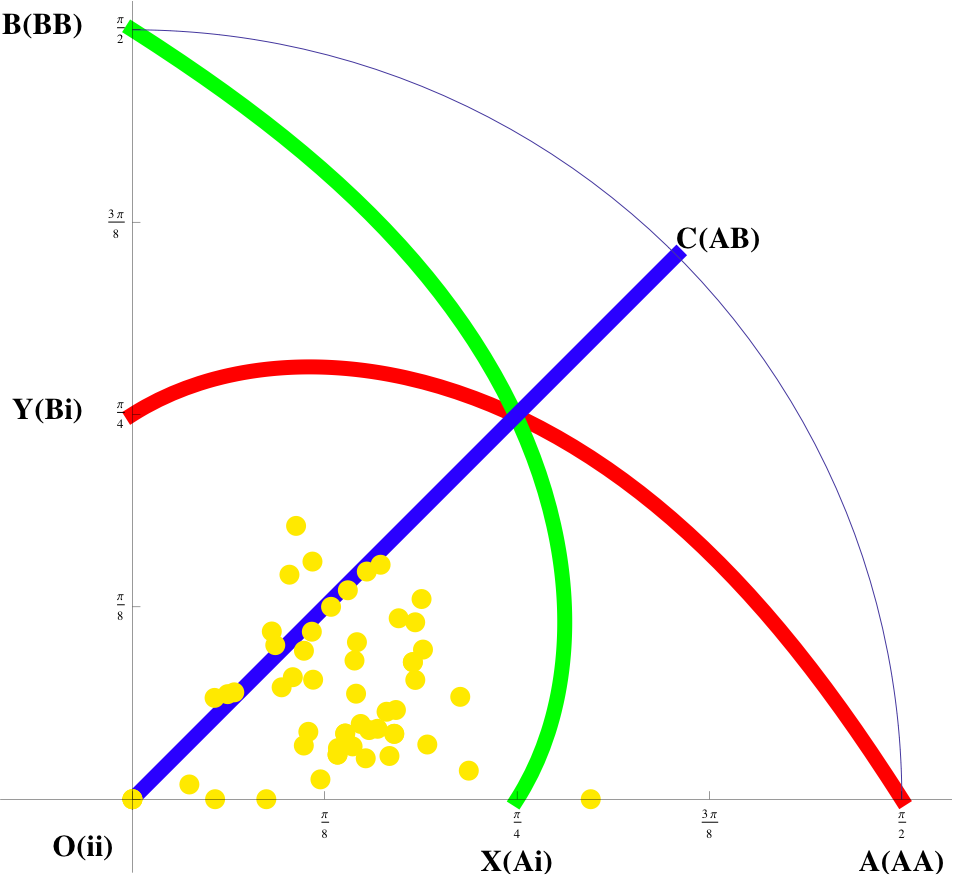} 
\end{flushleft}
\caption{  
The quadrant presented here is a 2-dimensional $(\theta_1,\theta_2)$ parameterization as in FIG.\ref{fig:1}.
We implement our theoretical prediction of the blood type ratio data of ethnic groups in the world from Table \ref{Table2}. We plot the theoretical prediction $(\theta_1,\theta_2)$ distribution of 
the world ethnic groups from Table \ref{Table2} as yellow dots.} 
\label{fig:2} 
\end{figure}

\center{{\bf FIG.\ref{fig:2}.} Quadrant with the Blood Type Population Ratio Distribution of {\bf World Ethnic Groups} from Table  \ref{Table2}.}

\newpage

\begin{figure}[!h] 
\begin{flushleft}
\includegraphics[scale=.65]{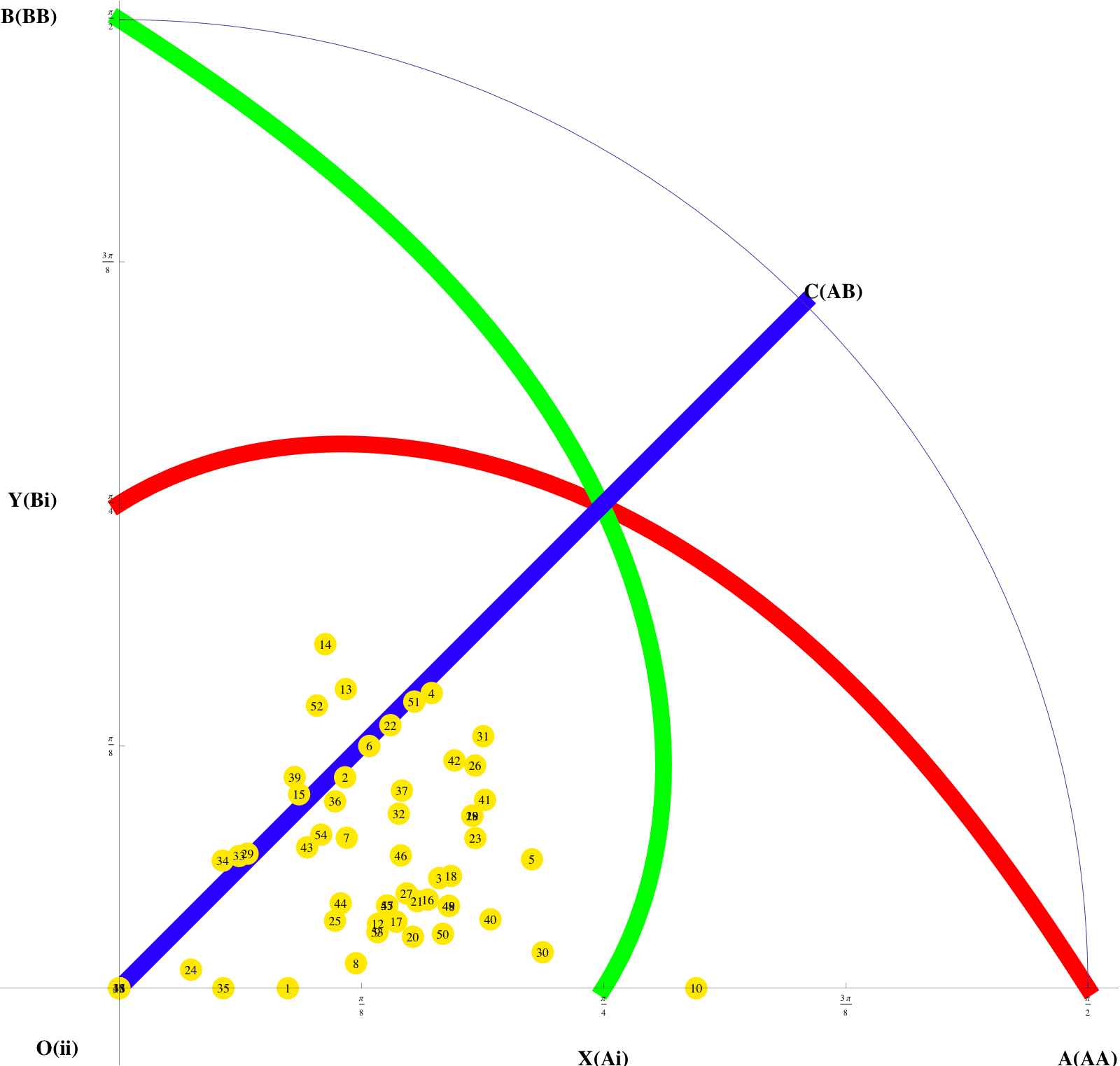} 
\end{flushleft}
\caption{Similar to FIG.\ref{fig:2}. The quadrant presented here is a 2-dimensional $(\theta_1,\theta_2)$ parameterization as in FIG.\ref{fig:1}.
We implement our theoretical prediction of the blood type ratio data of ethnic groups in the world from Table \ref{Table2}. We plot the theoretical prediction $(\theta_1,\theta_2)$ distribution of 
the world ethnic groups from Table \ref{Table2} as yellow dots. The numbers in the yellow dots specify the ethnic groups, numbered in the far-left column of Table \ref{Table2}.} 
\label{fig:3} 
\end{figure}

\center{{\bf FIG.\ref{fig:3}.} Quadrant with the Blood Type Population Ratio Distribution of {\bf World Ethnic Groups} from Table  \ref{Table2}.}

\begin{figure}[!h] 
\begin{flushleft}
\includegraphics[scale=.4]{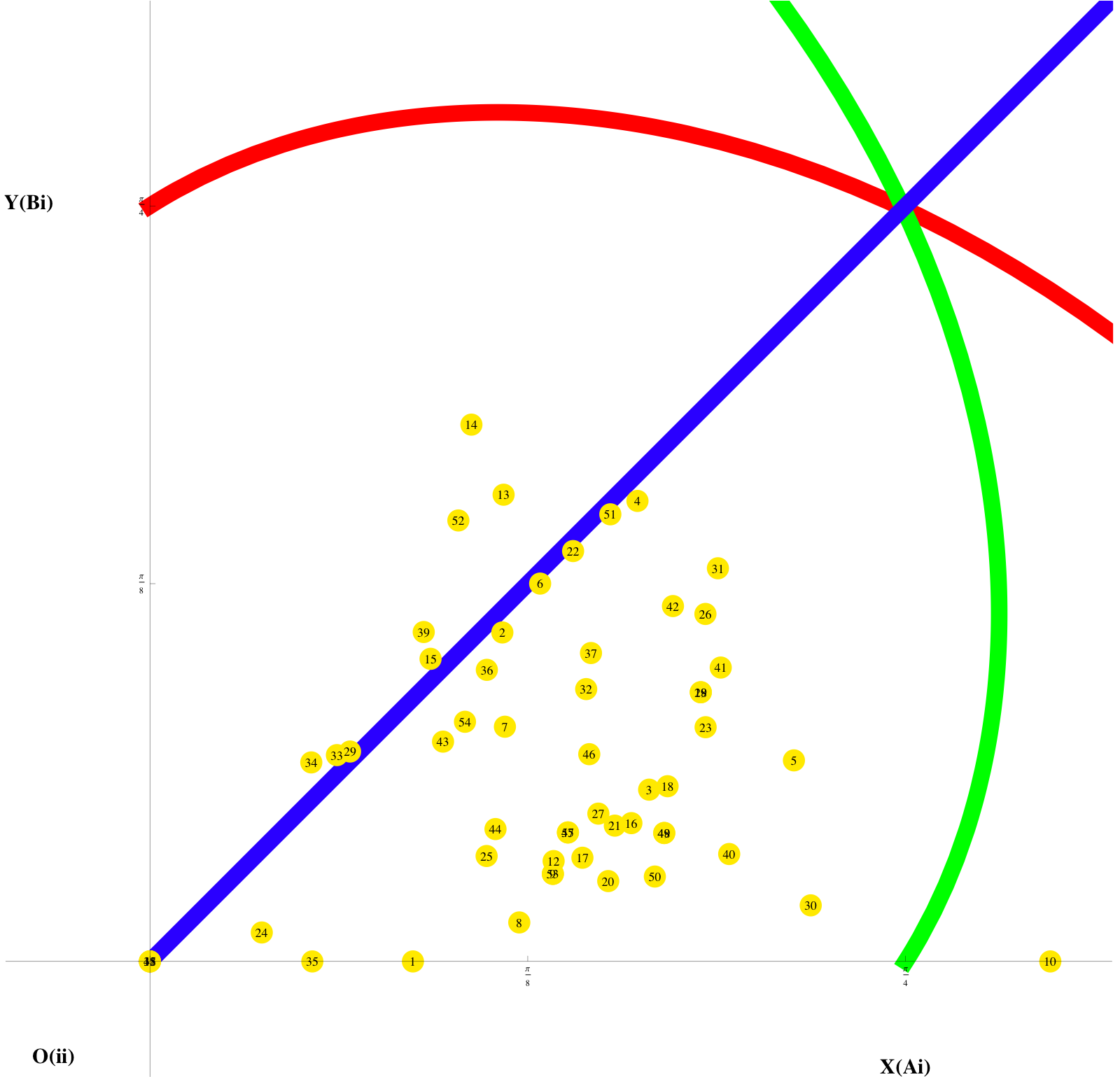} 
\includegraphics[scale=0.4]{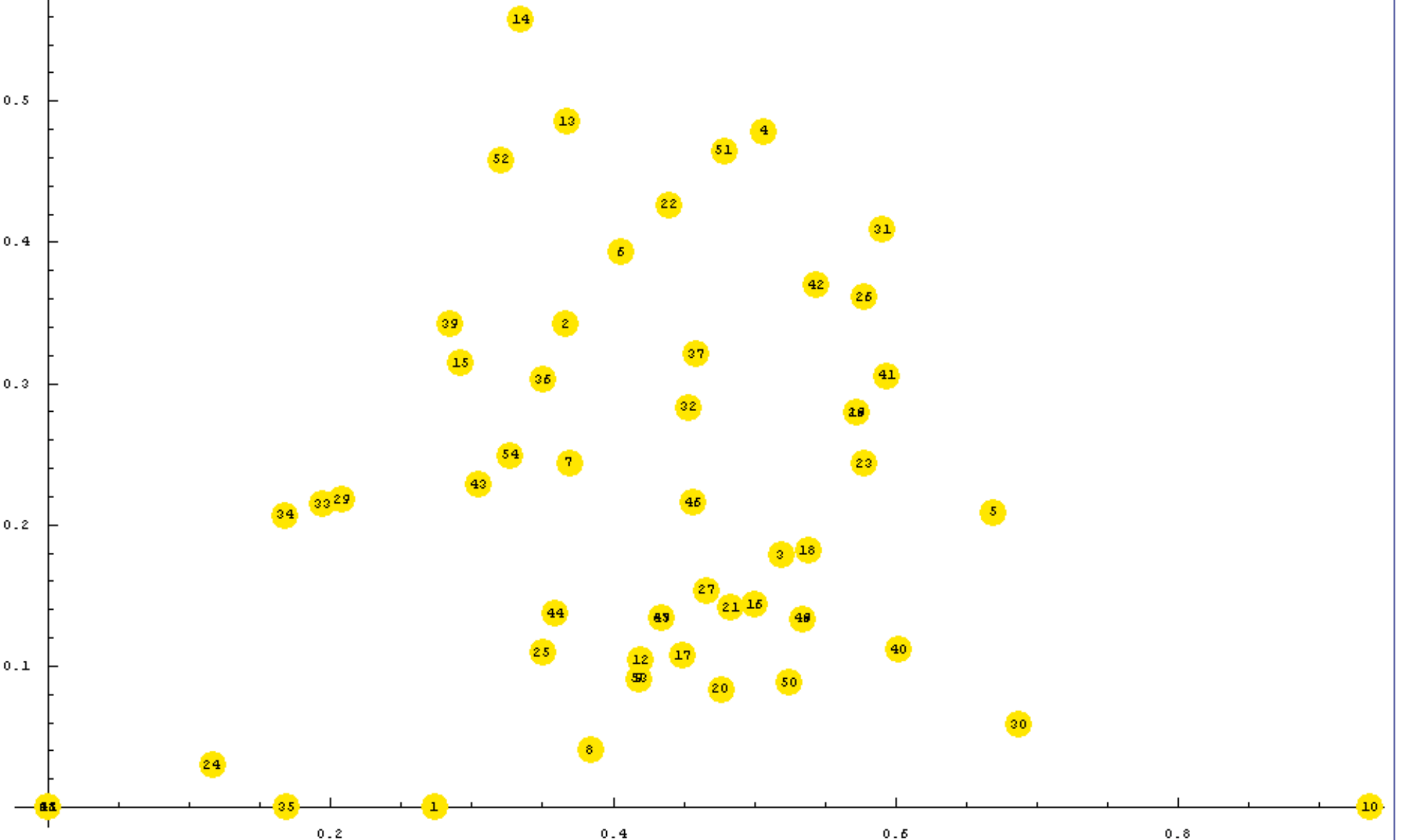} 
\end{flushleft}
\caption{The zoom-in view of FIG.\ref{fig:3}. The set up is the same as FIG.\ref{fig:3}. We implement our theoretical prediction of the blood type ratio data of ethnic groups in the world from Table \ref{Table2}. } 
\label{fig:4} 
\end{figure}

\newpage
\center{{\bf FIG.\ref{fig:4}.} Quadrant with the Blood Type Population Ratio Distribution of {\bf World Ethnic Groups} from Table  \ref{Table2}.}

\begin{figure}[!h] 
\begin{flushleft}
\includegraphics[scale=.95]{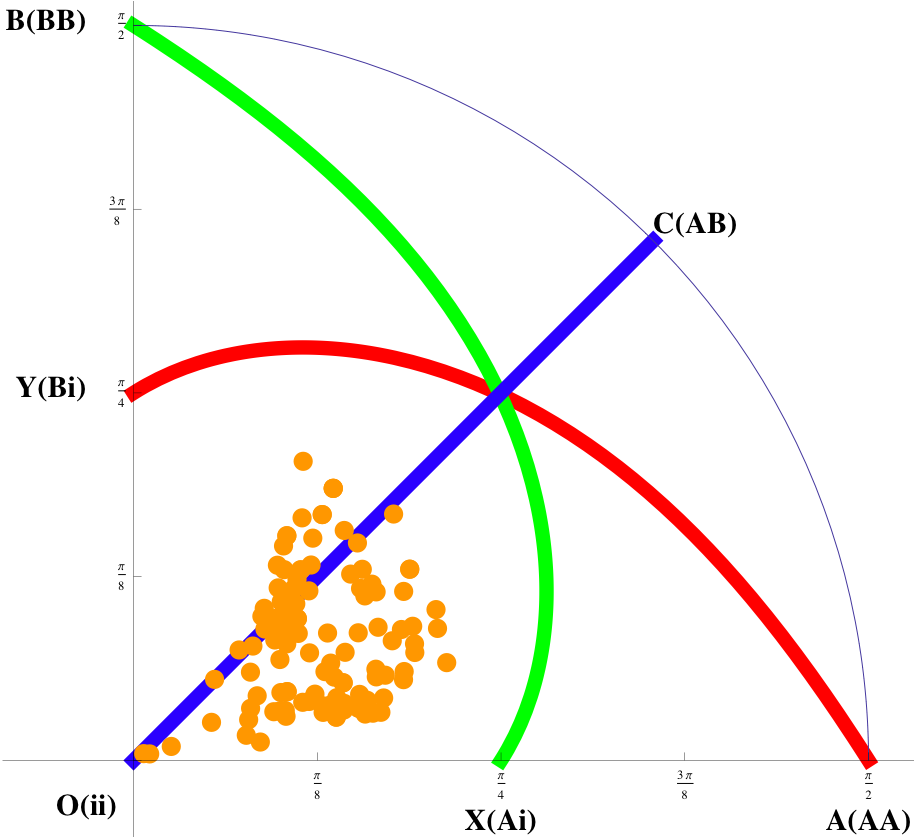} 
\end{flushleft}
\caption{The quadrant presented here is a 2-dimensional $(\theta_1,\theta_2)$ parameterization as in FIG.\ref{fig:1}.
We implement our theoretical prediction of the blood type ratio data of 114 countries in the world from Table \ref{Table5},\ref{Table6}. We plot the theoretical prediction $(\theta_1,\theta_2)$ distribution of 
the world ethnic groups from Table \ref{Table5},\ref{Table6} as orange dots.} 
\label{fig:5} 
\end{figure}

\newpage
\center{{\bf FIG.\ref{fig:5}.} Quadrant with the Blood Type Population Ratio Distribution by {\bf Countries} from Table \ref{Table5},\ref{Table6}.}

\newpage

\begin{figure}[!h] 
\begin{flushleft}
\includegraphics[scale=.65]{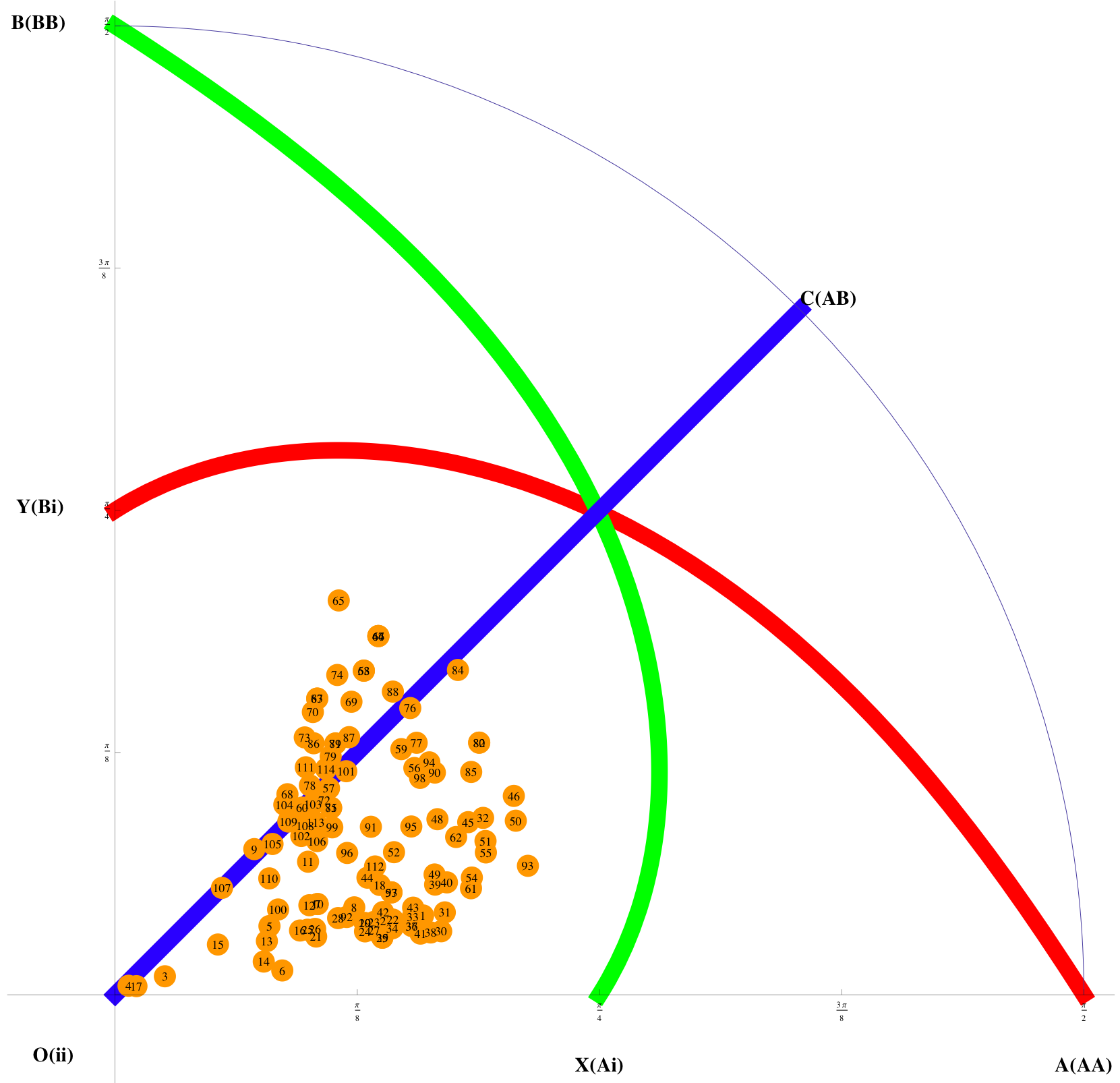} 
\end{flushleft}
\caption{
Similar to FIG.\ref{fig:5}. The quadrant presented here is a 2-dimensional $(\theta_1,\theta_2)$ parameterization as in FIG.\ref{fig:1}.
We implement our theoretical prediction of the blood type ratio data of 114 countries in the world from Table \ref{Table5},\ref{Table6}.   We plot the theoretical prediction $(\theta_1,\theta_2)$ distribution of 
the world ethnic groups from Table \ref{Table5},\ref{Table6} as orange dots. The numbers in the orange dots specify the countries, numbered in the far-left column of Table \ref{Table5},\ref{Table6}.
} 
\label{fig:6} 
\end{figure}

\center{{\bf FIG.\ref{fig:6}.} Quadrant with the Blood Type Population Ratio Distribution by {\bf Countries} from Table \ref{Table5},\ref{Table6}.}

\newpage

\begin{figure}[!h] 
\begin{flushleft}
\includegraphics[scale=.54]{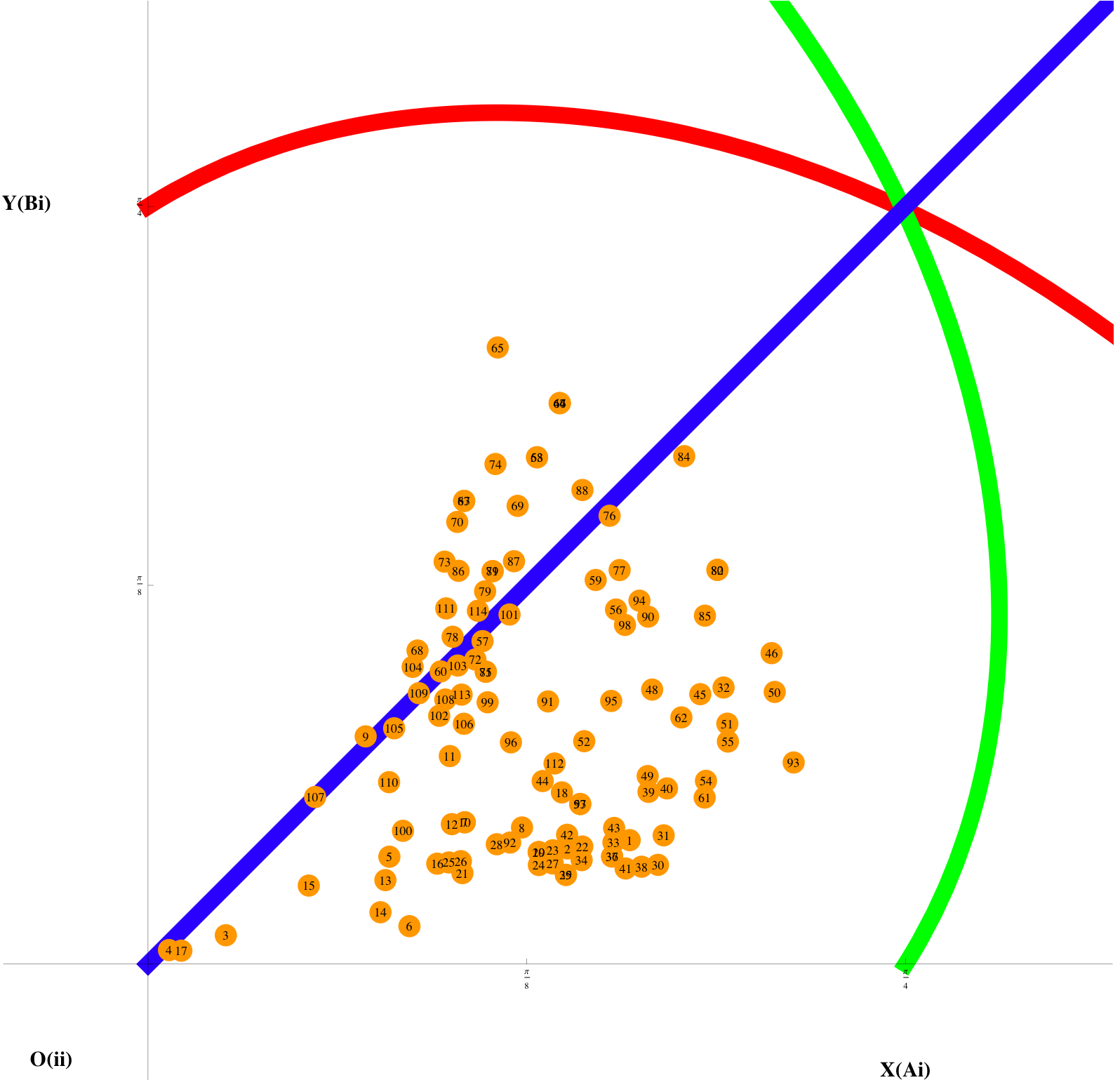} 
\includegraphics[scale=.55]{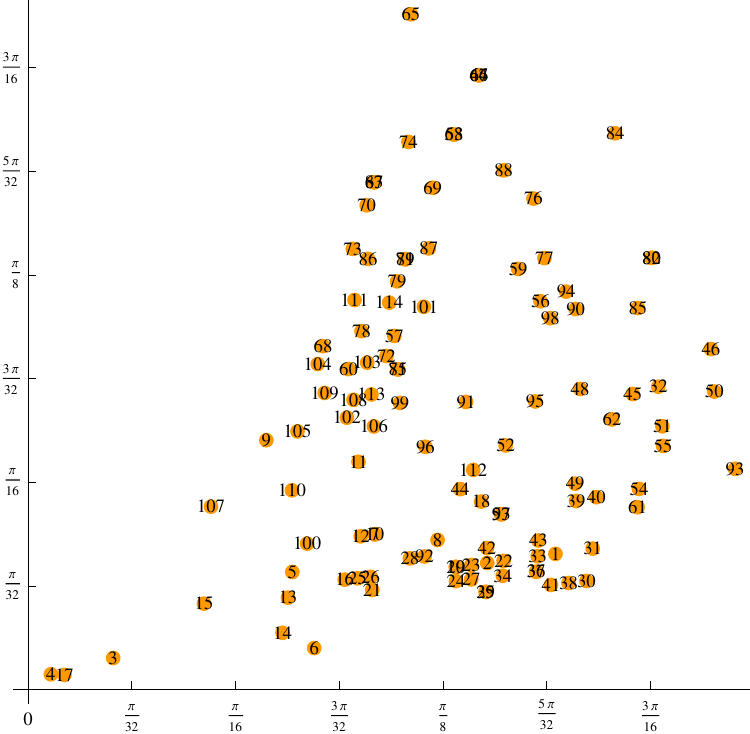} 
\end{flushleft}
\caption{The zoom-in view of FIG.\ref{fig:6}. The set up is the same as FIG.\ref{fig:6}. We implement our theoretical prediction of the blood type ratio data of ethnic groups in the world from Table \ref{Table5},\ref{Table6}. } 
\label{fig:7} 
\end{figure}

\center{{\bf FIG.\ref{fig:7}.} Quadrant with the Blood Type Population Ratio Distribution by {\bf Countries} from Table \ref{Table5},\ref{Table6}.}

\begin{figure}[!h] 
\begin{flushleft}
\includegraphics[scale=.85]{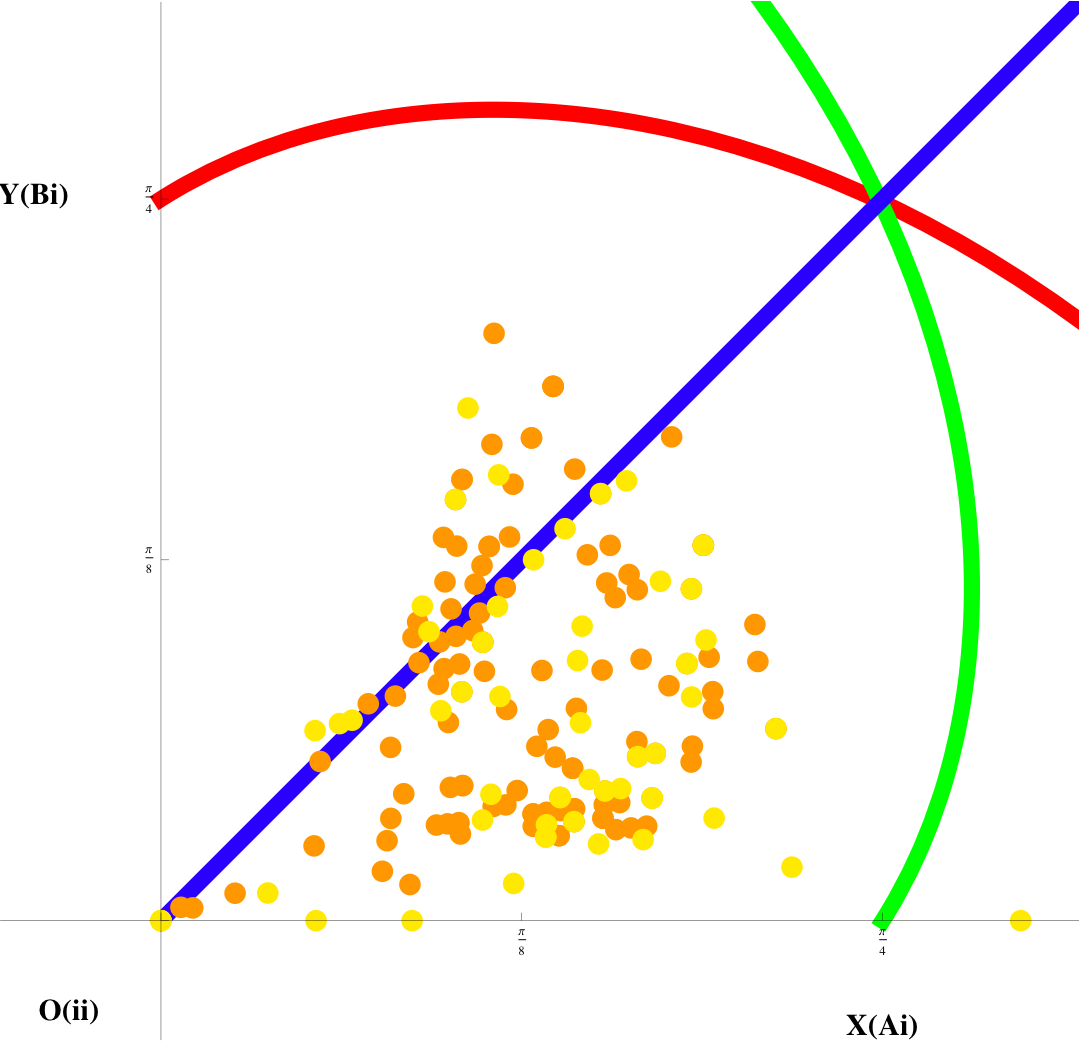} 
\end{flushleft}
\caption{
This is a combined figure of FIG.\ref{fig:2} and FIG.\ref{fig:3}.
The quadrant presented here is a 2-dimensional $(\theta_1,\theta_2)$ parameterization as in FIG.\ref{fig:1}.
We implement our theoretical prediction of the blood type ratio data of ethnic groups in the world from Table \ref{Table2}. We plot the theoretical prediction $(\theta_1,\theta_2)$ distribution of 
the world ethnic groups from Table \ref{Table2} as yellow dots. The numbers in the yellow dots specify the ethnic groups, numbered in the far-left column of Table \ref{Table2}.
We implement our theoretical prediction of the blood type ratio data of 114 countries in the world from Table \ref{Table5},\ref{Table6}. We plot the theoretical prediction $(\theta_1,\theta_2)$ distribution of 
the world ethnic groups from Table \ref{Table5},\ref{Table6} as orange dots.} 
\label{fig:8} 
\end{figure}

\newpage
\center{{\bf FIG.\ref{fig:8}.} 
Blood Type Population Ratio Distribution of both {\bf Ethnics} and {\bf Countries} from Table \ref{Table2},\ref{Table5},\ref{Table6}.}

\begin{figure}[!h] 
\includegraphics[scale=0.5]{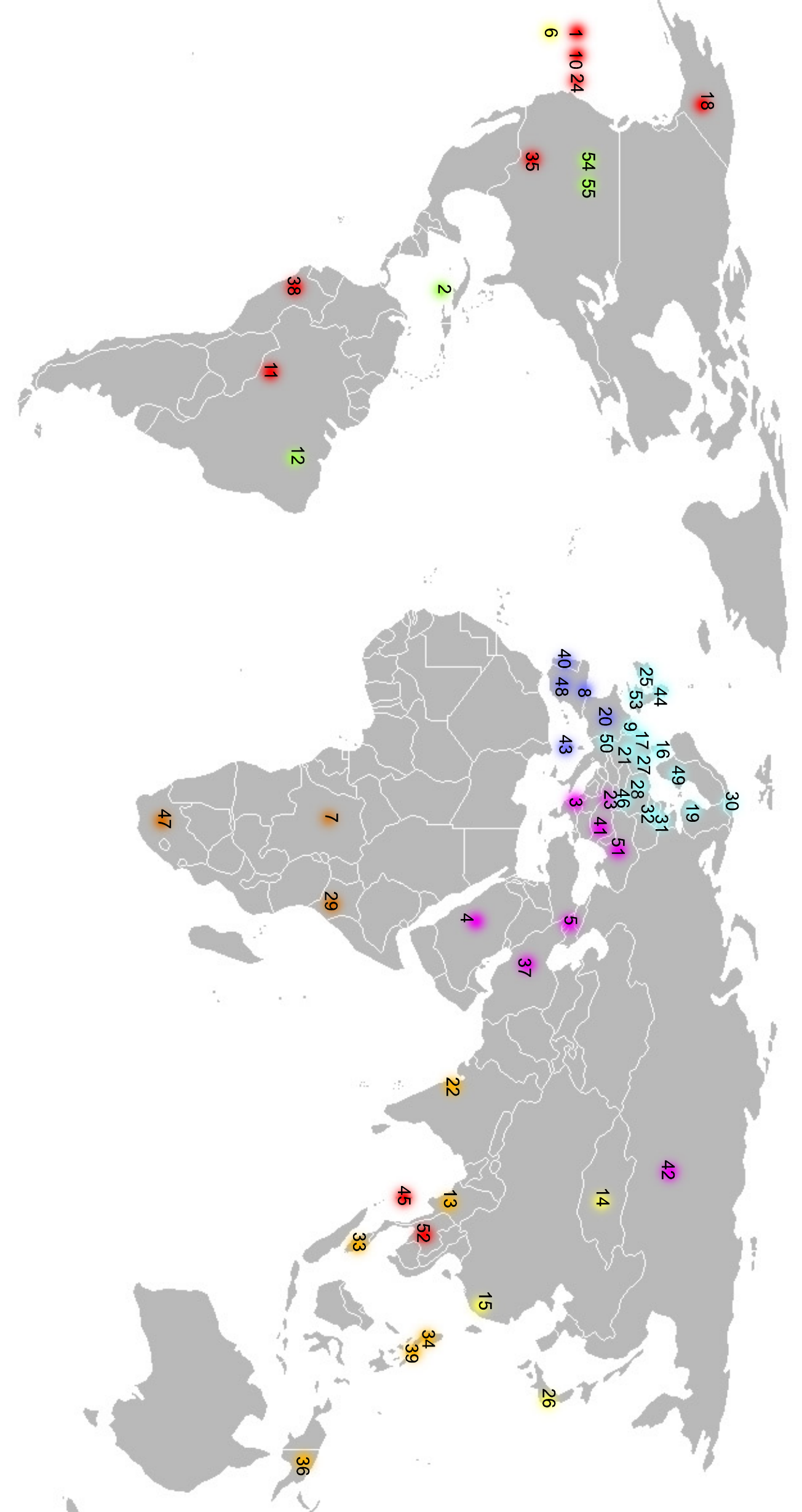} 
\caption{World map with ethnic group data labeled by the numbers in TABLE \ref{Table2}.
The numbers in the colored dots specify the ethnic groups, numbered in the far-left column of Table \ref{Table2}.} 
\label{fig:9} 
\end{figure}

\newpage
\center{{\bf FIG.\ref{fig:9}.}  Geographical Distribution of {\bf World Ethnic Groups} from Table  \ref{Table2}.}

\newpage

\begin{figure}[!h] 
\begin{flushleft}
\includegraphics[scale=.8]{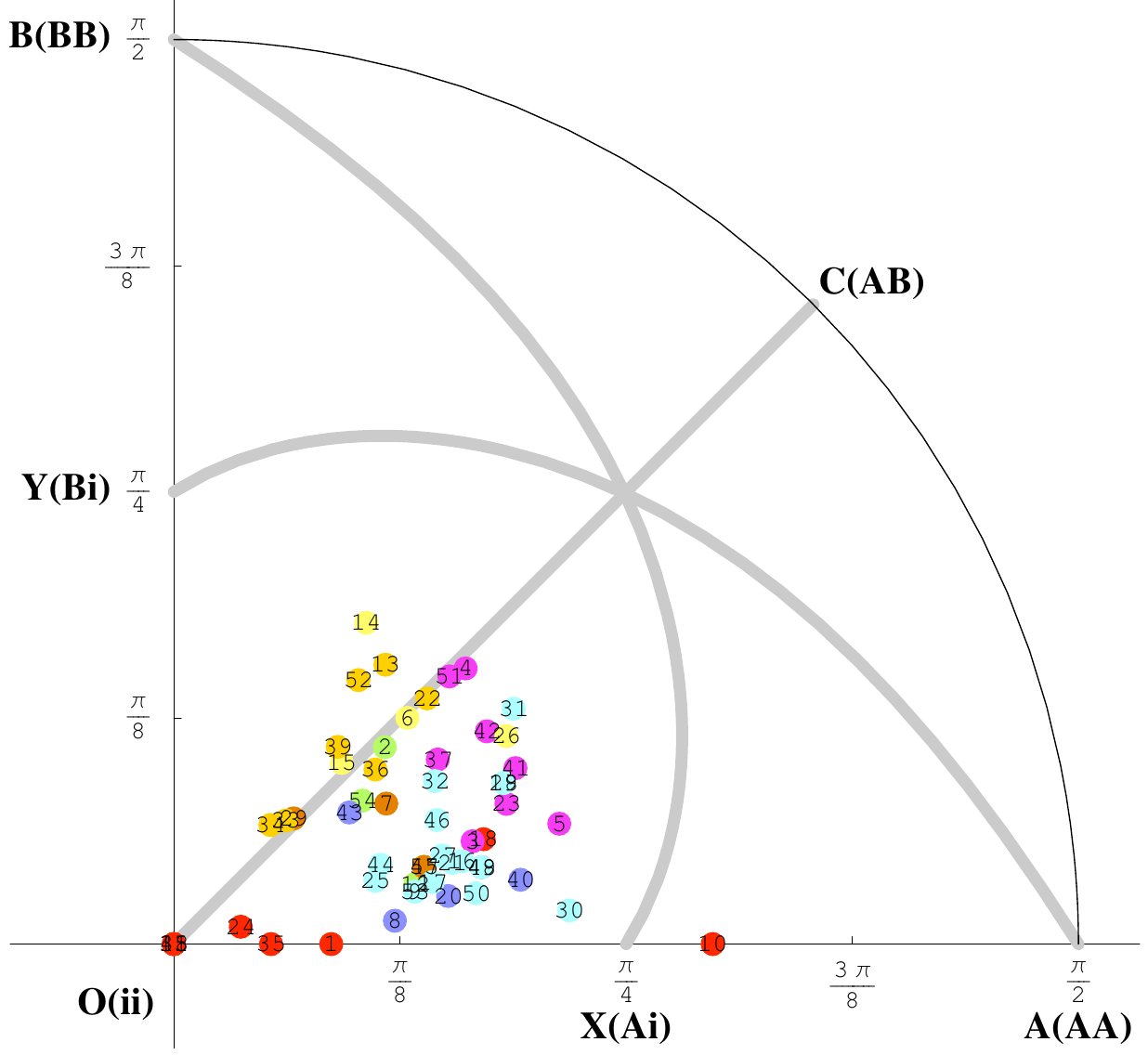} 
\end{flushleft}
\caption{
The quadrant presented here is a 2-dimensional $(\theta_1,\theta_2)$ parameterization as in FIG.\ref{fig:1}.
We implement our theoretical prediction of the blood type ratio data of ethnic groups in the world from Table \ref{Table2}. We plot the  
theoretical prediction $(\theta_1,\theta_2)$ distribution of 
the world ethnic groups from Table \ref{Table2} as colored dots. The numbers in the colored dots specify the ethnic groups, numbered in the far-left column of Table \ref{Table2}.
This FIG.\ref{fig:10} is an almost-equivalent figure as FIG.\ref{fig:3}. 
The only difference is that we implement the colored dots corresponding to the same colored dots geographically of FIG.\ref{fig:9}, for a better visualization and comparison to the 
geographical distribution of ethnic groups in FIG.\ref{fig:9}. 
} 
\label{fig:10} 
\end{figure}

\center{{\bf FIG.\ref{fig:10}.} Quadrant with the Blood Type Population Ratio Distribution of {\bf World Ethnic Groups} from Table  \ref{Table2}.}
%

\newpage
\center{{\bf FIG.\ref{fig:10}.} Quadrant with the Blood Type Population Ratio Distribution of {\bf World Ethnic Groups} from Table  \ref{Table2}.}
\begin{figure}[!h] 
\begin{flushleft}
\includegraphics[scale=.82]{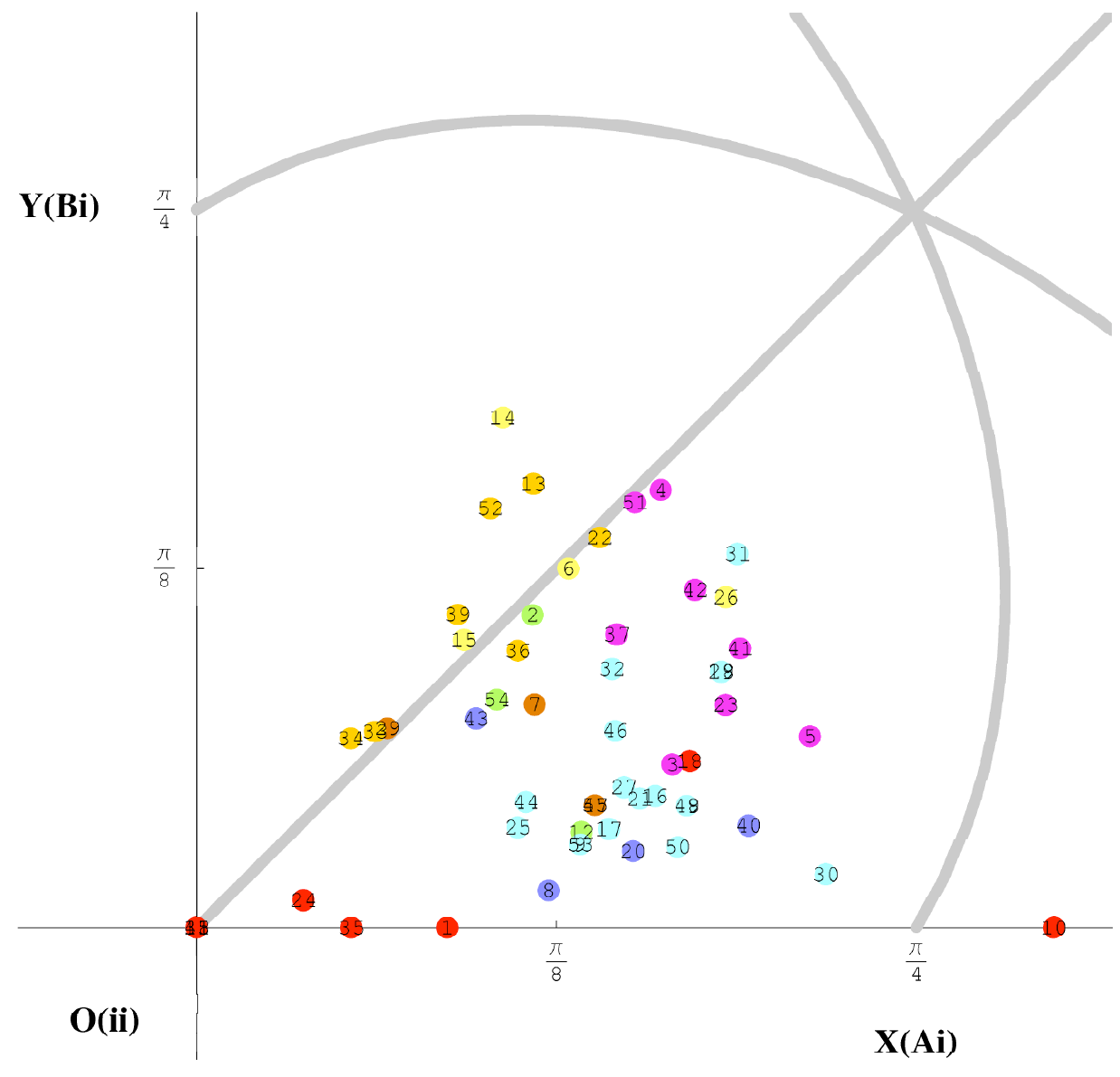} 
\end{flushleft}
\caption{
The zoom-in view of FIG.\ref{fig:10}. The set up is the same as FIG.\ref{fig:10}. We implement our theoretical prediction of the blood type ratio data of ethnic groups in the world from Table \ref{Table2}. 
} 
\label{fig:11} 
\end{figure}



\end{document}